\documentclass[12pt]{article}
\usepackage{mathtools,amssymb,mathrsfs,microtype,ytableau,tikz,slashed}
\usepackage[
 linktoc=all,
 colorlinks=true,
 linkcolor=blue,
 urlcolor=blue,
 citecolor=red,
 ]{hyperref}
\usepackage[backend=bibtex,style=numeric-comp,sorting=none,maxbibnames=99, minbibnames=99]{biblatex}

\usetikzlibrary{decorations.pathmorphing}
\usetikzlibrary{patterns}

\ytableausetup{centertableaux,boxsize=.5em}

\makeatletter\@addtoreset{equation}{section}\makeatother

\DeclareMathOperator{\tr}{tr}

\DeclareMathOperator{\rank}{rank}

\renewcommand{\title}[1]{\vbox{\center\LARGE{#1}}\vspace{5mm}}
\renewcommand{\author}[1]{\vbox{\center\large#1}\vspace{5mm}}
\newcommand{\address}[1]{\vbox{\center\em#1}}

\begin{document}

\begin{titlepage}
\begin{center}
\vspace{5mm}
%\hfill {\tt HU-EP-09/40}\\
\hfill {\tt }\\
\vspace{8mm}
\title{\makebox[\textwidth]{\fontsize{30}{40}\selectfont RG flows in 2d QCD}}
\vspace{10mm}
Diego Delmastro,${}^a$ \footnote{\href{mailto:ddelmastro@scgp.stonybrook.edu}{\tt ddelmastro@scgp.stonybrook.edu}}
Jaume Gomis${}^b$ \footnote{\href{mailto:jgomis@perimeterinstitute.ca}{\tt jgomis@perimeterinstitute.ca}}
\vskip 7mm
\address{
${}^a$Simons Center for Geometry and Physics,\\
SUNY, Stony Brook, NY 11794, USA}
\address{
${}^b$Perimeter Institute for Theoretical Physics,\\
Waterloo, Ontario, N2L 2Y5, Canada}

\end{center}

\vspace{5mm}
\abstract{
\noindent $2d$ QCD, Yang-Mills theory with gauge group $G$ and massless quarks in representations $(R_\ell, R_r)$ of $G$, flows in the infrared to a CFT or a TQFT depending on whether spectrum is gapless or gapped. We identify the infrared effective theory and construct the explicit RG flow map between the QCD operators in the UV and the IR, in particular identifying which operators create massive excitations and decouple in the IR, and those that create massless excitations or vacua and remain as nontrivial IR  operators. We determine the exact scaling dimensions of the QCD operators that remain in the IR, which generically acquire large anomalous dimensions. We also study QCD deformed by various operators in the ultraviolet (such as mass terms or four-fermi interactions), and determine the theory that emerges at low energies. We conjecture and provide some evidence for QCD deformed by various operators developing a nonperturbative fermion condensate that dynamically breaks the gauge symmetry $G$, thus explicitly realizing an old idea known as ``tumbling" in an exactly soluble setting. 
}
 
\vfill\eject

%\vspace{20pt}

{\hypersetup{linkcolor=black}
\tableofcontents
\thispagestyle{empty}
}

\end{titlepage}

\section{Introduction}
\setcounter{footnote}{0}

Given a collection of degrees of freedom and interactions at short distances, captured by a microscopic Hamiltonian $H_\text{UV}$, one would like to answer the following increasingly detailed and complex questions:
\begin{enumerate}
    \item Is the spectrum of the Hamiltonian gapped or gapless?
    \item What is the  effective field theory description ${\mathcal T}_{\text{IR}}$ in the deep IR? Specifically,
    \begin{itemize}
        \item if gapless, what is the infrared conformal field theory (CFT$_{\text{IR}}$)?
        \item if gapped, what is the infrared topological field theory (TQFT$_{\text{IR}}$)?
    \end{itemize}
    
    \item What is the renormalization group flow map ${\mathcal O}_{\text{UV}}\rightarrow {\mathcal O}_{\text{IR}}$ between UV and IR operators?
     \begin{itemize}
     \item Which operators ${\mathcal O}_{\text{UV}}$ are in the kernel of this map and thus decouple in the IR?
     \item  What is the IR dimension $\Delta_{\text{IR}}$ of the operator ${\mathcal O}_{\text{IR}}\in{\mathcal T}_{\text{IR}}$ in the image of ${\mathcal O}_{\text{UV}}$?
    \end{itemize}
    \item What is the IR description of the  Hamiltonian deformed by a UV operator $H_{\text{UV}}+\lambda  {\mathcal O}_{\text{UV}}$?
\end{enumerate}
Answering these questions reliably is often an out of reach, nonperturbative problem.
\smallskip

Building on previous work~\cite{Delmastro:2021otj}, in this paper we answer these questions for $2d$ QCD theories,\footnote{See~\cite{Donahue:2019adv,Cherman:2019hbq,Komargodski:2020mxz,Dempsey:2021xpf,Anand:2021qnd,Popov:2022vud,Cherman:2022ecu,Dempsey:2022nys,Dempsey:2022uie} for related recent work on two-dimensional gauge theories.} where a $2d$ QCD theory is specified by a choice of gauge group $G$ and a pair of representations $R_\ell$ and $R_r$ of $G$ acting on left and right chiral massless quarks. Furthermore, we argue that QCD deformed by various operators ${\mathcal O}_{\text{UV}}$ dynamically generates a fermion condensate that breaks the gauge symmetry $G$, in a two-dimensional version of the ``tumbling" picture in~\cite{RABY1980373}.

In QCD, the gauge coupling $g$ triggers a renormalization group flow from a conformal field theory of free quarks in the UV, which we denote $\text{CFT}_\text{UV}$, to a scale invariant theory in the deep IR that we denote ${\mathcal T}_\text{IR}$:
\begin{figure}[!h]
\centering
\begin{tikzpicture}
\node[right] at (5,5) {$\text{CFT}_\text{UV}$};
\node[right,align=left] at (5,3.2) {${\mathcal T}_\text{IR}=\begin{cases}
\text{CFT}_\text{IR}\\[+2pt]
\text{TQFT}_\text{IR}
\end{cases}
$};
%\text{CFT}_\text{IR}$\\($\text{TQFT}_\text{IR}$)};
%\node[right] at (5.6,4.2) {$\text{RG flow}$};
\draw[thick,->,>=stealth,decoration={snake,amplitude=1.5pt,post length=3pt},decorate] (5.4,5-.3) -- (5.4,3.3+.3);
%\node[right] at (9,5) {$\varphi^\text{UV}$};
%\draw[thick,->,>=stealth,decoration={snake,amplitude=1.5pt,post length=3pt},decorate] (5.4+4,5-.3) -- (5.4+4,1+.3);
%\draw[thick,->,>=stealth,decoration={snake,amplitude=1.5pt,post length=3pt},decorate] (5.4+4+.2,5-.3) to[in=180,out=-90,looseness=1.5] (5.4+4+2,3+.3);
%\node[right] at (5+4,1) {$\varphi^\text{IR}$};
%\node[right] at (11.5,3.3) {decouple};
\end{tikzpicture}
\caption{QCD induces an RG flow from a free fermion $\text{CFT}_\text{UV}$ in the UV to ${\mathcal T}_\text{IR}$ in the IR. The infrared theory is a conformal field theory, or a topological field theory if there is a mass gap.}
\label{fig:RG_flow}
\end{figure}

\noindent
Given a QCD theory, the IR theory ${\mathcal T}_\text{IR}$ is either a $\text{CFT}_\text{IR}$ or a $\text{TQFT}_\text{IR}$, depending on whether the QCD Hamiltonian is gapless or gapped. The classification of $2d$ QCD theories that are gapped (and therefore also of those that are gapless) was derived in~\cite{Delmastro:2021otj} by analyzing the temporal and lightcone Hamiltonians of QCD.

Given an operator ${\mathcal O}_{\text{UV}}$ in CFT$_{\text{UV}}$ of dimension $\Delta_{\text{UV}}$, we would like to know what happens to ${\mathcal O}_{\text{UV}}$ in the IR under the renormalization group flow triggered by the QCD gauge coupling. This is controlled by the large distance behavior of its connected two-point function 
\begin{equation}
\langle{\cal O}^\dagger_{\text{UV}}(x){\cal O}_{\text{UV}}(0)\rangle_{\text{c}}\xrightarrow{x\to\infty} \begin{cases}
e^{-mx}\,, &\text{decouples in the IR}\\[+4pt]
x^{-2\Delta_\text{IR}}\,,&\text{survives as a nontrivial operator in the IR.} 
\end{cases}
\label{largedistance}
\end{equation}
When the correlator exhibits power law behavior,   ${\mathcal O}_{\text{UV}}$ flows to    a nontrivial operator ${\mathcal O}_{\text{IR}}$ of dimension $\Delta_{\text{IR}}$ in ${\mathcal T}_\text{IR}$. On the other hand, when the decay is exponential,  ${\mathcal O}_{\text{UV}}$ decouples in the IR, and maps to the trivial operator in ${\mathcal T}_\text{IR}$. This defines the renormalization group map 
\begin{equation}
({\mathcal O}_{\text{UV}},\Delta_{\text{UV}}) \rightsquigarrow ({\mathcal O}_{\text{IR}}, \Delta_{\text{IR}})\,.
\label{RGmap}
\end{equation}
When the theory is gapped, there are finitely-many operators ${\mathcal O}_{\text{IR}}$  in $\text{TQFT}_\text{IR}$, all have  $\Delta_{\text{IR}}=0$ and  are   relevant in the IR,  even though most flowed from highly irrelevant operators in the UV.\footnote{We will encounter many examples of this, in particular  adjoint QCD, which flows to a TQFT with an exponentially large number of operators, which flow from UV operators that we explicitly identify (see section~\ref{sec:adjointQCD}).} When the theory is gapless, the number of operators ${\mathcal O}_{\text{IR}}$  in $\text{CFT}_\text{IR}$ is infinite, as they furnish representations of a chiral algebra ${\cal A}_\text{IR}$ and antichiral algebra ${\overline {\cal A}}_\text{IR}$ that contain at least the Virasoro algebra. The spectrum of dimensions $\{\Delta_{\text{IR}}\}$ is part of the  $\text{CFT}_\text{IR}$ data that we would like to determine. In general,  UV operators acquire large anomalous dimensions along the RG flow. %\footnote{In the RG flow triggered by the QCD gauge coupling, we show that $\Delta_{\text{UV}}\geq \Delta_{\text{IR}}$.}
 Of course, computing the large distance behavior of the correlators~\eqref{largedistance} involves strongly coupled dynamics and is generally out of reach.\footnote{An example where this computation can be   carried explicitly and   the derivation of the operator map using correlators  is possible is for $2d$ $U(1)$ gauge theory with $N$ Dirac fermions of charges $q_i$. See for example~\cite{Georgi:2019tch,Georgi:2020jik,Georgi:2022sdu} for some recent work.}
 
The map $({\mathcal O}_{\text{UV}},\Delta_{\text{UV}}) \rightsquigarrow ({\mathcal O}_{\text{IR}}, \Delta_{\text{IR}})$, for an arbitrary $2d$ QCD theory, can be obtained as follows. The free fermion theory $\text{CFT}_\text{UV}$ in  the UV admits a chiral algebra ${\mathcal A}_{\text{UV}}=SO(\dim(R_\ell))_1$ and an antichiral algebra ${\overline{\mathcal A}}_{\text{UV}}=SO(\dim(R_r))_1$~\cite{Witten:1983ar} (see also~\cite{Ji:2019ugf}). An operator $\mathcal O_\text{UV}$ in $\text{CFT}_\text{UV}$ can be decomposed into its chiral and antichiral parts, namely, it can be written as the product of chiral fermions $O_\text{UV}$ and antichiral fermions $\overline O_\text{UV}$:
\begin{equation}
\mathcal O_\text{UV}=O_\text{UV}\times \overline O_\text{UV}\,.
\end{equation}
Since QCD is obtained by gauging the symmetry $G$ acting on the free fermions of $\text{CFT}_\text{UV}$, we have a group embedding
\begin{equation}
G\subset SO(\dim(R_\ell))\,,\qquad G\subset SO(\dim(R_r))\,.  
\end{equation}
Importantly, this embedding can be extended to an embedding of chiral algebras,
\begin{equation}\label{eq:intro_embed}
G_{I(R_\ell)}\subset SO(\dim(R_\ell))_1\,,\qquad G_{I(R_r)}\subset SO(\dim(R_r))_1\,,  
\end{equation}
where $I(R)$ is the Dynkin index of the representation $R$.\footnote{The index is defined as $\tr(t^a_Rt^b_R)=I(R)\delta^{ab}$, with $t_R$ the generators of $G$ in the representation $R$. Cancellation of gauge anomalies in $2d$ QCD requires that $I(R_\ell)=I(R_r)$.} The algebra $SO(\dim R)_1$ is generated by the free chiral quarks  $\psi$, and the subalgebra $G_{I(R)}$ is generated by the gauge currents $J=\psi^\dagger t_R \psi$. From now on, we shall focus on the chiral half only, as the discussion goes through verbatim for the antichiral part. At the end, one must combine both chiralities and construct gauge invariant operators.

An operator $O_\text{UV}$ constructed out of chiral fermions corresponds to a state in the  module of the $SO(\dim R)_1$ current algebra. A crucial fact in what follows is that an arbitrary operator $O_\text{UV}\in SO(\dim R)_1$ admits  a \emph{unique} decomposition as the product of an operator in $G_{I(R)}$, times an operator in the coset chiral algebra\footnote{The fact that $2d$ QCD has an $SO(\dim R)_1/G_{I(R)}$ chiral algebra at any value of the gauge coupling can be derived in multiple ways, including by analyzing the Hamiltonians~\cite{Delmastro:2021otj,Kutasov:1994xq}, by showing that the generators of the current algebra remain chiral for $g  \neq 0$ in the quantum theory, or by realizing $2d$ QCD as a $3d$ theory on a slab with suitable boundary conditions. We discuss this latter perspective in appendix~\ref{ap:3d}.} $SO(\dim R)_1/G_{I(R)}$
 \begin{equation}
\begin{alignedat}{4}
&\qquad  O_\text{UV}&&=\qquad\quad  O&&\ \ \times &&\quad ~~ \tilde{  O} \\[-1ex]
&\qquad\tikz{\node[rotate=270]{$\in$};}&&\hspace{48pt} \tikz{\node[rotate=270]{$\in$};}&&&&\quad~\, \tikz{\node[rotate=270]{$\in$};}\\[-1.5ex]
&SO(\dim(R))_1&& ~~~~\frac{SO(\dim(R))_1}{G_{I(R)}} &&\ \ \times &&\quad ~~ G_{I(R)}
\end{alignedat}
\label{decomposa}
\end{equation}
where $\tilde O$ is an operator constructed out of copies of the gauge current $J=\psi^\dagger t_R \psi$ only, and $O$ is the remainder. This decomposition allows us to make a sharp statement about  which  operators $O_\text{UV}$ decouple  at large distances. By studying the Hamiltonians of QCD,  it was shown in~\cite{Delmastro:2021otj} that an operator $O_\text{UV}$ creates massless excitations if and only if it is annihilated by the gauge current $J$, that is, if and only if it is a primary with respect to the $G_{I(R)}$ subalgebra. Phrasing this statement in terms of the decomposition~\eqref{decomposa}, we conclude that
\begin{equation}\label{eq:OUV_decouple_survive}
%\begin{aligned}
%{ O}_{\text{UV}}&={ O} \times {\tilde { O}}=(\text{anything} \times \text{primary})  \overset{\text{IR}}{\longrightarrow}  { O_{\text{IR}}}\,,\\[+9pt]
%{O}_{\text{UV}}&={O} \times {\tilde { O}}=(\text{anything} \times \text{descendant)}  \overset{\text{IR}}{\longrightarrow} \text{decouples in IR}\,. 
O_\text{UV}=O \times \tilde{O}\overset{\text{IR}}{\longrightarrow}\begin{cases}O_{\text{IR}} & \tilde O=1\\ \text{decouples} & \tilde O\neq1\end{cases} 
%\end{aligned}
\end{equation}
Any  operator  ${O}_{\text{UV}}$ that involves one or more currents of the gauged $G_{I(R)}$ current algebra in its decomposition~\eqref{decomposa} decouples at large distances. Conversely, any operator ${O}_{\text{UV}}$ that contains no factors of $J$ survives as a nontrivial operator in the IR. The information about the specific $G_{I(R)}$ primary operator for a given $O_\text{UV}$ is carried by the factor $O$, while $\tilde O$ determines whether $O_\text{UV}$ decouples or not.

In~\eqref{decomposa}, the quotient $SO/G$ denotes a gauged WZW model, also known as a Goodard-Kent-Olive coset model~\cite{Goddard:1986ee}. The operator decomposition~\eqref{decomposa} is neatly encoded in the branching rules of the embedding $G_{I(R)}\subset SO(\dim R)_1$,
\begin{equation}\label{eq:intro_branch_rules}
d(q)=\sum_{\lambda} b_{\lambda}(q)  \chi_\lambda(q)\,.
\end{equation}
Here, $d(q)$ is the character of $SO(\dim R)_1$,\footnote{$SO(\dim R)_1$ is a holomorphic spin CFT and has a unique character for each spin structure.} while $\chi_\lambda(q)$ is a character of $G_{I(R)}$, with $\lambda$ running over (a subset of) the integrable representations thereof. The coefficients $b_\lambda(q)$ in the decomposition are the characters of the chiral algebra $SO(\dim R)_1/G_{I(R)}$. Each state in the module of $SO(\dim R)_1/G_{I(R)}$ labeled by $\lambda$ gives rise a nontrivial operator $O_{\text{IR}}$. The $q$-expansion of $b_\lambda(q)$ then classifies the chiral operators $O_\text{IR}$. The spectrum of infrared operators $\mathcal O_\text{IR}$ involves combining the chiral and antichiral halves.\footnote{The $SO(\dim R_\ell)_1/G_{I(R_\ell)}$ chiral algebra itself corresponds to $\mathcal O_\text{UV}$ operators with $\tilde O=\overline O=\overline {\tilde O}=1$ (same goes for the antichiral algebra $SO(\dim R_r)_1/G_{I(R_r)}$).}

This completely determines the RG flow map $ {\mathcal O}_{\text{UV}}  \rightsquigarrow  {\mathcal O}_{\text{IR}}$ and fixes the dimension $\Delta_{\text{IR}}$ of any operator that survives in the IR purely algebraically, without having to resort to Feynman diagrams or  other traditional methods. 
%The decomposition (\ref{decomposa}) implies that  $h(O_\text{UV})=h(O)+h(\tilde O)$, where $h$ denotes the (chiral) conformal dimension. 
The decomposition~\eqref{decomposa} implies that the conformal dimension of an operator ${\mathcal O}_{\text{UV}}$ that survives in the infrared is $\Delta_{\text{IR}}=\Delta_{\text{UV}}-\Delta_{(\lambda,\bar\lambda)}$, where $\lambda/\bar\lambda$ labels the $G_{I(R_\ell)}/G_{I(R_r)}$  primary operator in the decomposition of ${\mathcal O}_{\text{UV}}$. This in particular implies that in the RG flows triggered by QCD  the dimensions of operators that survive in the IR are strictly non-increasing, that is $\Delta_{\text{UV}}\geq \Delta_{\text{IR}}$.\footnote{The operators ${\mathcal O}_{\text{UV}}$ whose dimension does not change along the flow, that is $\Delta_{\text{UV}}= \Delta_{\text{IR}}$, are fermion operators such that $O_\text{UV}$ and $\overline O_\text{UV}$ are separately gauge invariant, that is, the product of a generator of $SO(\dim R_\ell)_1/G_{I(R_\ell)}$ with a generator of $SO(\dim R_r)_1/G_{I(R_r)}$.}

Are these \emph{all} the operators that remain in the IR? The analysis in~\cite{Delmastro:2021otj} correctly captures all the propagating massless degrees of freedom, but it may have missed some vacua. In other words, we cannot rule out the presence of a decoupled topological sector that is not described by the coset $SO(\dim R)_1/G_{I(R)}$. We will assume that this is not the case, i.e., we propose that the infrared effective description of QCD is given by the coset CFT/TQFT
 \begin{equation}\label{eq:intro_conjecture}
{\mathcal T}_\text{IR}=\frac{SO(\dim(R_\ell))_1}{G_{I(R_\ell)}}\times\frac{SO(\dim(R_r))_1}{G_{I(R_r)}}\,.
\end{equation}
This claim is correct under the assumption that the deep IR limit of a QCD theory is the strict $g^2\rightarrow \infty$ limit, as discussed in~\cite{Delmastro:2021otj} (see also~\cite{Kutasov:1994xq,Komargodski:2020mxz}).\footnote{This is equivalent to the RG interface being trivial in the $3d$ picture realization of $2d$ QCD (see appendix~\ref{ap:3d}).} The idea is that, in the $g^2\rightarrow \infty$ limit, the Lagrangian of QCD becomes $\mathcal L_\text{IR}=\bar\psi\slashed D\psi$, and  the gauge field simply behaves as a Lagrange multiplier setting the gauge currents to zero. It is well-known that this is the gauged WZW model description of the coset theory~\eqref{eq:intro_conjecture}~\cite{Witten:1983ar,Goddard:1986ee,Kent:1985bn,Kutasov:1994xq}.

Before moving on, let us present a simple example that illustrates the discussion so far. If we consider the QCD theory with gauge group $G_2$ and a Majorana fermion in the seven-dimensional representation, then the spectrum is gapless~\cite{Delmastro:2021otj}, and the infrared CFT is the (fermionic) $c=7/10$ minimal model, i.e., %(spin) tri-critical Ising model:\footnote{Summing over spin structures it yields the tricritical Ising model $M(5,4)$.}
\begin{equation}
\mathcal T_\text{IR}=\frac{SO(7)_1}{(G_2)_1}=\hbox{spin tri-critical Ising.}
\end{equation}

\begin{table}
\begin{equation*}
\begin{array}{c|c|c}
O_\text{IR}&\phi_{r,s}&h\\ \hline
1&\phi_{1,1}&0\\
\epsilon&\phi_{1,2}&1/10\\
\epsilon'&\phi_{1,3}&3/5\\
\epsilon''&\phi_{1,4}&3/2
\end{array}\qquad
\begin{aligned}
&\begin{array}{c|c|c|c}
O_\text{IR}&\phi_{r,s}&h\\ \hline
\sigma&\phi_{2,1}&7/16\\
\sigma'&\phi_{2,2}&3/80
\end{array}\\
&\vphantom{\int}
\end{aligned}
\end{equation*}
\caption{Neveu-Schwarz and Ramond sectors of the fermionic tri-critical Ising model, with $c=7/10$. $O_\text{IR}$ are the primaries of the chiral algebra, also denoted as $\phi_{r,s}$ in the Ka\v{c} table; $h$ is the scaling weight.}
\label{fig:tri_critical}
\end{table}

We now consider the mapping of operators~\eqref{RGmap} under the RG flow. By decomposing the $SO(7)_1$ characters into $(G_2)_1$ characters (see section~\ref{sec:tri_critical} for the details) we find that the chiral fermion composites $O_\text{UV}=\psi^a$, with $a=0,1,2,3$, flow at large distances to the Virasoro primaries $O_\text{IR}=\phi_{1,a+1}$ (see table~\ref{fig:tri_critical}). The full spectrum of Neveu-Schwarz operators follows by combining both chiralities, and is:\footnote{One can also look at the Ramond sector, but such operators are not genuine local operators, but rather they are live at the end of the topological line $(-1)^F$.} %These operators descend from non-local UV operators that involve the fermion zero modes $\oint \psi$.}
\begin{equation}
\begin{alignedat}{6}
&\text{bosonic:}\ &&\quad 1 && \rightsquigarrow\ 1 \qquad
\ \psi_\ell\psi_r&&\rightsquigarrow \epsilon\bar\epsilon \qquad
\psi^2_\ell\psi^2_r&&\rightsquigarrow \epsilon'\bar\epsilon' \qquad
\psi^3_\ell\psi^3_r&&\rightsquigarrow \epsilon''\bar\epsilon''\\[+4pt]
&\text{fermionic:}\quad && \psi^2_\ell\psi_r&&\rightsquigarrow \epsilon'\bar\epsilon \qquad
\psi_\ell\psi^2_r&&\rightsquigarrow \epsilon\bar\epsilon' \qquad
\ \psi^3_\ell&&\rightsquigarrow \epsilon'' \qquad
\quad \psi^3_r&&\rightsquigarrow \bar\epsilon''\,.
\end{alignedat}
\label{mapppp}
\end{equation}
Any other UV operator either decouples at large distances, or it flows to a Virasoro descendant of one of the fields above. For example,  $D_+\psi_\ell\psi_r\rightsquigarrow \partial_+\epsilon\bar\epsilon$ is a descendant, while the gauge invariant operators $\psi_\ell^4,\psi_r^4$ both decouple (as they can be written in terms of the gauge currents as $J_\ell^2,J_r^2$). Using this map, one can compute the renormalized scaling dimension of operators, such as for example $\Delta_\text{IR}(\psi_\ell\psi_r)=1/5$. We note the interesting fact that, for $\psi_\ell^3$ and $\psi_r^3$, the scaling dimension is not renormalized $\Delta_\text{IR}=\Delta_\text{UV}=3/2$. The scaling dimension of these operators is protected because they are chiral. Furthermore, in the IR these operators can be interpreted as a supercurrent (the  spin tri-critical Ising model is known to be the first element of the $\mathcal N=1$ minimal series). This means that this gauge theory has emergent supersymmetry at large distances! We shall discuss other gauge theories with emergent supersymmetry in section~\ref{sec:tri_critical}.

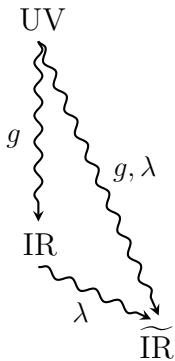
\begin{figure}[!h]
\centering
\begin{tikzpicture}

\node[right] at (5,5) {$\text{UV}$};
\node[right] at (5.05,1+1) {$\text{IR}$};
\node[right] at (6.6,-.3+1) {$\widetilde{\text{IR}}$};
\node[left,scale=.9] at (5.3,3.4) {$g$};
\node[left,scale=.9] at (7.1,3) {$g,\lambda$};
\node[left,scale=.9] at (6.2,1.1) {$\lambda$};
\draw[thick,->,>=stealth,decoration={snake,amplitude=1.5pt,post length=3pt},decorate] (5.4,5-.3) -- (5.4,1+.3+1);

\draw[thick,->,>=stealth,decoration={snake,amplitude=1.5pt,post length=3pt},decorate] (5.4,5-.3) -- (7,.05+1);

\draw[thick,->,>=stealth,decoration={snake,amplitude=1.5pt,post length=3pt},decorate] (5.4,1.7) -- (6.9,0+1);

\end{tikzpicture}
\caption{RG-flow from an initial $\text{CFT}_\text{UV}$. The first flow is triggered by the gauge coupling constant $g$, and the end result is $\mathcal T_\text{IR}$. The second flow is triggered by the gauge coupling constant plus a deformation $\lambda\mathcal O_\text{UV}$, and the end result is a new effective theory $\widetilde{\mathcal T}_\text{IR}$. For small deformation, this is equivalent to first flowing to $\mathcal T_\text{IR}$, and then turning on the infrared image of $\lambda\mathcal O_\text{UV}$, namely $\lambda\mathcal O_\text{IR}$.}
\label{fig:deformation_flow}
\end{figure}

We are now ready to tackle the last of the questions we set forth above: what happens to a QCD theory if we deform the UV Hamiltonian by some UV operator? We would like to determine what is the IR limit of the theory described by the deformed Hamiltonian
\begin{equation}
{H}_{\text{QCD}}+\lambda {\mathcal O}_{\text{UV}}\,.
\label{deformedQCD}
\end{equation}
We answer this question in the regime where  $|\lambda|$ is small at the scale set by the gauge coupling $g$. Then the endpoint of the RG flow of the deformed QCD theory~\eqref{deformedQCD} is the IR limit of 
\begin{equation}
{\mathcal T}_\text{IR}  +\lambda {\mathcal O}_{\text{IR}}\,,
\label{deformedQCDIR}
\end{equation}
where ${\mathcal O}_{\text{IR}}$ is the image of ${\mathcal O}_{\text{UV}}$ in ${\mathcal T}_\text{IR}$. Effectively, the RG flow occurs in two steps, first the QCD theory flows to ${\mathcal T}_\text{IR}$, and then  ${\mathcal T}_\text{IR}$ is deformed by ${\mathcal O}_{\text{IR}}$, see figure~\ref{fig:deformation_flow}. We can leverage the knowledge about the endpoint of RG flows of coset theories deformed by various relevant operators to give an explicit answer to what happens in the IR of deformed QCD.

We go further and conjecture that the endpoint of deformed QCD RG flows as determined reliably by the RG flow of $ {\mathcal T}_\text{IR}  +\lambda {\mathcal O}_{\text{IR}}$ is explained by QCD developing a nonperturbative condensate. We demonstrate in various examples that there always exists a channel where  a fermion condensate breaks the gauge group $G$ and gives mass to degrees of freedom in such a way that what remains in the infrared is precisely the IR limit of~\eqref{deformedQCDIR}. Although we cannot derive the condensate from microscopics, we show via multiple examples that there is always a condensate that reproduces the rigorously derived answer. 

In the $G_2$ gauge theory example discussed above, which flows to the spin tri-critical Ising model,  there are two relevant infrared directions, corresponding in the UV gauge theory to the mass term $\psi_\ell\psi_r$ and the four-fermi term $\psi_\ell^2\psi_r^2$. These map to the infrared CFT operators $\epsilon\bar\epsilon$ and $\epsilon'\bar\epsilon'$, respectively. The effect of the mass term is rather simple  --- it gaps the gauge theory --- which precisely matches the answer provided by the IR CFT. Indeed, the tri-critical Ising model deformed by $\epsilon\bar\epsilon$ is known to be gapped~\cite{Zamolodchikov:1987jf}.% and whose RG flow is described by the massive $E_7$ integrable theory. 

The fate of the gauge theory deformed by the four-fermi operator is more interesting. This is described by the endpoint of the spin tri-critical Ising model deformed by $\epsilon'\bar\epsilon'$. It is known that a minimal model deformed by $\phi_{1,3}$ (for one choice of sign of the coupling) flows to the next minimal model~\cite{osti_6989163,PhysRevB.30.3908,LUDWIG1987687}. In the case at hand, the spin tri-critical Ising model deformed by $\epsilon'\bar\epsilon'$ flows to the spin critical Ising model, i.e., a free Majorana fermion (recall that the Majorana CFT summed over spin structures is equivalent to the Ising CFT). This means that the original QCD theory, with the four-fermi interaction turned on, becomes a free fermion at large distances. This has a nice interpretation as the Goldstino of the spontaneously broken emergent supersymmetry discussed above. This free fermion can also be understood directly in terms of the UV variables: if we assume that a $G_2$-charged fermion bilinear condenses, $\langle\psi_\ell\psi_r\rangle\neq0$, this breaks the gauge group down to its stabilizer. If we take the condensate, for example, in the direction $\boldsymbol 7\subset\boldsymbol 7\times\boldsymbol 7$,\footnote{We study other possible symmetry breaking patters in section~\ref{sec:tri_critical_def}. For example, deforming by $\phi_{1,3}$ with the opposite sign (which is known to lead to a gapped theory) is explained by the UV theory via a condensate in the direction $\boldsymbol{27}\subset\boldsymbol 7\times\boldsymbol 7$. This Higgses the gauge group down to $SU(2)^2$ with a massless fermion in the representation $(\boldsymbol1,\boldsymbol3)$, which is a gapped gauge theory~\cite{Delmastro:2021otj}.} then the symmetry breaking pattern is $G_2\to SU(3)$, while the matter field breaks into a triplet and a singlet of $SU(3)$. The former picks up a mass and decouples, while the latter is identified with the free fermion we found before. Thus the RG flow of the deformed gauge theory can be explained by the gauge theory developing a gauge-symmetry-breaking condensate. We find that such a mechanism explains all the relevant deformations of QCD we study.

It is also important to consider what happens if we turn on deformations that are marginal or irrelevant in the infrared. The latter are rather straightforward: irrelevant deformations do not affect the large distance behaviour of the theory. An interesting case to study is the $T\bar T$ deformation which, despite being irrelevant, leads to a solvable and intricate structure in the theory~\cite{Smirnov:2016lqw,Cavaglia:2016oda}. For QCD theories, the UV operator $(T\bar T)_\text{UV}$ flow in the IR to $(T\bar T)_\text{IR}$ operator, which means $T\bar T$ deformations are easy to follow along the RG flow. It will be interesting to study such deformations in the context of QCD in more details in future work.

Finally, let us mention exactly marginal deformations. These are very exceptional and are not expected to exist unless they come from a flavor symmetry. Indeed, a theory with a flavor symmetry has spin $1$ currents all along the RG flow, which allow for the construction of exactly marginal operators. Intuitively, such deformations should correspond to changes in the shape and size of the infrared sigma model for the flavor symmetry. For example, theories with $U(1)$ flavor symmetries have a marginal deformation corresponding to changes in the radius of the target space circle. As we shall argue in section~\ref{sec:marginal_def}, the general case is rather similar, with marginal deformations essentially acting on the maximal torus of the flavor symmetry.

The plan for the rest of this work is as follows. In section~\ref{eq:IR_ops} we explain how to classify the operators that survive the infrared limit in several illustrative examples. This involves writing down the list of the operators that cannot be factorized into products of the gauge currents; we work out several examples in a brute-force way, and also show how the branching rules of the embedding~\eqref{eq:intro_embed} give rise to the same answer in a more systematic way. We find that some gauge theories have emergent supersymmetry at large distances, and make some conjectures about this phenomenon in more general theories. In section~\ref{sec:deformation} we analyze the problem of deforming the ultraviolet theory by some of its operators, such as adding mass terms or four-fermi interactions. Again, we go through multiple examples that demonstrate the general idea. In all these examples, we argue that the ``tumbling'' heuristic of~\cite{RABY1980373} correctly captures the rigorously derived answer. In section~\ref{sec:chiral} we turn our attention to chiral theories, i.e., QCD theories where the left- and right-moving quarks transform under distinct representations of the gauge group. While in many aspects these theories are essentially identical to their vector-like analogues, some features are new. For example, we find that gapped theories are intimately related to the existence of (fermionic) holomorphic modular invariants of the infrared CFT. Finally, we leave some technical details to appendices;~\ref{ap:3d} describes a $3d$ construction of our $2d$ theories that clarifies some of its properties, and~\ref{ap:spin_ops} extends the analysis of section~\ref{sec:qcd_fund}.

\section{Operator mapping under RG flow}\label{eq:IR_ops}

In this section we describe the problem of figuring out which short-distance operators decouple at long distances, and which survive. For the latter, we will also discuss their image in the infrared CFT/TQFT.

Let $\mathcal O_\text{UV}$ be an operator defined in the UV. We would like to determine whether this operator survives in the infrared limit or not. As discussed in the previous section, the operator $\mathcal O_\text{UV}$ survives if and only if it is a primary with respect to the subalgebra $G_{I(R)}$ generated by the gauge current $J^a={:}\psi^\dagger t^a_R\psi{:}$, where $t_R$ are the generators of $G$ in the representation $R$ under which the matter transforms. If we factor the operator into its chiral components $\mathcal O_\text{UV}=O_\text{UV}\times\overline O_\text{UV}$, then it survives at long distances if and only if the OPEs
\begin{equation}
J_\ell^a(x)O_\text{UV}(0),\qquad J_r^a(\bar x)\overline O_\text{UV}(0)
\end{equation}
have simple poles only, and no higher order singularities. Equivalently, $\mathcal O_\text{UV}$ decouples if and only if either of its chiral components can be written as
\begin{equation}\label{eq:JO}
O_\text{UV}={:}J^a_\ell O^a{:},\qquad \overline O_\text{UV}={:}J^a_r \overline O{}^a{:}
\end{equation}
for some operators $O^a,\overline O{}^a$.

Consider, for example, QCD with gauge group $G=Spin(N)$ and a fermion in the spinor representation (say, with $N$ odd). Then, any operator quadratic in $\psi$ can be written as a linear combination of terms of the form $\psi \gamma^{A_1A_2\cdots A_p}\psi$, where $\gamma^A$ are the gamma matrices of $Spin(N)$ (with $A$ a vector index) and $\gamma^{A_1A_2\cdots A_p}:=\gamma^{[A_1}\gamma^{A_2}\cdots \gamma^{A_p]}$. Of course, these operators are all independent for $p=0,1,\dots,\lfloor N/2\rfloor$.\footnote{The operator $\psi \gamma^{A_1A_2\cdots A_p}\psi$ actually vanishes if $\gamma^{A_1A_2\cdots A_p}$ happens to be symmetric, due to fermi statistics. Using the reality properties of the gamma matrices, it is easily checked that this matrix is symmetric (anti-symmetric) if and only if $\frac{1}{8} (N^2+4 N p+4 p^2+8 p+7)$ is odd (even).} The operator with $p=2$, i.e., $\psi \gamma^{A_1A_2}\psi$, is in fact the gauge current, because $\gamma^{A_1A_2}$ is the generator of the spinor representation of $Spin(N)$. Therefore, we conclude that the gauge invariant, Lorentz scalar quartic operator
\begin{equation}
\mathcal O_\text{UV}=\psi_\ell \gamma^{A_1A_2\cdots A_p}\psi_\ell\,\psi_r \gamma_{A_1A_2\cdots A_p}\psi_r
\end{equation}
can be written in terms of the gauge currents if and only if $p=2$. The $p=2$ operator thus decouples in the infrared, while the rest $p\neq2$ all survive the infrared limit. This also follows from a simple OPE computation, since one can easily check that $\mathcal O_\text{UV}$ has a second-order pole $\sim \tr(\gamma^{A_1A_2\cdots A_p}\gamma^{B_1B_2})/x^2$, which is non-zero for $p=2$ only. Hence, only $p=2$ is a descendant, and the rest $p\neq2$ are all $Spin(N)$-primaries.

One can perform a similar analysis for other gauge theories. Using the group-theoretic properties of the representation $R$ of $G$, one can often figure out whether a given operator $\mathcal O_\text{UV}$ can be written in terms of the gauge current or not, just by looking at the quantum numbers of the different pieces inside $\mathcal O_\text{UV}$. Alternatively, one can simply compute the OPE $J^a(x)\mathcal O_\text{UV}(0)$ and check whether there are higher order poles or not. That being said, for operators with many insertions of $\psi$, this can become quite cumbersome. As described in the introduction, the branching rules~\eqref{eq:intro_branch_rules} of the embedding
\begin{equation}\label{eq:GinSO}
SO(\dim R)_1\supseteq G_{I(R)}
\end{equation}
encode the spectrum of $G_{I(R)}$ primaries inside $SO(\dim R)_1$. Therefore, one can also use these rules to classify all UV operators that survive the IR limit, in a rather algorithmic way. In the remainder of this section we will illustrate these considerations by working out several examples.

\subsection{QCD with fundamental matter}\label{sec:qcd_fund}

Consider QCD with gauge group $U(N)$ and $N_F$ massless Dirac fermions in the fundamental representation. The flavor symmetry of the system is $SU(N_F)_\ell\times SU(N_F)_r$.\footnote{More precisely, the symmetry is $(SU(N_F)_\ell\times SU(N_F)_r)/\mathbb Z_{N_F}$. We will not be careful with such discrete quotients in this work as they play no significant role.} We denote the left-moving fermions as $\psi_\ell$, and the right-moving ones as $\psi_r$; under the flavor symmetry they transform as $(\ydiagram1,\boldsymbol1)$ and $(\boldsymbol1,\overline{\ydiagram1})$, respectively. Under the gauge group, they both transform as $\ydiagram1$ with charge $+1$. As we shall see, this theory flows in the infrared to the CFT ${\mathcal T}_{\text{IR}}=U(NN_F)_1/U(N)_{N_F}\equiv SU(N_F)_N$ WZW model.

What we would like to obtain here is the list of operators that survive the macroscopic limit, i.e., the operators of the infrared CFT. We will do this in two ways: once using the criterion~\eqref{eq:JO}, and once using the embedding~\eqref{eq:GinSO}. In other words, first we will write down all operators, and manually remove those that have a gauge current inside; and second, we will enumerate the operators of $U(NN_F)_1/U(N)_{N_F}$. The second method is entirely equivalent to the first one, except that all the combinatorics happen under the hood, i.e., the gauge currents are automatically removed for us by the algebraic properties of gauged WZW model.

We begin with the first approach, namely we list the first few gauge invariant operators that do not include any gauge currents. In order to keep the discussion focused we will only include spinless operators but in appendix~\ref{ap:spin_ops} we discuss spinning operators too. In appendix~\ref{ap:gen_fun} we also discuss an alternative method to classify operators using generating functions.

We will use the notation $\{\psi\cdots\psi\}$ to denote a generic operator of shape $\psi\cdots\psi$, with some structure of gauge/flavor index contractions. For example, $\{\psi^2\}$ denotes an operator that is quadratic in $\psi$.

The first non-trivial gauge singlet has $\Delta_\text{UV}=1$, namely $\{\psi_\ell\psi_r^\dagger\}$. Under $SU(N)\times SU(N_F)_\ell\times SU(N_F)_r$, the quantum numbers of $\{\psi_\ell\psi_r^\dagger\}$ are classified by
\begin{equation}
[\ydiagram1,\ydiagram1,\boldsymbol1]\times[\overline{\ydiagram1},\boldsymbol1,\ydiagram1]=[\boldsymbol1,\ydiagram1,\ydiagram1]+\cdots
\end{equation}
where $\cdots$ denote other non-gauge invariant operators. This means that there is one gauge-invariant operator of the form $\{\psi_\ell\psi_r^\dagger\}$, which has quantum numbers $(\ydiagram1,\ydiagram1)$ under the flavor symmetry $SU(N_F)_\ell\times SU(N_F)_r$. It is manifest that this operator does not contain any gauge currents $\psi^\dagger_\ell t^a\psi_\ell$ (nor $\psi^\dagger_r t^a\psi_r$), hence it survives the infrared limit. Naturally, one also has the conjugate operator $\{\psi_\ell^\dagger\psi_r\}$, with quantum numbers $(\overline{\ydiagram1},\overline{\ydiagram1})$.

The first non-trivial operator that contains a gauge current appears at $\Delta_\text{UV}=2$, namely $\{\psi_\ell\psi^\dagger_\ell\psi_r\psi^\dagger_r\}$. Under $SU(N)\times SU(N_F)_\ell\times SU(N_F)_r$, the quantum numbers of this operator are classified by
\begin{equation}
\begin{aligned}
\overbrace{\biggl([\ydiagram1,\ydiagram1,\boldsymbol1]\times[\overline{\ydiagram1},\overline{\ydiagram1},\boldsymbol1]\biggr)}^{\psi_\ell\psi_\ell^\dagger}&\times \overbrace{\biggl([\ydiagram1,\boldsymbol1,\overline{\ydiagram1}]\times [\overline{\ydiagram1},\boldsymbol1,\ydiagram1]\biggr)}^{\psi_r\psi^\dagger_r}\\
&=
\biggl(\overbrace{[\boldsymbol1,\boldsymbol1,\boldsymbol1]}^{J_\ell^{U(1)}}+\overbrace{[\ydiagram{1+1,1},\boldsymbol1,\boldsymbol1]}^{J_\ell^{SU(N)}}+[\boldsymbol1,\ydiagram{1+1,1},\boldsymbol1]+[\ydiagram{1+1,1},\ydiagram{1+1,1},\boldsymbol1]\biggr)\times\\
&\qquad\biggl(\overbrace{[\boldsymbol1,\boldsymbol1,\boldsymbol1]}^{J_r^{U(1)}}+\overbrace{[\ydiagram{1+1,1},\boldsymbol1,\boldsymbol1]}^{J_r^{SU(N)}}+[\boldsymbol1,\boldsymbol1,\ydiagram{1+1,1}]+[\ydiagram{1+1,1},\boldsymbol1,\ydiagram{1+1,1}]\biggr)
\end{aligned}
\end{equation}
where $\ydiagram{1+1,1}$ denotes the adjoint representation. Therefore, the operators of the form $\{\psi_\ell\psi^\dagger_\ell\psi_r\psi^\dagger_r\}$ and \emph{without gauge currents} are classified by
\begin{equation}
\bigl([\boldsymbol1,\ydiagram{1+1,1},\boldsymbol1]+[\ydiagram{1+1,1},\ydiagram{1+1,1},\boldsymbol1]\bigr)\times\bigl([\boldsymbol1,\boldsymbol1,\ydiagram{1+1,1}]+[\ydiagram{1+1,1},\boldsymbol1,\ydiagram{1+1,1}]\bigr)=2[\boldsymbol1,\ydiagram{1+1,1},\ydiagram{1+1,1}]+\cdots
\end{equation}
where $\cdots$ denote other non-gauge invariant operators. This means that there are two gauge-invariant operator of the form $\{\psi_\ell\psi^\dagger_\ell\psi_r\psi^\dagger_r\}$ that survive the infrared limit, and they have quantum numbers $(\ydiagram{1+1,1},\ydiagram{1+1,1})$ under the flavor symmetry $SU(N_F)_\ell\times SU(N_F)_r$. There are other gauge-invariant operators, but they include factors of the gauge currents, and hence they decouple at large distances.

One can repeat the same analysis for higher order operators. One writes down all composites $\{\psi\psi\cdots\psi^\dagger\psi^\dagger\}$, possibly with derivative insertions, and then removes any of them that has any gauge currents inside. The resulting list is the set of all operators that survive in the infrared. For example, the list up to $\Delta_\text{UV}\le3$ reads:

\paragraph{Scaling dimension $\boldsymbol{\Delta_\text{UV}=0}$.} The only operator is the identity, with quantum numbers $(\boldsymbol1,\boldsymbol1)$.

\paragraph{Scaling dimension $\boldsymbol{\Delta_\text{UV}=1}$.} We have $\{\psi_\ell\psi^\dagger_r\}$ (plus conjugate), whose quantum numbers are $(\ydiagram1,\ydiagram1)$.

\paragraph{Scaling dimension $\boldsymbol{\Delta_\text{UV}=2}$.} We have $\{\psi_\ell\psi_\ell\psi^\dagger_r\psi^\dagger_r\}$ (plus conjugate), whose quantum numbers are $(\ydiagram2,\ydiagram2)+(\ydiagram{1,1},\ydiagram{1,1})$. We also have operators $\{\psi_\ell\psi^\dagger_\ell\psi_r\psi^\dagger_r\}$, with quantum numbers $2(\ydiagram{1+1,1},\ydiagram{1+1,1})$.

\paragraph{Scaling dimension $\boldsymbol{\Delta_\text{UV}=3}$.}  We have $\{\psi_\ell\psi_\ell\psi_\ell\psi^\dagger_r\psi^\dagger_r\psi^\dagger_r\}$ (plus conjugate),  with quantum numbers $(\ydiagram{1,1,1},\ydiagram{1,1,1})+(\ydiagram{2,1},\ydiagram{2,1})+(\ydiagram3,\ydiagram3)$. We also have $\{\psi_\ell\psi_\ell\psi^\dagger_\ell\psi_r\psi^\dagger_r\psi^\dagger_r\}$ (plus conjugate), classified by $2(\ydiagram{1+1,1+1,1},\ydiagram{1+1,1+1,1})+2(\ydiagram{1+2,1},\ydiagram{1+2,1})+(\ydiagram{1+1,1+1,1},\ydiagram{1+2,1})+(\ydiagram{1+2,1},\ydiagram{1+1,1+1,1})$. Similarly, we also have $\{D_+\psi^\dagger_\ell D_-\psi_r\}$ (plus conjugate), whose quantum numbers are $(\ydiagram1,\ydiagram1)$. Finally, we have $\{\psi_\ell\psi_\ell\psi^\dagger_\ell D_-\psi^\dagger_r\}$ (plus conjugate, plus $\ell\leftrightarrow r$), classified by $(\ydiagram{1+1,1+1,1},\ydiagram1)+(\ydiagram{1+2,1},\ydiagram1)$.

Whatever the infrared CFT is, it must contain operators with these quantum numbers. Their scaling dimension $\Delta_\text{IR}$ will typically get renormalized. All we can say, without further analysis, is that their scaling dimension will be $\Delta_\text{IR}=\Delta_\text{UV}+\mathcal O(1/N_F)$ in the large $N_F$ limit. For example, the UV operator $\{\psi_\ell\psi^\dagger_r\}$ will map to a non-trivial IR operator with quantum numbers $(\ydiagram1,\ydiagram1)$ under the flavor symmetry, and scaling dimension $\Delta_\text{IR}=1+\mathcal O(1/N_F)$.

We now move on to the gauged WZW approach. Namely, the space of infrared operators of QCD is equivalent to the space of operators of $U(NN_F)_1/U(N)_{N_F}$. This set of operators is rather straightforward: by level-rank duality, this coset is identical to the WZW model $SU(N_F)_N$. The operators of this rational CFT are enumerated by the torus partition function,
\begin{equation}
Z(q,\bar q)=\sum_{\lambda} |\chi_\lambda(q)|^2
\end{equation}
where $\lambda$ denotes the integrable representations of $SU(N_F)_N$, and $\chi_\lambda$ is the associated chiral characters. The first few characters are
\begin{equation}
\begin{aligned}
\chi_{\boldsymbol1}(q)&=\boldsymbol1+\ydiagram{1+1,1}\,q+(\boldsymbol1+ 2\,\ydiagram{1+1,1} + \ydiagram{1+1,1+1,1,1}+\ydiagram{2+2,2}\,)q^2+\cdots\\
\chi_{\ydiagram1}(q)&=\ydiagram1+(\ydiagram1+\ydiagram{1+1,1+1,1}+\ydiagram{1+2,1}\,)q+\cdots\\
\chi_{\ydiagram{1,1}}(q)&=\ydiagram{1,1}+(\,\ydiagram{1,1} +\ydiagram2 +\ydiagram{1+1,1+1,1+1,1}+\ydiagram{1+2,1+1,1}\,)q+\cdots\\
\chi_{\ydiagram2}(q)&=\ydiagram2+(\,\ydiagram{1,1} +\ydiagram2+\ydiagram{1+2,1+1,1}+\ydiagram{1+3,1}\,)q+\cdots\\
\chi_{\ydiagram{1+1,1}}(q)&=\ydiagram{1+1,1}+(\boldsymbol1+2\,\ydiagram{1+1,1}+\ydiagram{1+1,1+1,1,1}+\ydiagram{1+2,1,1}+\ydiagram{2+1,2+1,2}+\ydiagram{2+2,2}\,)q+\cdots
%\chi_\ell(q)&=\wedge^\ell\ydiagram1+\cdots
\end{aligned}
\end{equation}
where we omit the global power $q^{h-c/24}$ to simplify the notation.

In order to make contact with the previous discussion, we list the operators graded by the classical scaling dimension dimension $\Delta_\text{UV}:=\lim_{N_F\to\infty}(h+\bar h)$. For primaries, $\Delta_\text{UV}$ is equal to the number of boxes in its Young diagram. We restrict ourselves to spinless operators here, relegating the discussion of the full spectrum to appendix~\ref{ap:spin_ops}.

\paragraph{Scaling dimension $\boldsymbol{\Delta_\text{UV}=0}$.} The only operator is the identity,
\begin{equation}
[q^0\bar q^0]|\chi_{\boldsymbol1}|^2=(\boldsymbol1,\boldsymbol1)
\end{equation}

\paragraph{Scaling dimension $\boldsymbol{\Delta_\text{UV}=1}$.} The operators are
\begin{equation}
[q^0\bar q^0]|\chi_{\ydiagram1}|^2+\text{c.c.}=(\ydiagram1,\ydiagram1)+\text{c.c.}
\end{equation}

\paragraph{Scaling dimension $\boldsymbol{\Delta_\text{UV}=2}$.} The operators are 
\begin{equation}
\begin{aligned}
[q^0\bar q^0]\bigl(|\chi_{\ydiagram{1,1}}|^2+|\chi_{\ydiagram2}|^2+\text{c.c.}\bigr)&=(\ydiagram{1,1},\ydiagram{1,1})+(\ydiagram{2},\ydiagram{2})+\text{c.c}\\
[q^0\bar q^0]|\chi_{\ydiagram{1+1,1}}|^2&=(\ydiagram{1+1,1},\ydiagram{1+1,1})\\
[q^1\bar q^1]|\chi_{\boldsymbol1}|^2&=(\ydiagram{1+1,1},\ydiagram{1+1,1})
\end{aligned}
\end{equation}
\paragraph{Scaling dimension $\boldsymbol{\Delta_\text{UV}=3}$.}  The operators are
\begin{equation}
\hspace*{-1cm}\begin{aligned}
[q^0&\bar q^0]\bigl(|\chi_{\ydiagram{1,1,1}}|^2+|\chi_{\ydiagram{1+1,1+1,1}}|^2+|\chi_{\ydiagram{2,1}}|^2+|\chi_{\ydiagram3}|^2+|\chi_{\ydiagram{1+2,1}}|^2+\text{c.c.}\bigr)\\
&=(\ydiagram{1,1,1},\ydiagram{1,1,1})+(\ydiagram{1+1,1+1,1},\ydiagram{1+1,1+1,1})+(\ydiagram{2,1},\ydiagram{2,1})+(\ydiagram3,\ydiagram3)+(\ydiagram{1+2,1},\ydiagram{1+2,1})+\text{c.c.}\\
[q^1&\bar q^1]|\chi_{\ydiagram1}|^2+\text{c.c.}=(\ydiagram1+\ydiagram{1+1,1+1,1}+\ydiagram{1+2,1},\ydiagram1+\ydiagram{1+1,1+1,1}+\ydiagram{1+2,1})+\text{c.c.}
\end{aligned}
\end{equation}

Unsurprisingly, we get the exact same list of operators as in our brute force approach from before. What's more, if we compare the lists of operators obtained by these two approaches, we get the explicit map of UV to IR operators. For example, the UV operator $\{\psi_\ell\psi^\dagger_r\}$ maps to the IR operator corresponding to the $SU(N_F)_N$ primary $(\ydiagram1,\ydiagram1)$, and hence $\Delta_\text{IR}(\psi_\ell\psi_r^\dagger)=\frac{N_F^2-1}{N_F(N+N_F)}$.\footnote{This value of $\Delta_\text{IR}$ can be used to test other, more traditional strong-coupling methods. For example, it would be interesting to see if this scaling dimension can be observed on the lattice. Similarly, a Veneziano-like limit $N,N_F\to\infty$ with $\lambda=N/N_F$ fixed, should be able to reproduce $\Delta_\text{IR}(\psi_\ell\psi_r^\dagger)=\frac{1}{1+\lambda}$.} Note that, indeed, $\Delta_\text{IR}=1+\mathcal O(1/N_F)$ in the large $N_F$ limit.

More generally, the UV to IR map of operators is
\begin{equation}
\begin{aligned}
1&\rightsquigarrow (\boldsymbol1,\boldsymbol1)\\
\psi_\ell\psi^\dagger_r&\rightsquigarrow (\ydiagram1,\ydiagram1)\\
\psi_\ell\psi_\ell\psi^\dagger_r\psi^\dagger_r&\rightsquigarrow(\ydiagram2,\ydiagram2)+(\ydiagram{1,1},\ydiagram{1,1})\\
\psi_\ell\psi^\dagger_\ell\psi_r\psi^\dagger_r&\rightsquigarrow(\ydiagram{1+1,1},\ydiagram{1+1,1}) + \text{descendants of $(\boldsymbol1,\boldsymbol1)$}\\
\psi_\ell\psi_\ell\psi_\ell\psi^\dagger_r\psi^\dagger_r\psi^\dagger_r&\rightsquigarrow(\ydiagram{1,1,1},\ydiagram{1,1,1})+(\ydiagram{2,1},\ydiagram{2,1})+(\ydiagram3,\ydiagram3)\\
\psi_\ell\psi_\ell\psi^\dagger_\ell\psi_r\psi^\dagger_r\psi^\dagger_r&\rightsquigarrow(\ydiagram{1+1,1+1,1},\ydiagram{1+1,1+1,1})+(\ydiagram{1+2,1},\ydiagram{1+2,1})+\text{descendants of $(\ydiagram1,\ydiagram1)$}\\
%(\ydiagram{1+1,1+1,1},\ydiagram{1+1,1+1,1})+(\ydiagram{1+2,1},\ydiagram{1+2,1})+(\ydiagram{1+1,1+1,1},\ydiagram{1+2,1})+(\ydiagram{1+2,1},\ydiagram{1+1,1+1,1})\\
D_+\psi^\dagger_\ell D_-\psi_r&\rightsquigarrow \text{descendants of $(\ydiagram1,\ydiagram1)$}\\
%(\ydiagram1,\ydiagram1)\\
\psi_\ell\psi_\ell\psi^\dagger_\ell D_-\psi^\dagger_r&\rightsquigarrow \text{descendants of $(\ydiagram1,\ydiagram1)$}\\
%(\ydiagram{1+1,1+1,1},\ydiagram1)+(\ydiagram{1+2,1},\ydiagram1)
&\text{etc.}
\end{aligned}
\end{equation}
from where generic scaling dimensions $\Delta_\text{IR}$ can be computed. The first few entries in this map were also obtained in~\cite{He:2021xvg}.

We make one final interesting comment. For large $N_F$ the RG-flow is very short and the infrared theory is very close to the ultraviolet free fermion theory. Thus, for large $N_F$, the theory $\mathcal T_\text{IR}$ should approach a CFT of $NN_F$ free Dirac fermions. This means that $SU(N_F)_N$ WZW becomes a free fermion theory for large $N_F$, as has already been discussed in e.g.~\cite{Kiritsis:2010xc}.

\subsection[$Spin(7)$ plus matter in $\boldsymbol8$]{$\boldsymbol{Spin(7)}$ plus matter in $\boldsymbol8$.}

Consider QCD with gauge group $Spin(7)$ and a Majorana fermion in the eight dimensional representation $\boldsymbol8$. As in the previous section, we would like to obtain here the list of operators that survive the macroscopic limit, i.e., the operators of the infrared CFT. We do this in two ways: first, by writing down all operators, and manually removing those that have a gauge current inside; and second, by looking at the operators of $SO(8)_1/Spin(7)_1$. As we shall see, this gauge theory flows to the  Ising CFT in the IR.

\paragraph{Chiral operators.} For now, we work at the chiral level, and we simply denote the field as $\psi$, which could stand for either chirality; we will combine the two halves later on. In what follows we shall classify generic operators that can be written in terms of $\psi$. We use the notation $\{\psi\cdots\psi\}$ to denote the set of operators of the form $\psi\cdots\psi$, with all possible gauge-index structures. For example, $\{\psi^2\}$ denotes all operators that are quadratic in $\psi$.

The quark field $\psi$ transforms in the spinor representation of the gauge group, namely $\boldsymbol8$. Therefore, the irreducible components of a quadratic operator of the form $\{\psi^2\}$ are classified by the anti-symmetrized product $\wedge^2\boldsymbol8=\boldsymbol{7}+\boldsymbol{21}$. Specifically, the two components are
\begin{equation}
{:}\psi \gamma^A\psi{:},\qquad {:}\psi\gamma^{AB}\psi{:}
\end{equation}
where $A,B=1,\dots,7$ are vector indices of $Spin(7)$, and $\gamma^{A_1A_2\cdots A_p}:=\gamma^{[A_1}\gamma^{A_2}\cdots \gamma^{A_p]}$.  The component $\psi\gamma^{AB}\psi$ has a simple interpretation -- it is the gauge current $J^a={:}\psi t^a\psi{:}$, since $\gamma^{AB}$ is the generator of the spinor representation. The other component $\psi\gamma^A\psi$ cannot be written in terms of the current $J^a$. So, all in all, there are two operators of the form $\{\psi^2\}$, one of them being the gauge current, and the other an operator that \emph{cannot} be written in terms of the gauge current. The former decouples at large distances, while the latter does not.

We can introduce the notation
\begin{equation}
\epsilon^A:={:}\psi \gamma^A\psi{:},\qquad J^{AB}:={:}\psi\gamma^{AB}\psi{:}
\end{equation}
to stand for the vector $\boldsymbol 7$ and the gauge current $\boldsymbol{21}$, respectively. In this notation, we can write
\begin{equation}
\{\psi^2\}\leftrightarrow J\oplus\epsilon
\end{equation}
to indicate the fact that there are two irreducible operators of the form $\{\psi^2\}$, corresponding to $J$ and $\epsilon$.

One can do the same analysis for operators with more spinors. For example, let us discuss operators of the form $\{\psi^3\}$ and $\{\partial\psi\}$. We consider both options together since they have the same scaling dimension and therefore they may mix under the RG-flow. The operators $\{\partial\psi\}$ are straightforward: there is a single operator of this form, namely $\partial\psi_\alpha$, transforming as $\boldsymbol 8$ under the gauge group. Here, $\alpha=1,\dots,8$ is a spinor index under $Spin(7)$.

On the other hand, the operators of the form $\{\psi^3\}$ are classified by $\wedge^3\boldsymbol8=\boldsymbol8+\boldsymbol{48}$. It is easy to see what these stand for. In order to construct cubic operators we can take the quadratic operators $\{\psi^2\}$ from before, and multiply them by another factor of $\psi$,
\begin{equation}
{:}\psi \gamma^A\psi\,\psi_\alpha{:},\qquad{:}\psi\gamma^{AB}\psi\,\psi_\alpha{:}
\end{equation}
Actually, we do not need to consider both these expressions, since they are linearly-dependent:
\begin{equation}\label{eq:psi3_current}
{:}\psi\gamma^A\psi\,\psi_\alpha{:}\equiv\frac{1}{120}{:}\psi\gamma_{BC}\psi\, (\gamma^{ABC}\psi)_\alpha{:}
\end{equation}
Let us, then, look at the operator $\psi \gamma^A\psi\,\psi_\alpha$. This operator is not yet irreducible, since we can break it into a ``pure trace'' part and a  ``traceless'' part,
\begin{equation}\label{eq:irr_components_psi3}
\begin{aligned}
t_\alpha&:={:}\psi\gamma^A\psi\,(\gamma_A\psi)_\alpha{:}\\
t^A_\alpha&:={:}\psi\gamma^A\psi\,\psi_\alpha{:}-\tfrac17(\gamma^At)_\alpha
\end{aligned}
\end{equation}
such that $\gamma_A t^A\equiv 0$. These operators correspond to the two irreducible components in $\wedge^3\boldsymbol8=\boldsymbol8+\boldsymbol{48}$. Indeed, they have $8$ and $7\times 8-8=48$ components, respectively. So, all in all, there are two operators of the form $\{\psi^3\}$, given by the two expressions in~\eqref{eq:irr_components_psi3}.

Note that $t_\alpha$ transforms as $\boldsymbol 8$, and therefore this component may mix with $\partial\psi_\alpha$. The operator that survives the infrared limit, if any, can be some non-trivial linear combination of $t_\alpha$ and $\partial\psi_\alpha$. On the other hand, $t^A_\alpha$ is the only operator at this scaling dimension that transforms as $\boldsymbol{48}$, so this operator will not mix with anything -- it either survives the infrared limit by itself, or it decouples.

With this in mind, we must now ask whether the three operators $t_\alpha,t^A_\alpha,\partial\psi_\alpha$ can be written in terms of the gauge current $J^{AB}={:}\psi\gamma^{AB}\psi{:}$. If so, they will decouple in the IR, and if not, they will survive the IR limit. Concretely, we ask whether there exist operators $\varphi$ such that $\psi^3,\partial\psi\equiv {:}J\varphi{:}$. By fermi statistics and dimensional analysis, there is only one candidate, namely $\varphi\propto\psi$, so we must compute ${:}J^{AB}\psi_\alpha{:}$. A straightforward computation shows that
\begin{equation}\label{eq:OPE_Ising_Jphi}
\begin{aligned}
t_\alpha+42\partial\psi_\alpha&=\tfrac18{:}J_{AB}(\gamma^{AB}\psi)_\alpha{:}\\
t^A_\alpha&=\tfrac{1}{28}{:}J_{BC}((\tfrac{1}{15}\gamma^{ABC}-\delta^{A[B}\gamma^{C]})\psi)_\alpha{:}
\end{aligned}
\end{equation}
The consequence is that the operator $t^A_\alpha$ can be written as $\sim{:}J\varphi{:}$, hence it decouples. On the other hand, the linear combination $t_\alpha+42\partial\psi_\alpha$ can also be written in terms of the gauge currents, and thus it also decouples. There is another linear combination of $t_\alpha$ and $\partial\psi_\alpha$ that \emph{cannot} be written in terms of the gauge currents. This linear combination survives the large distance limit. One can take, for example,
\begin{equation}\label{eq:def_Phi_ising}
\Phi_\alpha:=t_\alpha-6\partial\psi_\alpha
\end{equation}
which satisfies
\begin{equation}
J^{AB}(x)\Phi_\alpha(0)=-\frac{2}{x}(\gamma^{AB}\Phi)_\alpha+\text{finite}
\end{equation}
which shows that $\Phi_\alpha$ is a $Spin(7)_1$-primary, transforming in the spinor representation.

%Note: the only three non-trivial OPEs at this scaling dimension are
%\begin{equation}
%\begin{aligned}
%\tfrac18{:}J_{AB}(\gamma^{AB}\psi)_\alpha{:}&=t_\alpha+42\partial\psi_\alpha\\
%\tfrac{1}{120}{:}J_{BC}(\gamma^{ABC}\psi)_\alpha{:}&=t^A_\alpha+\tfrac17(\gamma^At)_\alpha+6(\gamma^A\partial\psi)_\alpha\\
%\tfrac14{:}J_{BC}\delta^{A[B}(\gamma^{C]}\psi)_\alpha{:}&=-5t^A_\alpha+\tfrac27(\gamma^At)_\alpha+12(\gamma^A\partial\psi)_\alpha
%\end{aligned}
%\end{equation}
%They are redundant and can be reduced to the two equations in~\eqref{eq:OPE_Ising_Jphi}.

The conclusion is that the operator $t^A$, and the linear combination $t+42\partial\psi$, are both proportional to the gauge current. On the other hand, the linear combination $t-6\partial\psi$ cannot be written in terms of the current. We schematically write the former as $J\cdot \sigma$, and the latter as $\sigma$, where $\sigma$ stands for the representation $\boldsymbol 8$. In this notation, we can write
\begin{equation}
\{\psi^3\}\oplus\{\partial\psi\}\leftrightarrow \{J\cdot \sigma\}\oplus \sigma
\end{equation}
The two operators in $\{J\cdot \sigma\}$ decouple in the IR, while $\sigma$ survives.

Finally, let us discuss quartic operators, i.e., operators of the form $\{\psi^4\}$. At this scaling dimension we must also consider $\{\psi\partial\psi\}$ since these can mix under the RG-flow. We begin by the former; these operators are classified by $\wedge^4\boldsymbol8=\boldsymbol1+\boldsymbol7+\boldsymbol{27}+\boldsymbol{35}$. We can write all such operators as $\psi\Gamma\psi\,\psi\Gamma'\psi$, where $\Gamma,\Gamma'=\gamma^A,\gamma^{AB}$, subject to suitable irreducibility conditions. A possible choice is
\begin{equation}\label{eq:basis_psi4}
%\psi\gamma_A\psi \,\psi\gamma^A\psi,\quad\psi\gamma_A\psi \,\psi\gamma^{AB}\psi,\quad\psi\gamma^{(A}\psi \,\psi\gamma^{B)}\psi-\text{trace},\quad\psi\gamma^{[A}\psi\,\psi \gamma^{BC]}\psi
\begin{aligned}
t&:={:}\psi\gamma_A\psi \,\psi\gamma^A\psi{:}\\
t^B&:={:}\psi\gamma_A\psi \,\psi\gamma^{AB}\psi{:}\\
t^{AB}&:={:}\psi\gamma^{(A}\psi \,\psi\gamma^{B)}\psi-\text{trace}{:}\\
t^{ABC}&:={:}\psi\gamma^{[A}\psi\,\psi \gamma^{BC]}\psi{:}
\end{aligned}
\end{equation}
although it is not unique. We can also write these operators in a different form, using the gamma matrices identities
\begin{equation}\label{eq:redundant_psi4}
\begin{aligned}
{:}\psi\gamma_A\psi \,\psi\gamma^A\psi{:}&\equiv\frac{1}{8} {:}\psi \gamma_{AB}\psi\,\psi \gamma^{AB}\psi{:}\\
{:}\psi\gamma^{(A}\psi \,\psi\gamma^{B)}\psi-\text{trace}{:}&\equiv\frac{1}{20}{:}\psi \gamma_C\gamma^{(A}\psi\,\psi \gamma^{B)}\gamma^C\psi-\text{trace}{:}\\
{:}\psi\gamma^{[A_1}\psi\,\psi \gamma^{A_2A_3]}\psi{:}&\equiv\frac{1}{16i}\epsilon^{A_1\cdots A_7}{:}\psi \gamma_{A_4A_5}\psi\,\psi \gamma_{A_6A_7}\psi{:}
\end{aligned}
\end{equation}

Similarly, operators of the form $\{\psi\partial\psi\}$ are classified by $\boldsymbol 8\times\boldsymbol 8=\boldsymbol1+\boldsymbol7+\boldsymbol{21}+\boldsymbol{35}$, namely
\begin{equation}\label{eq:psi_partial_psi_basis}
%\psi\partial\psi,\quad \psi\gamma^A\partial\psi,\quad \psi\gamma^{AB}\partial\psi,\quad \psi\gamma^{ABC}\partial\psi
\begin{aligned}
s&:={:}\psi\partial\psi{:}\\
s^A&:={:}\psi\gamma^A\partial\psi{:}\\
s^{AB}&:={:}\psi\gamma^{AB}\partial\psi{:}\\
s^{ABC}&:={:}\psi\gamma^{ABC}\partial\psi{:}
\end{aligned}
\end{equation}
Note that $t$ and $s$ both transform as singlets of $Spin(7)$; $t^A$ and $s^A$ both as $\boldsymbol 7$; and $t^{ABC}$ and $s^{ABC}$ both as $\boldsymbol{35}$. Thus, these three pairs of operators may mix under the RG-flow, and the operators that survive the infrared limit, if any, will be suitable linear combinations thereof. On the other hand, $t^{AB}$ and $s^{AB}$ transform as $\boldsymbol{27}$ and $\boldsymbol{21}$, respectively, and cannot mix; they will either decouple or survive by themselves.

As before, we must ask whether there are operators $\varphi$ such that $s,t\sim {:}J\varphi{:}$. By fermi statistics and dimensional analysis, the candidates are $\varphi\propto\{\psi^2\},\partial$. By $\partial$ we mean the operator $\partial J^{AB}$. Actually, it is easy to see that $s^{AB}$ can be written in this form, $s^{AB}=\frac12\partial J^{AB}$. For $\{\psi^2\}$, we need the contractions ${:}J^{AB}\psi_\alpha\psi_\beta{:}$. The result is:
\begin{equation}\label{eq:four_fermi_map}
\begin{aligned}
t+84s&=\tfrac18{:}J^{AB}J_{AB}{:}\\
t^B-48s^B&={:}J^{AB}\epsilon_A{:}\\
t^{BC}&=\tfrac{1}{20}\delta_{AD}{:}J^{A(B}J^{C)D}-\text{trace}{:}\\
s^{BC}&=\tfrac12\partial J^{BC}\\
t^{ABC}&=\frac{1}{24} (8 {:}J^{[AB}\epsilon^{C]}{:}-i \epsilon^{ABCD_1D_2D_3D_4}{:}J_{D_1D_2}J_{D_3D_4}{:} )\\
s^{ABC}&=-\frac{1}{192} (16 {:}J^{[AB}\epsilon^{C]}{:}+i \epsilon^{ABCD_1D_2D_3D_4}{:}J_{D_1D_2}J_{D_3D_4}{:} )
%\frac{1}{16i}\epsilon^{ABCD_1D_2D_3D_4}{:}J_{D_1D_2}J_{D_3D_4}{:}&=t^{ABC}+4s^{ABC}\\
%{:}J^{[AB}v^{C]}{:}&=t^{ABC}-8s^{ABC}
\end{aligned}
\end{equation}
All these operators can be written in terms of the gauge currents, and thus they all decouple. The operators that survive are a suitable linear combination of $t,s$, and a suitable linear combination of $t^A,s^A$. One can take, for example,
\begin{equation}\label{eq:four_fermi_IR}
\begin{aligned}
\Phi&:=t-12 s\\
\Phi^A&:=t^A+48 s^A
\end{aligned}
\end{equation}
which satisfy
\begin{equation}
\begin{aligned}
J^{AB}(x)\Phi(0)&=\text{finite}\\
J^{AB}(x)\Phi^C&=-\frac{8}{x}\delta^{C[A}\Phi^{B]}+\text{finite}
\end{aligned}
\end{equation}
This shows that $\Phi$, $\Phi^A$ are $Spin(7)_1$-primaries, transforming in the singlet and vector representations, respectively.

In conclusion, all operators of the form $\{\psi^4\}\oplus\{\psi\partial\psi\}$ can be written in terms of the currents, except for $\Phi$ and $\Phi^A$, which are given by certain linear combinations of $s,t$ and $s^A,t^A$. The operators that can be written in terms of currents are schematically of the form $J^2\oplus \partial J\oplus J\cdot\epsilon$, while the operators that cannot are a singlet $\boldsymbol1$ and a vector $\boldsymbol 7$. We express this as
\begin{equation}
\{\psi^4\}\oplus\{\psi\partial\psi\}\leftrightarrow \{J^2\}\oplus\partial J\oplus \{J\cdot\epsilon\}\oplus \epsilon\oplus\boldsymbol1
\end{equation}
where $\epsilon$ stands for the vector representation $\boldsymbol 7$.

The summary of the discussion up to scaling dimension $h=2$ is contained in the following map:
\begin{equation}\label{eq:ising_map_nb}
\begin{aligned}
\{1\}&\leftrightarrow 1\\
\{\psi\}&\leftrightarrow\sigma\\
\{\psi^2\}&\leftrightarrow J\oplus \epsilon\\
\{\psi^3\}\oplus\{\partial\psi\}&\leftrightarrow \{J\cdot\sigma\}\oplus\sigma\\
\{\psi^4\}\oplus\{\psi\partial\psi\}&\leftrightarrow \{J^2\}\oplus\partial J\oplus \{J\cdot\epsilon\}\oplus \epsilon\oplus\boldsymbol1\\
&\ \ \vdots
\end{aligned}
\end{equation}

The claim is that all operators that include $J$ will decouple at large distances, while those that do not, survive the infrared limit and become non-trivial operators in the low-energy effective theory.  Up to the order described so far, the only operators with no currents are $\psi$ itself, the vector $\epsilon^A={:}\psi\gamma^A\psi{:}$ inside $\{\psi^2\}$, the spinor $\Phi_\alpha$ inside $\{\psi^3\}\oplus\{\partial\psi\}$, and the singlet $\Phi$ and vector $\Phi^A$ in $\{\psi^4\}\oplus\{\psi\partial\psi\}$. In other words, a basis of infrared operators (at the chiral level) is
\begin{equation}\label{eq:IR_basis_ising}
1,\qquad \psi_\alpha,\qquad {:}\psi\gamma^A\psi{:},\qquad \Phi_\alpha,\qquad \Phi,\qquad\Phi^A,\quad\dots
\end{equation}

Let us make one final comment. Recall that the two operators $t={:}\psi\gamma_A\psi \,\psi\gamma^A\psi{:}$ and $s={:}\psi\partial\psi{:}$ mix under the RG-flow. The correct linear combinations are
\begin{equation}
\begin{aligned}
\Phi&=t-12s\\
{:}J^2{:}&=t+84s
\end{aligned}
\end{equation}
cf.~\eqref{eq:four_fermi_map},~\eqref{eq:four_fermi_IR}. Of course, we can invert this to yield
\begin{equation}
\begin{aligned}
t&=\tfrac{1}{8}(7\Phi+{:}J^2{:})\\
s&=\tfrac{1}{96}(-\Phi+{:}J^2{:})
\end{aligned}
\end{equation}
In the infrared we set $J^a\equiv0$, while $\Phi\neq0$. This means that both $t,s$ survive the infrared limit, but become the same operator, namely $t\to \frac78\Phi$, $s\to\frac{-1}{96}\Phi$.\footnote{The phenomenon where two operators $t,s$ exist at large distances but the difference $t-12s$ decouples was dubbed ``conformal coalescence'' in~\cite{Georgi:2019tch}. This is a very common phenomenon in $2d$.} In order to write down a basis of infrared operators, we can use $\Phi$ directly, but we may just as well use either of $t,s$. The same thing can be said about $\Phi_\alpha$: instead of using this combination to characterize the infrared operators, we can use either of its components $t_\alpha,\partial\psi_\alpha$. Both components survive the infrared limit, but converge to the same operator. Similarly, we can use $\Phi^A$ or either of its components $t^A,s^A$. With this in mind, and to keep the notation as simple as possible, we can declare that, up to dimension $h=2$, a basis of infrared operators is
\begin{equation}
1,\qquad\psi_\alpha,\qquad{:}\psi\gamma^A\psi{:},\qquad\partial\psi_\alpha,\qquad {:}\psi\partial\psi{:},\qquad{:}\psi\gamma^A\partial\psi{:},\quad\dots
\end{equation}
instead of~\eqref{eq:IR_basis_ising}.

\paragraph{Branching rules.} The previous analysis can be simplified significantly by exploiting the branching rules of $SO(8)_1/Spin(7)_1$, which automatically handle all the combinatorics that decide whether a given UV operator is a $Spin(7)_1$-primary or a descendant.

We begin by writing down the character of the free fermion CFT in the UV $SO(8)_1$:\footnote{We look at the NS-NS character for concreteness; the NS-R character takes the same form but half-integral powers of $q$ carry a negative sign.}
\begin{equation}
\begin{aligned}
d_\text{NS-NS}&=q^{-1/6}\bigl(\boldsymbol1+\boldsymbol8\,q^{1/2}+(\boldsymbol 7+\boldsymbol{21})q\\
&+(2\,\boldsymbol 8+\boldsymbol{48})q^{3/2}+(2 \,\boldsymbol1+ 2\,\boldsymbol7 + \boldsymbol{21} + \boldsymbol{27} + 2\,\boldsymbol{35} ) q^2+\cdots\bigr)
\end{aligned}
\end{equation}
This character classifies the operators that we can construct out of $\psi$. For example, $\boldsymbol1$ stands for the identity operator $1$; $\boldsymbol8$ stands for the fermion $\psi$ itself, which transforms in the eight-dimensional representation of $Spin(7)$; $\boldsymbol7+\boldsymbol{21}\equiv\wedge^2\boldsymbol8$ stands for the quadratic $\{\psi^2\}$; etc. We can also write this character in a more schematic notation as follows:
\begin{equation}
\begin{aligned}
d_\text{NS-NS}&=q^{-1/6}\bigl(1+\psi \,q^{1/2}+\{\psi^2\} q+(\{\psi^3\}\oplus\{\partial\psi\})q^{3/2}+(\{\psi^4\}\oplus\{\psi\partial\psi\})q^2\\
&+(\{\psi^5\}\oplus\{\psi^2\partial\psi\}\oplus\{\partial^2\psi\})q^{5/2}+\cdots\bigr)
\end{aligned}
\end{equation}
where at each order $q^h$ we write all combinations of $\psi$ and its derivatives whose weights add up to $h$.

Next, we write the characters of the gauge group $Spin(7)_1$:
\begin{equation}
\begin{aligned}
\chi_1&=q^{-7/48}(\boldsymbol1+\boldsymbol{21}\, q+(\boldsymbol1 +\boldsymbol{21}+\boldsymbol{27}+\boldsymbol{35})q^2+\cdots)\\
\chi_\epsilon&=q^{17/48}(\boldsymbol 7+(\boldsymbol7+\boldsymbol{35})q+(2\,\boldsymbol7+\boldsymbol{21}+\boldsymbol{35}+\boldsymbol{105})q^2+\cdots)\\
\chi_\sigma&=q^{7/24}(\boldsymbol8+(\boldsymbol8+\boldsymbol{48})q+(2\,\boldsymbol8+2\,\boldsymbol{48}+\boldsymbol{112})q^2+\cdots)
\end{aligned}
\end{equation}
As before, these characters classify operators constructed out of the gauge current $J^a$. For example, in the vacuum character $\chi_1$, $\boldsymbol1$ stands for the identity operator $1$; $\boldsymbol{21}$ stands for the current $J$ itself (which is an adjoint under the gauge group); $\boldsymbol1+\boldsymbol{27}+\boldsymbol{35}\equiv S^2\boldsymbol{21}\ominus \boldsymbol{168}$ (where $\boldsymbol{168}$ is a null operators) stands for $\{J^2\}$; etc. Similarly, in the vector character $\chi_\epsilon$, $\boldsymbol 7$ stands for the vector representation of $Spin(7)$, namely the primary $v_\epsilon$, while $\boldsymbol 7+\boldsymbol{35}\equiv \boldsymbol{21}\times\boldsymbol 7$ stands for the action of the current $J$ on this primary, i.e., $\{J\cdot v_\epsilon\}$. With this in mind, we can also write this character in a more schematic notation as follows:
\begin{equation}
\begin{aligned}
\chi_1&=q^{-7/48}(v_1+\{J\cdot v_1\}q+\{(\partial J\oplus J^2)\cdot v_1\}q^2+\cdots)\\
\chi_\epsilon&=q^{17/48}(v_\epsilon+\{J\cdot v_\epsilon\}q+\{(\partial J\oplus J^2)\cdot v_\epsilon\}q^2+\cdots)\\
\chi_\sigma&=q^{7/24}(v_\sigma+\{J\cdot v_\sigma\} q+\{(\partial J\oplus J^2)\cdot v_\sigma\}q^2+\cdots)
\end{aligned}
\end{equation}
where $v_1,v_\epsilon,v_\sigma$ are the three primaries of $Spin(7)_1$, corresponding to the integrable representations $\boldsymbol1,\boldsymbol7,\boldsymbol8$ respectively, and $\{J\cdot v_\lambda\}$ denotes the action of the gauge currents on them. As before, at each order $q^h$ we write all combinations of $J$ and its derivatives whose conformal weights add up to $h$.

We are finally in position to write down the operators of the coset $SO(8)_1/Spin(7)_1$. The branching rules take the form
\begin{equation}\label{eq:branch_spin_7_ising}
d_\text{NS-NS}(q)=b_1(q)\chi_1(q)+b_\epsilon(q)\chi_\epsilon(q)+b_\sigma(q)\chi_\sigma(q)
\end{equation}
for some coefficients $b_\lambda(q)$ that we are to determine. We write them in the general form
\begin{equation}
b_\lambda(q)=q^{m_\lambda}\sum_{j\ge0}\mu_{\lambda,j}q^j
\end{equation}
for some exponents $m_\lambda$ and some integer coefficients $\mu_{\lambda,j}$. Plugging this into~\eqref{eq:branch_spin_7_ising}, and using the known expressions for $d_\text{NS-NS}$ and $\chi_\lambda$, we get
\begin{equation}
\begin{aligned}
q^{-1/6}(1+\psi \,q^{1/2}+\{\psi^2\} q+\cdots)&=q^{-7/48+m_1}(v_1+\{J\cdot v_1\}q+\cdots)(\mu_{1,0}+\cdots)\\
&+q^{17/48+m_\epsilon}(v_\epsilon+\{J\cdot v_\epsilon\}q+\cdots)(\mu_{\epsilon,0}+\cdots)\\
&+q^{7/24+m_\sigma}(v_\sigma+\{J\cdot v_\sigma\} q+\cdots)(\mu_{\sigma,0}+\cdots)
\end{aligned}
\end{equation}
Now, we match powers of $q$ on both sides and solve for the exponents $m_\lambda$ and coefficients $\mu_{\lambda,j}$.

The first representation on the \emph{l.h.s.}~is $\boldsymbol1$, corresponding to the leading term. This is the same representation as $v_1$, and therefore these two terms must match:
\begin{equation}
q^{-1/6}\equiv q^{-7/48+m_1} v_1\mu_{1,0}
\end{equation}
from where we find $m_1=-1/48$ and $\mu_{1,0}=1$. The next non-trivial representation is $\boldsymbol 8$, provided by $\psi$. This is the same representation as $v_\sigma$, hence these two terms again must match:
\begin{equation}
\psi \,q^{1/3}\equiv q^{7/24+m_\sigma} v_\sigma \mu_{\sigma,0}
\end{equation}
from where $m_\sigma=1/24$ and $\mu_{\sigma,0}=1$. Next, we look at the term $\{\psi^2\}$, which stands for $\boldsymbol7+\boldsymbol{21}$. The $\boldsymbol 7$ is carried by $v_\epsilon$, while the $\boldsymbol{21}$ by $\{J\cdot v_1\}$, hence we have
\begin{equation}
\{\psi^2\} \,q^{5/6}\equiv q^{5/6} \{J\cdot v_1\} \mu_{1,0}\oplus v_1\mu_{1,1}\oplus q^{17/48+m_\epsilon} v_\epsilon\mu_{\epsilon,0}
\end{equation}
from where we find $m_\epsilon=23/48$, $\mu_{1,1}=0$, and $\mu_{\sigma,0}=1$. If we continue this way, we get the following branching functions:
\begin{equation}\label{eq:branch_ising_b}
\begin{aligned}
b_1(q)&=q^{-1/48}(1+q^2+q^3+2q^4+\cdots)\\
b_\epsilon(q)&=q^{23/48}(1+q+q^2+q^3+\cdots)\\
b_\sigma(q)&=q^{1/24}(1+q+q^2+2q^3+\cdots)
\end{aligned}
\end{equation}
%which, indeed, we recognize as the characters of the Ising CFT.
Furthermore, at the level of operators, we get the following map:
\begin{equation}\label{eq:chiral_ising_match}
\begin{aligned}
1&\leftrightarrow \mu_{1,0}v_1\\
\psi&\leftrightarrow \mu_{\sigma,0}v_\sigma\\
\{\psi^2\}&\leftrightarrow  \mu_{1,0}\{J\cdot v_1\}\oplus \mu_{\epsilon,0}v_\epsilon\\
%\psi^3+\partial\psi&\leftrightarrow \mu_{\sigma,0}J\cdot v_\sigma+\mu_{\sigma,1}v_\sigma\\
\{\psi^3\}\oplus \{\partial\psi\}&\leftrightarrow \mu_{\sigma,0}\{J\cdot v_\sigma\}\oplus \mu_{\sigma,1}v_\sigma\\
\{\psi^4\}\oplus\{\psi\partial\psi\}&\leftrightarrow \mu_{1,0}\{(\partial J+J^2)\cdot v_1\} \oplus\mu_{1,2}v_1\oplus \mu_{\epsilon,0}\{J\cdot v_\epsilon\}\oplus\mu_{\epsilon,1}v_\epsilon\\
&\ \ \vdots 
\end{aligned}
\end{equation}

This reproduces, in a systematic way, the brute-force classification that we obtained at the beginning of this section by doing group theory, cf.~\eqref{eq:ising_map_nb}.

We remark that the coset $SO(8)_1/Spin(7)_1$ is actually the Ising CFT. Indeed, the branching functions~\eqref{eq:branch_ising_b} agree with the Ising characters. The operators of the gauge theory that survive the infrared limit (i.e., those that are primaries with respect to $Spin(7)_1$) are in bijection with the operators of the Ising theory.

\paragraph{Combining chiralities.} After classifying operators at the chiral level, we will now combine the two chiralities and select the gauge-invariant operators in the UV that survive the infrared limit. To keep the discussion focused, we restrict attention to spinless operator, although the general case is entirely analogous.

For concreteness, we will order the operators by their classical scaling dimension, where $\Delta_\text{UV}(\psi_\ell)=\Delta_\text{UV}(\psi_r)=1/2$. The true dimension in the infrared is typically quite different. Unlike before, here we do not have a parameter (such as $1/N_F$) which, when small, makes $\Delta_\text{IR}\approx\Delta_\text{UV}$. That being said, scaling dimensions satisfy $\Delta_\text{IR}\le\Delta_\text{UV}$, so the labelling by the UV dimension is still useful. Note also that here we do not have any continuous flavor symmetries. There is a discrete $\mathbb Z_2$ symmetry that acts as $\psi_\ell\to-\psi_\ell$.

Recall that a possible basis of chiral operators that survive the infrared limit is
\begin{equation}
1,\qquad\psi_\alpha,\qquad\psi\gamma^A\psi,\qquad\partial\psi_\alpha,\qquad \psi\partial\psi,\qquad\psi\gamma^A\partial\psi,\quad\dots
\end{equation}
where $\psi$ refers to either $\psi_\ell$ or $\psi_r$. Consequently, the first few gauge invariant operators without spin are
\begin{equation}
\begin{aligned}
\text{$\mathbb Z_2$-even:}&\quad1,\quad \psi_\ell\gamma^A\psi_\ell\,\psi_r\gamma_A\psi_r,\quad \psi_\ell D_+\psi_\ell\,\psi_r D_-\psi_r,\quad \psi_\ell\gamma^AD_+\psi_\ell\, \psi_r\gamma_AD_-\psi_r,\ \dots\\
\text{$\mathbb Z_2$-odd:}&\quad \psi_\ell\psi_r,\quad D_+\psi_\ell D_-\psi_r ,\ \dots
\end{aligned}
\end{equation}
These operators have power-law behavior at large distances instead of decaying exponentially. We will next determine the IR image of these operators; they will be in one-to-one correspondence with the operators of Ising, from where we can determine, for example, the infrared scaling dimensions $\Delta_\text{IR}$.

The map of UV to IR operators is contained in~\eqref{eq:chiral_ising_match}. For example, the correspondence $\psi\leftrightarrow\mu_{\sigma,0}v_\sigma$ means that the operator $\psi_\ell\psi_r$ maps into the Ising primary operator $\phi_{\frac{1}{16},\frac{1}{16}}$. Similarly, the correspondence $\{\psi^2\}\leftrightarrow \mu_{1,0}\{J\cdot v_1\}\oplus\mu_{\epsilon,0}v_\epsilon$ means that the operator $\psi_\ell\gamma^A\psi_\ell\,\psi_r\gamma_A\psi_r$ descends into the Ising primary operator $\phi_{\frac12,\frac12}$. Finally, the correspondence $\{\psi^3\}\oplus\{\partial\psi\}\leftrightarrow \mu_{\sigma,0}\{J\cdot v_\sigma\}\oplus \mu_{\sigma,1}v_\sigma$ implies that the operator $D_+\psi_\ell \psi_r $ maps to the Ising operator $L_{-1}\phi_{\frac{1}{16},\frac{1}{16}}$, etc. The next few operators, spinless or otherwise, are:
\begin{equation}
\begin{aligned}
1&\rightsquigarrow\phi_{0,0}\\
\psi_\ell D_+\psi_\ell&\rightsquigarrow L_{-2}\phi_{0,0}\\
\psi_\ell D^2_+\psi_\ell&\rightsquigarrow L_{-3}\phi_{0,0}\\
\psi_\ell D_+^3\psi_\ell&\rightsquigarrow L_{-4}\phi_{0,0}\\
D_+\psi_\ell D_+^2\psi_\ell&\rightsquigarrow L_{-2}^2\phi_{0,0}\\
\psi_\ell D_+\psi_\ell\,\psi_r D_-\psi_r&\rightsquigarrow L_{-2}\bar L_{-2}\phi_{0,0}\\
&\ \ \vdots
\end{aligned}
\end{equation}
for descendants of the vacuum, and
\begin{equation}
\begin{aligned}
\psi_\ell\gamma^A\psi_\ell\,\psi_r\gamma_A\psi_r&\rightsquigarrow\phi_{\frac12,\frac12}\\
\psi_\ell\gamma^AD_+\psi_\ell\,\psi_r\gamma_A\psi_r&\rightsquigarrow L_{-1}\phi_{\frac12,\frac12}\\
D_+\psi_\ell\gamma^AD_+\psi_\ell\,\psi_r\gamma_A\psi_r&\rightsquigarrow L_{-2}\phi_{\frac12,\frac12}\\
\psi_\ell\gamma^AD_+\psi_\ell\,\psi_r\gamma_AD_-\psi_r&\rightsquigarrow L_{-1}\bar L_{-1}\phi_{\frac12,\frac12}\\
D_+\psi_\ell\gamma^AD_+^2\psi_\ell\,\psi_r\gamma_A\psi_r&\rightsquigarrow L_{-3}\phi_{\frac12,\frac12}\\
D_+\psi_\ell\gamma^AD_+\psi_\ell\,\psi_r\gamma_AD_-\psi_r&\rightsquigarrow L_{-2}\bar L_{-1}\phi_{\frac12,\frac12}\\
&\ \ \vdots
\end{aligned}
\end{equation}
for descendants of the energy field, and
\begin{equation}
\begin{aligned}
\psi_\ell\psi_r&\rightsquigarrow\phi_{\frac{1}{16},\frac{1}{16}}\\
D_+\psi_\ell\psi_r&\rightsquigarrow L_{-1}\phi_{\frac{1}{16},\frac{1}{16}}\\
D_+^2\psi_\ell\psi_r&\rightsquigarrow L_{-2}\phi_{\frac{1}{16},\frac{1}{16}}\\
D_+\psi_\ell D_-\psi_r&\rightsquigarrow L_{-1}\bar L_{-1}\phi_{\frac{1}{16},\frac{1}{16}}\\
D_+^3\psi_\ell\psi_r&\rightsquigarrow L_{-3}\phi_{\frac{1}{16},\frac{1}{16}}\\
\psi_\ell\gamma^A\psi_\ell D_+^2\psi_\ell\gamma_A\psi_r&\rightsquigarrow L_{-2}L_{-1}\phi_{\frac{1}{16},\frac{1}{16}}\\
D_+^2\psi_\ell D_-\psi_r&\rightsquigarrow L_{-2}\bar L_{-1}\phi_{\frac{1}{16},\frac{1}{16}}\\
&\ \ \vdots
\end{aligned}
\end{equation}
for descendants of the spin field.

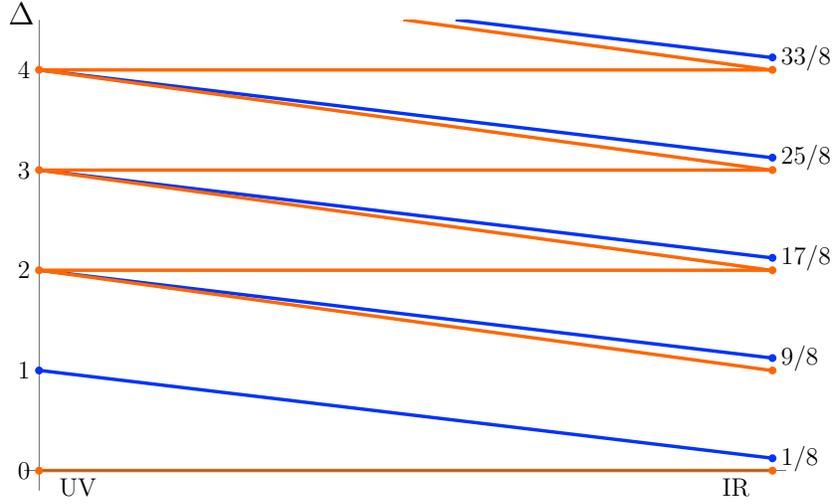
\begin{figure}[h!]
\centering
\begin{tikzpicture}
\definecolor{myorange}{rgb}{1,0.4,0};
\definecolor{myblue}{rgb}{0,0.2,1};
\node at (0,0) {\includegraphics[scale=.8]{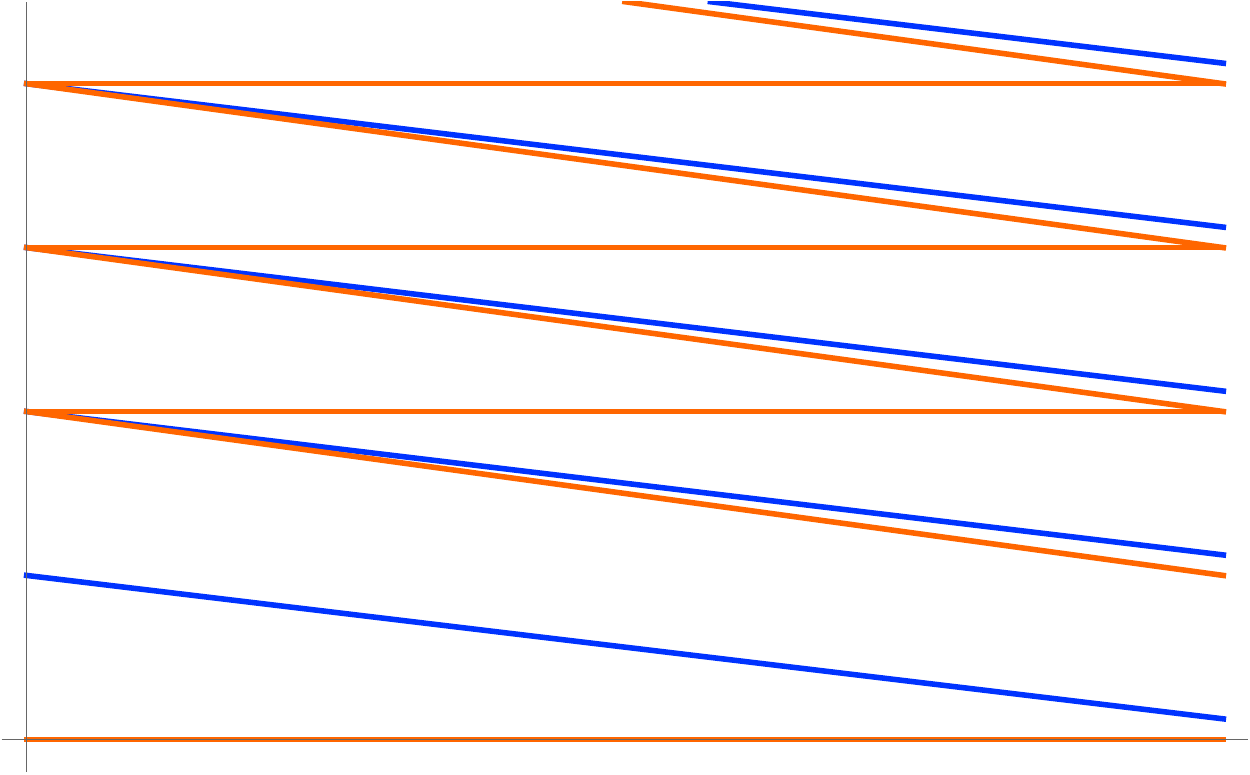}};
\fill[fill=myorange] (-4.865,-2.865) circle (1.5pt);
\node[anchor=east,scale=.8] at (-4.865,-2.865) {$0$};
\node[anchor=north] at (-5.1,3.5) {$\Delta$};
%\node[anchor=north,scale=.8] at (-4.25,-2.85) {$g^2\to0$};
%\node[anchor=north,scale=.8] at (4.3,-2.85) {$g^2\to\infty$};
\node[anchor=north,scale=.8] at (-4.35,-2.85) {UV};
\node[anchor=north,scale=.8] at (4.4,-2.85) {IR};
\fill[fill=myorange] (-4.865+9.75,-2.865) circle (1.5pt);
\fill[fill=myblue] (-4.865,-2.865+.5*2.66666) circle (1.5pt);
\node[anchor=east,scale=.8] at (-4.865,-2.865+.5*2.66666) {$1$};
\fill[fill=myblue] (-4.865+9.75,-2.7) circle (1.5pt);
\node[anchor=west,scale=.8] at (-4.865+9.75,-2.7) {$1/8$};
\fill[fill=myblue] (-4.865+9.75,-2.7+.5*2.66666) circle (1.5pt);
\node[anchor=west,scale=.8] at (-4.865+9.75,-2.7+.5*2.66666) {$9/8$};
\fill[fill=myblue] (-4.865+9.75,-2.7+2.66666) circle (1.5pt);
\node[anchor=west,scale=.8] at (-4.865+9.75,-2.7+2.66666) {$17/8$};
\fill[fill=myblue] (-4.865+9.75,-2.7+1.5*2.66666) circle (1.5pt);
\node[anchor=west,scale=.8] at (-4.865+9.75,-2.7+1.5*2.66666) {$25/8$};
\fill[fill=myblue] (-4.865+9.75,-2.7+2*2.66666) circle (1.5pt);
\node[anchor=west,scale=.8] at (-4.865+9.75,-2.7+2*2.66666) {$33/8$};
\fill[fill=myorange] (-4.865,-2.865+2.66666) circle (1.5pt);
\node[anchor=east,scale=.8] at (-4.865,-2.865+2.66666) {$2$};
\fill[fill=myorange] (-4.865+9.75,-2.865+2.66666) circle (1.5pt);
\fill[fill=myorange] (-4.865+9.75,-2.865+.5*2.66666) circle (1.5pt);
\fill[fill=myorange] (-4.865,-2.865+1.5*2.66666) circle (1.5pt);
\node[anchor=east,scale=.8] at (-4.865,-2.865+1.5*2.66666) {$3$};
\fill[fill=myorange] (-4.865+9.75,-2.865+1.5*2.66666) circle (1.5pt);
\fill[fill=myorange] (-4.865,-2.865+2*2.66666) circle (1.5pt);
\node[anchor=east,scale=.8] at (-4.865,-2.865+2*2.66666) {$4$};
\fill[fill=myorange] (-4.865+9.75,-2.865+2*2.66666) circle (1.5pt);
\end{tikzpicture}
\caption{Renormalization of scaling dimensions in the gauge theory $Spin(7)+\boldsymbol8$. In orange, the $\mathbb Z_2$-even operators and in blue, the $\mathbb Z_2$-odd operators.}
\label{fig:scaling_dim_ising}
\end{figure}

Of course, the spin and $\mathbb Z_2$ charge are preserved under the RG-flow. The scaling dimension gets renormalized, see figure~\ref{fig:scaling_dim_ising}.

\subsection[$G_2$ and matter in $\boldsymbol 7$]{$\boldsymbol{G_2}$ and matter in $\boldsymbol 7$.}\label{sec:tri_critical}

Here we would also like to look at the theory with gauge group $G_2$ (the exceptional simple group) with a quark in the seven-dimensional representation. The basic idea is almost identical to the theory in the previous section so we only highlight the key equations. %This gauge theory flows to the  spin tri-critical Ising CFT in the IR. 

The low-energy coset algebra is $SO(7)_1/G_{2,1}$. Its central charge is $c=7/10$, and therefore this is the tri-critical Ising model. More precisely, $(-1)^F$ is not part of the gauge group, which means that the theory is fermionic; the infrared CFT is the \emph{fermionic} tri-critical Ising theory~\cite{Hsieh:2020uwb,Kulp:2020iet}.

The branching rules take the form
\begin{equation}
\begin{aligned}
d_{\text{NS}-\pm}(q)&=(\chi_{1,1}(q)\pm\chi_{3,1}(q))\chi_{\boldsymbol 1}(q)+(\chi_{3,2}(q)\pm\chi_{3,3}(q))\chi_{\boldsymbol 7}(q)\\
d_\text{R-NS}(q)&=\sqrt2\chi_{2,1}(q)\chi_{\boldsymbol 1}(q)+\sqrt2\chi_{2,2}(q)\chi_{\boldsymbol 7}(q)
\end{aligned}
\end{equation}
where $\chi_{\boldsymbol1},\chi_{\boldsymbol 7}$ are the characters of $G_{2,1}$ and $\chi_{r,s}$ are the $c=7/10$ Virasoro characters
\begin{equation}
\chi_{r,s}(q)=\frac{1}{\eta(q)}\sum_{n\in\mathbb Z}q^{(40 n+5 r-4 s)^2/80}-q^{(40 n+5 r+4 s)^2/80}
\end{equation}
The combinations $\chi_{1,1}\pm\chi_{3,1}$, etc.~are indeed the result of fermionizing the bosonic $c=7/10$ minimal model (see e.g.~\cite{Smith:2021luc}).

Expanding the branching rules, and comparing powers of $q$ on each side, we derive the map of operators. We find the following map of primaries:
\begin{equation}
\begin{aligned}
1&\rightsquigarrow\phi_{0,0}\\
\psi_\ell\psi_r&\rightsquigarrow\phi_{\frac{1}{10},\frac{1}{10}}\\
\psi^2_\ell\psi^2_r&\rightsquigarrow \phi_{\frac{3}{5},\frac{3}{5}}\\
\psi^3_\ell\psi^3_r&\rightsquigarrow \phi_{\frac{3}{2},\frac{3}{2}}\\
\psi^2_\ell\psi_r&\rightsquigarrow \phi_{\frac{3}{5},\frac{1}{10}}\\
\psi_\ell\psi^2_r&\rightsquigarrow \phi_{\frac{1}{10},\frac{3}{5}}\\
\psi^3_\ell&\rightsquigarrow \phi_{\frac{3}{2},0}\\
\psi^3_r&\rightsquigarrow \phi_{0,\frac{3}{2}}
\end{aligned}
\end{equation}
Virasoro descendants are obtained by inserting derivatives into these primaries.

We highlight the fact that the operators $\psi_\ell^3,\psi_r^3$ do not acquire an anomalous dimension. This was actually to be expected, since they are holomorphic and hence protected. We would also like to remark on the fact that their dimension is $3/2$, which is the dimension a supercurrent would have. Indeed, the fermionic tri-critical Ising is the first model in the $\mathcal N=1$ minimal series. This means that the QCD theory $G_2+\boldsymbol 7$ has emergent supersymmetry in the infrared!

It is tempting to conjecture that any QCD theory that has a cubic invariant anti-symmetric symbol will become supersymmetric at large distances. Such symbol allows for the construction of gauge-invariant holomorphic operators $\mu_{ijk}\psi^i\psi^j\psi^k$, whose dimension is fixed to $3/2$ all along the flow. In the infrared, such operator will likely generate a supersymmetry (a possible exception would be a theory whose infrared CFT contains decoupled free fermions, i.e., operators of dimension $1/2$). In some cases, one can indeed show that such operator generates a super-Virasoro algebra~\cite{GODDARD1985226}.

For example, consider $SU(2)$ plus a real fermion with isospin $j\in \mathbb Z$. One can show that $\wedge^3R$ contains a singlet if and only if $j$ is odd, so we are led to expect emergent SUSY for $j=3,5,7,\dots$, and no SUSY for $j=4,6,8,\dots$. For $j=3$ the IR turns out to be again tri-critical Ising, which is indeed supersymmetric. For $j=4$, one can check by brute-force that there is a Ramond field with $h-c/24=-5/248$, which is negative; this indeed shows that the IR is \emph{not} supersymmetric. For $j=5$, one can check that the lightest Ramond field has $h-c/24=1/336$, which is positive; this is compatible with supersymmetry (and, furthermore, the Witten index is easily shown to vanish~\cite{Delmastro:2021xox}, as expected from $h\neq c/24$). These three examples are all consistent with the general conjecture above.

Another interesting example is $SU(N)$ with a fundamental quark, which was already briefly mentioned in~\cite{Delmastro:2021otj}. The infrared coset is $U(1)_N$. The UV theory has an anti-symmetric symbol if and only if $N=3$, since $\wedge^3\ydiagram1=\ydiagram{1,1,1}$. This is again compatible with the conjecture, since $U(1)_N$ is also supersymmetric if and only if $N=3$.\footnote{For $N$ even this is obvious, since $U(1)_N$ is naturally bosonic. For $N$ odd the theory is fermionic, and it can be constructed by extending $U(1)_{4N}$ by a fermionic vertex operator. The lightest Ramond-Ramond field has spin $h=\frac{1}{2\times(4N)}$, which means that $Z_\text{R-R}(q,\bar q)\sim2(q\bar q)^{\frac{1}{8N}-\frac{1}{24}}+\cdots$, which is a constant only for $N=3$.} %Proof: the RR partition function looks something like $Z_{RR}\sim 2(q\bar q)^{\frac{3-N}{24N}}+\cdots$, which can only be constant for $N=3$. Indeed, for $N=3$ we find $Z_{RR}=2$, so the index is two. A more conceptual proof: the $U(1)$ is actually the R-symmetry. An $\mathcal N=2$ SCFT has anomaly $k=3c$ for this symmetry. For $U(1)_N$ the central charge is $c=1$, so the anomaly is $k=3$, and only $U(1)_3$ has this anomaly.

Finally, consider QCD with $N_F$ adjoint quarks. The theory admits a cubic operator of the form $S^{abc}=f_{ijk}\psi^{ia}\psi^{jb}\psi^{kc}$, where $a,b,c$ are flavor indices and $i,j,k$ are color indices, with $f_{ijk}$ the structure constants of the gauge group. The operator $S^{acb}$ is symmetric in its flavor indices. Its traceless symmetric part survives in the infrared, while its trace part $\delta_{ab}S^{abc}\sim J_k\psi^{kc}$ contains the gauge current $J^k=\delta_{ab}f^{ijk}\psi^{ai}\psi^{bj}$ and hence decouples. Thus, we learn that the gauge theory has a spin $3/2$ operator in the infrared, that transforms as a rank-3 traceless symmetric tensor $\ydiagram3$ under the flavor $SO(N_F)$ symmetry. We propose that adjoint QCD has emergent supersymmetry at large distances, possibly as much as $\mathcal N=N_F$. The (admitedly weak) evidence we have for this extended SUSY is the following:
\begin{itemize}
\item The flavor symmetry $SO(N_F)$ is large enough to be the $R$-symmetry of an extended SUSY algebra.
\item When the gauge group is $SU(N)$ and the number of flavors is $N_F=2$, the theory is indeed known to develop $\mathcal N=2$ SUSY~\cite{Gopakumar:2012gd,Isachenkov:2014zua}.
\item When the gauge group is $SU(2)$ (and $N_F$ is arbitrary), the infrared CFT is $SO(N_F)_3$, which is known to have at least $\mathcal N=1$, with supercurrent transforming in the $\ydiagram3$ representation~\cite{Johnson-Freyd:2019wgb}, and is expected to actually have more supercharges.\footnote{We would like to thank Theo Johnson-Freyd for an interesting discussion regarding this point.}
\end{itemize}

For other theories that flow to WZW models in the IR, one can determine whether there is emergent SUSY by using the known (partial) classifications of WZW models that happen to be supersymmetric, see e.g.~\cite{Johnson-Freyd:2019wgb,Bae:2021lvk}. See also~\cite{Kikuchi:2022jbl,Duan:2022ltz} for more recent work on emergent supersymmetry in two dimensions.

\subsection{Adjoint QCD}\label{sec:adjointQCD}

In this section we would like to classify the infrared operators of adjoint QCD, namely a gauge theory with gauge group $G$ and a quark in the adjoint representation. As we have argued, the space of operators that survive the low-energy limit is in bijection with the coset $SO(\dim R)_1/G_h$. It is known~\cite{KAC1988156} that this coset is in fact a TQFT -- it has finitely-many operators -- which already implies that adjoint QCD is a gapped theory. More specifically, the number of infrared operators is $2^r$, where $r$ is the rank of $G$.

There is in fact a very simple argument for this, which does not require a detailed understanding of the coset $SO(\dim R)_1/G_h$. In QCD, the fermion fields $\psi_\ell,\psi_r$ take values in the adjoint of $\mathfrak g$ (times the algebra of Grassmann numbers). In other words, we can expand
\begin{equation}
\psi=\sum_{a=1}^{\dim(\mathfrak g)}\theta_a t^a
\end{equation}
where $\theta_a$ is a set of independent, real-valued Grassmann-odd numbers, and $t^a$ is a basis of $\mathfrak g$ (i.e., the generators of the adjoint representation). In the infrared we are to set $J^a\equiv 0$, a condition that translates to
\begin{equation}
0=\psi^2=\sum_{a>b}\theta_a\theta_b[t^a,t^b]
\end{equation}
or, in other words, $[t^a,t^b]\equiv0$. This means that, while in the UV $\psi$ takes values in the adjoint of $\mathfrak g$, the space of operators that survive the infrared limit are actually spanned by the Cartan subalgebra of $\mathfrak g$, to wit, the condition $J^a=0$ imposes
\begin{equation}
\psi=\sum_{i=1}^{\rank(\mathfrak g)}\theta_i h^i
\end{equation}
where $h^i$ is a basis of the Cartan of $\mathfrak g$. Given that we have two chiralities $\psi_\ell,\psi_r$, the end-result is a Clifford algebra of rank $2r$, which leads to a representation of dimension $2^r$, as claimed.

This simple argument, while quite elementary, does not really give a very explicit expression for the space of infrared operators. In order to be as explicit as possible, let us restrict our attention to $G=SU(N)$, although the analysis and main conclusions, for the other gauge groups, are rather similar.

In adjoint QCD, listing gauge invariant operators is quite simple. We think of $\psi_\ell$ and $\psi_r$ as $N\times N$ hermitian matrices. Any gauge singlet can then be written as a suitable multi-trace operator:
\begin{equation}
\mathcal O_\text{UV}=\tr(\psi_\ell\psi_\ell\psi_r\psi_\ell\cdots)\tr(\psi_r\psi_\ell\psi_r\cdots)\cdots
\end{equation}
where each trace consists of an arbitrary word in $\psi_\ell,\psi_r$. In this situation, removing gauge currents is also rather simple: in matrix notation, the left-moving current is nothing but the product $J_\ell=\psi_\ell^2$ (and the right-moving current $J_r=\psi_r^2$). Therefore, any operator that includes two consecutive left-movers, or two consecutive right-movers, can be written in terms of gauge currents and thus decouples. In other words, the most general gauge singlet that does not include currents is an operator of the form
\begin{equation}
\mathcal O_\text{UV}=\tr(\psi_\ell\psi_r\psi_\ell\psi_r\cdots\psi_\ell\psi_r)\tr(\psi_\ell\psi_r\psi_\ell\psi_r\cdots\psi_\ell\psi_r)\cdots
\end{equation}
where $\psi_\ell$ and $\psi_r$ always appear next to each other.

With this in mind, we learn that, while generic operators can be written in terms of traces of $\psi_\ell$ and $\psi_r$, the massless operators only depend on the fermions via the product $\psi_\ell\psi_r$. If we introduce the notation $U:=\psi_\ell\psi_r$, we see that infrared operators are labeled by integer tuples $n=(n_1,n_2,\dots)$, such that
\begin{equation}
\mathcal O_\text{UV}(n):=\prod_i\tr(U^{n_i})\,.
\end{equation}
Of course, the tuples $(n_1,n_2,\dots)$ and $(n_2,n_1,\dots)$ represent the same operator. Thus, we can take without loss of generality $n_1\ge n_2\ge\cdots$. We can in fact think of the operators $\mathcal O_\text{UV}(n)$ as being labelled by integer partitions of $|n|$:
\begin{equation}
\begin{alignedat}{4}
&|n|=0&&:\ 0&&\quad 1\\
&|n|=1&&:\ 1&&\quad \tr(U)\\
&|n|=2&&:\ 2&&\quad \tr(U^2)\\
&         &&:\ 1+1&&\quad \tr(U)^2\\
&|n|=3&&:\ 3&&\quad \tr(U^3)\\
&         &&:\ 2+1&&\quad \tr(U^2)\tr(U)\\
&         &&:\ 1+1+1&&\quad \tr(U)^3\\
&\qquad\vdots
\end{alignedat}
\end{equation}
Note that $|n|\equiv\Delta_\text{UV}$ is just the classical scaling dimension of $\mathcal O_\text{UV}(n)$. Therefore, there are $p(\Delta_\text{UV})$ operators with scaling dimension $\Delta_\text{UV}\in\mathbb N$, where $p(\,\cdot\,)=1,1,2,3,5,7,\dots$ denotes the partition function.

Were it not for fermi statistics, this would be the end of the story. But, as the fields are fermionic, if any of $n_i$ is sufficiently large, we automatically have $\mathcal O_\text{UV}(n)\equiv0$. This means that the tuple $n$ is actually bounded by above (by some function of $N$). The number of operators with scaling dimension $\Delta_\text{UV}$ is $p(\Delta_\text{UV})$ all the way up to $\Delta_\text{UV}\sim N$, but after that it actually decays and eventually becomes zero. For example, there are no operators with $\Delta_\text{UV}\gg N^2$, as we only have $\sim N^2-1$ fermionic degrees of freedom. That being said, the number of operators is quite large: there are, at least, $\sum_{\Delta_\text{UV}=0}^Np(\Delta_\text{UV})\sim e^{\pi\sqrt{2N/3}}/\sqrt{N}$ infrared operators.

Another complication is the following. While it is true that any massless operator can be written in terms of $U$ alone, the converse is not true: some operators that can be written in terms of $U$ still contain gauge currents, and hence decouple. This dependence on the gauge currents is not manifest, and it is a consequence of the fact that strings of fermions can generically be rearranged. For example, it is straightforward to check that, for $SU(2)$, the following relation holds:
\begin{equation}
\tr(U)^2\equiv \tr(\psi_\ell^2\psi_r^2)\,.
\end{equation}
Therefore, while the \emph{l.h.s.}~is written in terms of $U$ alone, and does not contain any manifest factors of the gauge currents, the \emph{r.h.s.}~does, and thus the operator $\tr(U)^2$ in fact decouples at large distances. In conclusion, the partition $1+1$ does not describe an infrared operator in adjoint $SU(2)$.

The two observations above can be summarized as follows: all infrared operators are labeled by integer partitions, but not every integer partition labels a valid infrared operator. Some integer partitions yield identically vanishing operators, and some others yield operators that can be rearranged into expressions that contain gauge currents and hence decouple. It turns out that the rule to decide whether a given partition is valid or not is quite simple: if we represent the partition $n=(n_1,n_2,\dots)$ as a Young diagram, then the diagram represents a valid operator if and only if the number of rows, plus the number of columns, is at most $N$. The generating function of such restricted partitions is well-known: the number of infrared operators with classical scaling dimension $\Delta_\text{UV}$ is
\begin{equation}
[q^{\Delta_\text{UV}}]\sum_{k=0}^N\binom{N}{k}_q-\binom{N-1}{k}_q
\end{equation}
where $\binom{N}{k}_q$ denotes the $q$-binomial coefficient. Note that this agrees with $p(\Delta_\text{UV})$ for all $\Delta_\text{UV}\le N$, while it decays for $\Delta_\text{UV}\gg N$ and eventually vanishes for $\Delta_\text{UV}>\lfloor N^2/4\rfloor$. The total number of operators is given by the generating function at $q=1$:
\begin{equation}
\sum_{k=0}^N\binom{N}{k}-\binom{N-1}{k}\equiv 2^N-2^{N-1}=2^{N-1}\,.
\end{equation}
This is precisely the expected number of operators, which generically reads $2^r$ with $r$ the rank of the gauge group. Therefore, we have obtained the complete list of UV operators that survive the IR limit. This list is finite, since the theory is gapped. All these operators have $\Delta_\text{IR}=0$.

We note that a similar counting, in a different but related context, was discussed in~\cite{Isachenkov:2014zua} (these authors were classifying chiral primaries in \emph{two} flavor adjoint QCD).

We can also describe explicitly the map of UV to IR operators. Recall that the infrared theory is described by the topological coset $SO(N^2-1)_1/SU(N)_N$, whose operators are labelled by integrable representations of $SU(N)_N$. The map from UV operators to IR operators is very simple:% the operator $\mathcal O_\text{UV}(n)$ labelled by the partition $n$ descends to the $SU(N)_N$ representation labelled by the Young diagram given by ${\ytableausetup{notabloids,boxsize=1em}\begin{ytableau}\none & n \\n^t & \none\end{ytableau}}\ytableausetup{boxsize=.5em}$, where ${\ytableausetup{notabloids,boxsize=1em}\begin{ytableau}n\end{ytableau}}\ytableausetup{boxsize=.5em}$ denotes the diagram associated to the partition $n$, and ${\ytableausetup{notabloids,boxsize=1em}\begin{ytableau}n^t\end{ytableau}}\ytableausetup{boxsize=.5em}$ is its transpose.
%The explicit operator mapping betweem UV and IR operators is 
\begin{equation}
\frac{1}{|n|!}\sum_{\sigma \in S_{|n|}} \chi_n(\sigma) \tr\bigl(\sigma\cdot U\bigr)\overset{\text{IR}}{\longrightarrow}SU(N)_N~\text{integrable representation}~  {\ytableausetup{notabloids,boxsize=1.1em}\begin{ytableau}\none & n \\n^t & \none\end{ytableau}}\ytableausetup{boxsize=.5em}\,,\\
%&\lambda=\text{Young diagram of partition }(n_i)  \hbox{{\bf como decir en tÔøΩrminos de R}}
\end{equation}
where $\chi_n(\sigma)$ are characters of the permutation group $S_{|n|}$. For example, the UV mass term $\tr(U)$ is labeled by the partition $n=(1)$, whose Young diagram is $\ydiagram 1$. This descends to the $SU(N)_N$ adjoint representation, $\ydiagram{1+1,1}$.

%\note{For example, the mass term $\tr(U)$ is labelled by the partition $n=(1)$, whose Young diagram is $\ydiagram 1$. Therefore, this operator maps to the $SU(N)_N$ representation with Young diagram $\ydiagram{1+1,1}$, i.e., the adjoint representation. Similarly, the UV operator $\tr(U^2)\tr(U)$ corresponds to the partition $n=(2,1)$, whose associated Young diagram is $\ydiagram{3,1}$; hence, this operator descends to the $SU(N)_N$ representation with Young diagram $\ydiagram{2+3,2+1,2,1,1}$ (or, in terms of Dynkin labels, $(2,1,0,\dots,0,1,0,1)$).}

\subsection{Schwinger model}

We close this section with one last example: QED. We consider a $U(1)$ gauge theory with a charge-$q$ quark. This theory can be solved exactly via bosonization. The infrared is gapped and described by a $\mathbb Z_q$ gauge theory of $q$ degenerate vacua. The branching rules of
\begin{equation}
\frac{U(1)_1}{U(1)_{q^2}}=\mathbb Z_q
\end{equation}
determine the $q$ vacua explicitly. We find that the $q$ operators that survive the infrared limit are
\begin{equation}
v_n:=\psi\partial\psi\partial^2\psi\cdots\partial^n\psi
\end{equation}
for $n=0,\dots,q-1$. More precisely, these are the operators at the chiral level; the non-chiral gauge-invariant operators are $v_{n,\ell}v^\dagger_{n,r}$. The operator $v_{2,\ell}v^\dagger_{2,r}$ was also considered in~\cite{Cherman:2022ecu}.

\section{Deformations and flows}\label{sec:deformation}

So far, we have only considered QCD as defined by massless quarks and the usual cubic gauge interaction $A^a J^a$. An important follow-up is to study deformations of this system, such as turning on masses for the quarks or other interactions, such as four-fermi terms $\sim\psi^4$. When the deformation is large, it is hard to say anything concrete: one has to solve the new problem from scratch. On the other hand, if the deformation is small enough, one can proceed perturbatively: turning on an operator $\mathcal O_\text{UV}$ with small coefficient $\lambda$ is equivalent to first flowing to the infrared, and then turning on the image $\mathcal O_\text{IR}$ in the IR CFT. This is illustrated in figure~\ref{fig:deformation_flow}.

Given the identification of the low-energy effective theory for a given QCD theory in the form of a GKO coset $\mathcal T_\text{IR}\equiv SO(\dim R)_1/G_{T(R)}$, and the explicit map of UV to IR operators as discussed in the previous section, these deformations of QCD are formally solvable as well, or at least reducible to a known problem in CFT. First, we identify the infrared image of whatever operator we wish to deform the UV theory by. Then, we study the effect of this deformation in the IR; this last step is hard in general, although some cases have been understood in the literature, and some others can be tackled using well-developed tools, such as conformal perturbation theory and the truncated conformal space approach.

Ideally, we would also like to be able to explain the flow from $\mathcal T_\text{UV}$ to $\widetilde{\mathcal T}_\text{IR}$ directly, without having to go through $\mathcal T_\text{IR}$. In all examples we have looked at, we find that such an explanation actually exists: if we assume that a suitable fermion bilinear condenses, this triggers a semiclassical flow from $\mathcal T_\text{UV}$ to $\widetilde{\mathcal T}_\text{IR}$ that can be followed explicitly. We have no first-principles argument as to why the corresponding operator condenses under the deformation, but we find it remarkable that such a mechanism exists in the first place, which is able to explain the deeply non-perturbative flow $\mathcal T_\text{UV}\to\widetilde{\mathcal T}_\text{IR}$. We leave it to future work to explore this in more detail.

\begin{figure}[!h]
\centering
\begin{tikzpicture}
\node[right] at (5.05,5) {$G+R$};
\node[right] at (4.5,1+1) {$\frac{SO(\dim R)_1}{G_{T(R)}}$};
\node[right] at (6,-.5+1) {$\frac{SO(\dim R')_1}{H_{T(R')}}$};
\draw[thick,->,>=stealth,decoration={snake,amplitude=1.5pt,post length=3pt},decorate] (5.4,5-.3) -- (5.4,1+.5+1);
\draw[thick,->,>=stealth,decoration={snake,amplitude=1.5pt,post length=3pt},decorate] (5.4,5-.3) -- (7,.05+1);
\draw[thick,->,>=stealth,decoration={snake,amplitude=1.5pt,post length=3pt},decorate] (5.4,1.6) -- (6.9,0+1);

\node[right] at (8,3) {$\begin{aligned}G&\supset H\\\langle\psi_\ell\psi_r\rangle&\mapsto \boldsymbol1+\cdots\end{aligned}$};
\end{tikzpicture}
\caption{In the examples we have studied, the effect of deforming QCD with gauge group $G$ and matter in $R$ can be reproduced by assuming that a fermion composite condenses. This condensation breaks $G\to H$ and $R\mapsto R'+\cdots$ according to the standard Higgs mechanism, namely $H$ is a subgroup of $G$ that leaves $\langle\psi_\ell\psi_r\rangle$ invariant. The new theory flows in the IR to the coset associated to the new gauge theory data $(H,R')$ according to the usual rules.}
\label{fig:deformation_flow_higgs}
\end{figure}
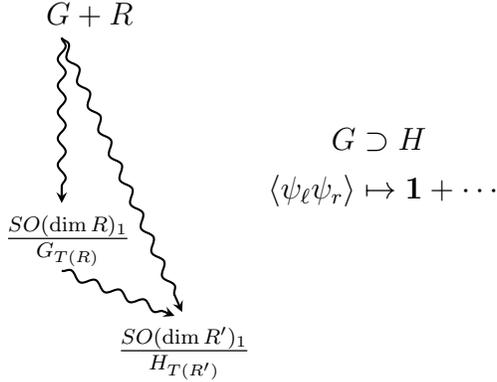

The basic idea is the following. Let us begin with some gauge group $G$ and quarks in some representation $R$. The dynamical assumption is that the deformations of this theory can be modeled by the assumption that some fermion composite condenses. The condensing field is always a singlet under the flavor symmetry (in accordance with the Coleman theorem), but it transforms non-trivially under the gauge group. Its condensation Higgses the group down $G\to H$, and gives mass to some fermions. The end-result is some new gauge theory, with smaller gauge group $H$ and massless quarks in some representation $R'$; this new theory flows to the corresponding GKO coset $SO(\dim R')_1/H_{T(R')}$. The claim is that the infrared of the deformed theory is precisely this GKO coset, see figure~\ref{fig:deformation_flow_higgs}.

In the remainder of this section we illustrate this discussion via several examples.

\subsection{QCD with fundamental matter}

Let us study deformations of $U(N)$ plus $N_F$ massless Dirac fermions in the fundamental representation. Recall that the infrared CFT for this theory was argued to be $SU(N_F)_N$.

The most relevant deformation one can consider is a mass term. If we turn on a sufficiently large mass for a given flavor, then we can integrate out the quark and we are left with $U(N)$ and $N_F-1$ fermions. The low-energy CFT is $SU(N_F-1)_N$. If the mass is small, it is unclear if this is still correct, since we cannot integrate out the fermion reliably. The infrared CFT offers an answer: the effective dynamics for a small mass is given by the mass deformation of the WZW model $SU(N_F)_N$. Recall that the mass term was argued to descend to the primary $(\ydiagram1,\ydiagram1)$, and therefore the model we are to look at is $SU(N_F)_N+m(\ydiagram1,\ydiagram1)$. This deformation is in fact rather straightforward, since it preserves an $SU(N_F-1)_\ell\times SU(N_F-1)_r$ algebra, and therefore the end result is indeed an $SU(N_F-1)_N$ WZW model, precisely as in the large mass case.

More generally, note that any QCD theory with a chiral symmetry $G_\ell\times G_r$ always has a $G_{\ell,k_\ell}\times G_{r,k_r}$ WZW subsector in the infrared, as the symmetries are generated by homomorphic currents (which are protected under RG flow). If the mass terms break $G_i\to H_i$, then the WZW subsector in the infrared CFT will accordingly break to $H_{\ell,k'_\ell}\times H_{r,k'_r}$ (where $k'_i=x_i k_i$, with $x_i$ the embedding index of $H_i\subset G_i$).

For a more interesting deformation, consider turning on a four-fermi operator of the form $\psi_\ell\psi_\ell^\dagger\psi_r\psi_r^\dagger$. This flows in the infrared to the $(\ydiagram{1+1,1},\ydiagram{1+1,1})$ primary of $SU(N_F)_N$. Therefore, QCD deformed by this four-fermi operator is equivalent to $SU(N_F)_N+\lambda(\ydiagram{1+1,1},\ydiagram{1+1,1})$. This deformation preserves a diagonal $SU(N_F)$ symmetry, which enhances to $SU(N_F)_{\ell,k}\times SU(N_F)_{r,k}$ at the end of the flow, for some level $k\le N$. Anomalies and other strong-coupling considerations suggest that $k=\gcd(N,N_F)$~\cite{Lecheminant:2015iga,Tanizaki:2018xto,Ohmori:2018qza}. Therefore, we conclude that our original QCD theory, deformed by this four-fermi operator, flows at large distances to the $SU(N_F)_k$ CFT, where the most natural conjecture for $k$ is $\gcd(N,N_F)$.

As a matter of fact, there is a semiclassical argument that can explain this flow directly in terms of the UV variables. We assume that, to a rough approximation, the sole effect of the deformation is to make some fermion composite condense. The simplest operator that is a singlet both under Lorentz and flavor is
\begin{equation}
\mathcal O^i{}_j=\psi^{i \alpha	\dagger}_\ell \psi_{\ell,k \alpha}\psi^{k \beta\dagger}_r\psi_{r,j\beta}
\end{equation}
where $i,j,k$ are color indices and $\alpha,\beta$ are flavor indices. If $\mathcal O^i{}_j$ -- or any other operator that is an adjoint under the gauge group -- condenses, then the gauge group will break from $U(N)$ to some smaller subgroup, one that leaves $\mathcal O^i{}_j$ invariant. The precise symmetry-breaking pattern depends on the eigenvalues of $\mathcal O$; generically, we expect
\begin{equation}\label{eq:UNtoUk}
U(N)\to U(k_1)\times U(k_2)\times\cdots\times U(k_n)
\end{equation}
with $\sum_i k_i=N$. Under this breaking, each fermion branches as $\ydiagram 1\mapsto\bigoplus_i (\boldsymbol1,\dots,\boldsymbol 1,\ydiagram1,\boldsymbol1,\dots,\boldsymbol 1)$, with $\ydiagram1$ at position $i$. All but one of these pick up a mass and decouple. The end-result is an effective theory
\begin{equation}
U(N)+N_F\ydiagram1\longrightarrow U(k)+N_F\psi
\end{equation}
for some $k\in\{k_1,\dots,k_n\}$, depending on which factor in~\eqref{eq:UNtoUk} remains massless. The infrared CFT of this gauge theory is $SU(N_F)_k$, precisely as we obtained before. It would be very interesting to find a mechanism that explains why the massless factor is $k=\gcd(N,N_F)$, if at all.

It would also be interesting to understand the flow triggered by the operators $\psi_\ell\psi_\ell\psi^\dagger_r\psi^\dagger_r$. In the infrared CFT, these map into the primaries $(\ydiagram{1,1},\ydiagram{1,1})$ and $(\ydiagram2,\ydiagram2)$. For large $N_F$, these operators are nearly marginal, $\Delta=2-\frac{2(N\pm1)}{N_F}+\mathcal O(1/N^2_F)$, and therefore the flow can likely be understood perturbatively.

Generically, we do not expect other interesting deformations in this theory, since the rest of operators in $SU(N_F)_N$ are all irrelevant.

\subsection[$Spin(7)$ and matter in $\boldsymbol 8$]{$\boldsymbol{Spin(7)}$ and matter in $\boldsymbol 8$.}

Let us now look at deformations of the gauge theory with group $Spin(7)$ and matter in the spinor representation. Recall that this theory was argued to flow to the Ising model.

The most relevant deformation is the mass term $\psi_\ell\psi_r$. For large mass, we can integrate out this field and we are left with pure Yang-Mills, which is trivially gapped. On the other hand, for small masses, we cannot integrate out the quark. Instead, we are to look at the deformed Ising model. Recall that the mass term was argued to flow in the infrared to the spin operator $\phi_{1/16}$. Hence, the massive gauge theory flows at large distances to the Ising model perturbed by the spin operator. The effect of such deformation is well-known: the operator trivially gaps the theory.\footnote{Furthermore, the flow is integrable and one ends up with eight massive particles~\cite{Fateev:1990hy}. It is tempting to speculate that these eight particles are related to the eight components of the quark in the UV theory.\label{fn:E8}} Therefore, we learn that the large mass and small mass phases are both trivially gapped. Although we cannot rule out a more complicated phase diagram, it is reasonable to assume that there are no phase transitions as a function of $m$ (other than the massless point $m=0$).

Another interesting deformation is that triggered by $\psi_\ell^2\psi_r^2$. This maps in the infrared to the energy operator $\phi_{1/2}$. The effect of this deformation is also understood: it gaps the system. For one sign of the coupling, the vacuum is non-degenerate, while for the other sign, there are two vacua. Therefore, the gauge theory deformed by the four-fermi operator is gapped, with either one or two vacua, depending on the sign of the interaction.

As before, there is a simple semiclassical argument that yields the same answers. Let us assume that, under the deformation, a condensate of the form $\langle \psi_\ell\psi_r\rangle$ is generated. The possible condensates are classified by $\boldsymbol 8\times\boldsymbol 8=\boldsymbol1+\boldsymbol7+\boldsymbol{21}+\boldsymbol{35}$. If a condensate in one of these four directions is generated, the gauge group will break into some subgroup that leaves that direction invariant. The maximal subgroups of $Spin(7)$ are $Spin(6)$, $Spin(5)\times Spin(2)$, $Spin(4)\times Spin(3)$ and $G_2$. For example, under $Spin(7)\supset Spin(6)$, the direction $\boldsymbol{21}$ branches as $\boldsymbol6+\boldsymbol{15}$, neither of which is a scalar, which means that $\boldsymbol{21}$ has no invariant component under $Spin(6)$. On the other hand, the direction $\boldsymbol 7$ branches as $\boldsymbol 1+\boldsymbol 6$, which does have a scalar component. This means that a condensate in the direction of $\boldsymbol 7$ may break the gauge group as $Spin(7)\to Spin(6)$, but a condensate in the direction of $\boldsymbol{21}$ cannot. More generally, the possible (maximal) breaking patterns are
\begin{itemize}
\item A condensate $\langle\psi_\ell\psi_r\rangle\in\boldsymbol 7$ Higgses the gauge group down to $Spin(6)$,
\item A condensate $\langle\psi_\ell\psi_r\rangle\in\boldsymbol{21}$ Higgses the gauge group down to $Spin(5)\times Spin(2)$,
\item A condensate $\langle\psi_\ell\psi_r\rangle\in\boldsymbol{35}$ Higgses the gauge group down to $Spin(4)\times Spin(3)$ or $G_2$.
\end{itemize}

Let us begin by looking the first possibility. Under this breaking, the quark branches as $\boldsymbol 8\mapsto\boldsymbol 4+\overline{\boldsymbol 4}$. The condensate corresponds to the singlet in $\boldsymbol 4\times\overline{\boldsymbol 4}=\boldsymbol1+\cdots$, which means that both quarks pick up a mass. The resulting theory is trivially gapped.

Similarly, if we assume that $\boldsymbol{21}$ condenses, then the gauge group becomes $Spin(5)\times Spin(2)$, under which $\boldsymbol8\mapsto (\boldsymbol 4,\boldsymbol 2)$. As before, the quark pick up a mass, and the resulting theory is trivially gapped. Finally, if we assume that $\boldsymbol{35}$ condenses, then the gauge group becomes $Spin(4)\times Spin(3)$ or $G_2$, under which $\boldsymbol 8\mapsto (\boldsymbol2,\boldsymbol2)+(\boldsymbol2',\boldsymbol2)$ and $\boldsymbol8\mapsto \boldsymbol1+\boldsymbol7$, respectively. Any of these components can in principle pick up a mass, so we consider all possibilities:
\begin{itemize}
\item If either of $(\boldsymbol2,\boldsymbol2)$, $(\boldsymbol2',\boldsymbol2)$ becomes massive, the infrared effective theory becomes $Spin(3)^2+(\boldsymbol2,\boldsymbol2)$, which has two vacua,
\item If $\boldsymbol 1$ becomes massive, the infrared effective theory becomes $G_2+\boldsymbol 7$, which flows to the tri-critical Ising model (see next section). Of course, it is not possible to flow from Ising to tri-critical Ising, so this condensate is not kinematically consistent,
\item If $\boldsymbol 7$ becomes massive, then the infrared effective theory becomes $G_2+\boldsymbol 1$, which is still the Ising model.
\end{itemize}

Summarizing, we see that all possible symmetry breaking patters predict either a single vacuum, or a pair of vacua, which is precisely the set of possible endpoints of all relevant flows of the Ising model, as summarized above. (There is also a breaking pattern that does not change the large distance behaviour, which can formally be identified with an irrelevant deformation.)

\subsection[$G_2$ and matter in $\boldsymbol 7$]{$\boldsymbol{G_2}$ and matter in $\boldsymbol 7$.}\label{sec:tri_critical_def}

Another interesting example is a $G_2$ gauge theory with a fermion in the seven-dimensional representation. The infrared coset of this theory is
\begin{equation}
\frac{SO(7)_1}{G_{2,1}}
\end{equation}
which is easily identified with the fermionic tri-critical Ising model (for one thing, the central charge is $c=\frac72-\frac{1}{1+4}\times14=\frac{7}{10}$). By analyzing the branching rules of this coset, one learns that the primaries map as follows:
\begin{equation}
\psi\leftrightarrow \phi_{1/10},\quad \psi^2\leftrightarrow \phi_{3/5},\quad \psi^3\leftrightarrow \phi_{3/2}\,.
\end{equation}
The only relevant deformations are, then, the mass term $\psi_\ell\psi_r$ and the four-fermi term $\psi_\ell^2\psi_r^2$, which map to the two relevant directions of the tri-critical ising model, $\phi_{1/10}$ and $\phi_{3/5}$, respectively.

The effect of the mass term is quite straightforward. In the original UV theory, the mass term gaps out the fermion, and we are left with pure Yang-Mills, which is trivially gapped. This is indeed reproduced by the IR coset, since the tri-critical Ising model deformed by $\phi_{1/10}$ is known to be trivially gapped for either sign of the deformation.\footnote{As in footnote~\ref{fn:E8}, this flow is again integrable and there are seven massive particles~\cite{Fateev:1990hy}, which again appears to be the number of quark components in the UV.}

The effect of the four-fermi term is more interesting. From the point of view of the IR theory, this term is known to trivially gap the theory for one sign of the deformation, while it triggers a flow to the free Majorana fermion CFT for the other sign (this being the fermionization of the Ising model). Therefore, we learn that the original gauge theory, deformed by the four-fermi term, is either gapped or a free fermion, depending on the sign of the deformation.

Interestingly, this same conclusion can be reached if we assume that the deformation induces a bilinear condensate
\begin{equation}
\langle\psi_\ell\psi_r\rangle\neq 0\,.
\end{equation}
The possible condensates are classified by $\boldsymbol 7\times\boldsymbol 7=\boldsymbol 1+\boldsymbol 7+\boldsymbol{14}+\boldsymbol{27}$. For concreteness, let us look at a condensate in the direction of $\boldsymbol 7$. This Higgses the gauge group down to $G_2\to SU(3)$, under which the fermion decomposes as $\boldsymbol 7\mapsto\boldsymbol 1+\boldsymbol{3}+\overline{\boldsymbol 3}$. The invariant direction is contained in $\boldsymbol{3}\times\overline{\boldsymbol 3}=\boldsymbol 1+\cdots$, which means that both $\boldsymbol3$ and $\overline{\boldsymbol 3}$ pick up a mass, leaving behind only the component $\boldsymbol 1$. This means that the low energy theory contains a single massless Majorana fermion, and nothing else, precisely as expected from the known flow discussed above.

%\subsection{Adjoint QCD}

%\note{
%Deformations of adjoint QCD are more boring than the other examples since the theory is gapped. There are $2^r$ topological operators, all of which are relevant in the IR. We can lift vacua just by deforming by these operators. It would be interesting but also challenging to figure out the precise pattern of lifting of vacua.

%In keeping in line with the philosophy of this section, we can mention one interesting possibility for this problem. We can assume that the lifting of vacua is realized by a fermion condensate, as in the previous examples. The simplest case to describe is a full Higgsing of the gauge group down to its maximal torus $U(1)^r$. Generically, this will lead to $\theta$ angles for these $U(1)$ factors. Any $\theta\neq\pi$ has a single vacuum, while $\theta=\pi$ leads to two-fold degeneracy. This means that the number of vacua descends from $2^r$ down to $2^{r'}$, where $r'$ is the number of $U(1)$ factors with $\theta=\pi$. It would be nice to understand which UV operators lead to this particular lifting $2^r\to 2^{r'}$.

%}

\subsection{Minimal models}

Another interesting example is the following. Consider the QCD theory with gauge group
\begin{equation}
SU(2)^3\times SO(k)
\end{equation}
for some $k\in\mathbb N$, and quarks in the representation
\begin{equation}
R_1\oplus R_2=(\boldsymbol2,\boldsymbol1,\boldsymbol2,\boldsymbol1)+(\boldsymbol2,\boldsymbol2,\boldsymbol1,\ydiagram1)
\end{equation}

This theory flows in the infrared to the coset
\begin{equation}
\frac{SO(4k+4)_1}{SU(2)_{k+1}\times SU(2)_k\times SU(2)_1\times SO(k)_4}
\end{equation}
By computing the central charge, it is easy to see that this coset realizes the $k$-th Virasoro minimal model $\mathcal M_k$, with $k=1,2,\dots$ corresponding to $c=\frac12,\frac{7}{10},\dots$.

A deformation of the minimal models that is particularly well understood is the $\phi_{1,3}$ field, as its spin is $\frac{k+1}{k+3}=1-\frac2k+\mathcal O(1/k^2)$, and hence the deformation is perturbatively close to marginality. There are many UV operators that flow to $\phi_{1,3}$:
\begin{equation}\label{eq:phi13_UV}
\begin{aligned}
\psi^{(2)}&\leftrightarrow\phi_{1,3} \times(\boldsymbol3,\boldsymbol1,\boldsymbol1,\boldsymbol1)+\cdots\\
\psi^{(4)}&\leftrightarrow\phi_{1,3} \times(\boldsymbol3,\boldsymbol3,\boldsymbol1,\ydiagram2)+\cdots\\
\psi^{(6)}&\leftrightarrow\phi_{1,3} \times(\boldsymbol3,\boldsymbol1,\boldsymbol1,\ydiagram4)+\cdots\\
&\ \ \vdots\\
\psi^{(k+1)}&\leftrightarrow\phi_{1,3}\times\big[(\boldsymbol k,\boldsymbol{k+1},\boldsymbol2,\boldsymbol1)+(\boldsymbol k,\boldsymbol{k-1},\boldsymbol2,\ydiagram2)\bigr]+\cdots
\end{aligned}
\end{equation}
where $\psi^{(n)}$ denotes an operator with $n$ factors of $\psi$ (and some index structure so as to reproduce the representation on the \emph{r.h.s.}).

If we deform the UV theory by any of these operators, the resulting flow corresponds to the IR theory $\mathcal M_k+\lambda\phi_{1,3}$. The effect of this deformation is well-known: it triggers a flow to the next minimal model $\mathcal M_{k-1}$. In conclusion, the original theory deformed by any of the operators in~\eqref{eq:phi13_UV} flows at large distances to the $(k-1)$-th minimal model.

%The UV operator that flows to this IR primary is
%\begin{equation}\label{eq:minimal_model_13_map}
%\psi^{k+1}\leftrightarrow\phi_{1,3}\times\bigg[(\boldsymbol k,\boldsymbol{k+1},\boldsymbol2,\boldsymbol1)+(\boldsymbol k,\boldsymbol{k-1},\boldsymbol2,\ydiagram2)\biggr]+\cdots
%\end{equation}
%where by $\psi^{k+1}$ we mean one copy of $R_1$, and $k$ copies of $R_2$:
%\begin{equation}\label{eq:minimal_model_tensor_product}
%R_1\times\wedge^k R_2=(\boldsymbol k,\boldsymbol{k+1},\boldsymbol2,\boldsymbol1)+(\boldsymbol k,\boldsymbol{k-1},\boldsymbol2,\ydiagram2)+\cdots
%\end{equation}
%where $\wedge^kR_2=(\boldsymbol{k+1},\boldsymbol{k+1},\boldsymbol1,\boldsymbol1)+(\boldsymbol{k-1},\boldsymbol{k-1},\boldsymbol1,\ydiagram2)+\cdots$. As a check of~\eqref{eq:minimal_model_13_map}, note that the $j$-th primary of $SU(2)_k$ has spin $\frac{j(j+1)}{k+2}$, while the symmetric representation of $SO(n)_k$ has spin $\frac{n}{n+k-2}$. Therefore, the two terms on the \emph{r.h.s.}~of~\eqref{eq:minimal_model_13_map} have spins
%\begin{equation}
%\begin{alignedat}{5}
%&&\frac{k+1}{k+3}&+\frac{k^2-1}{4(k+3)}\,+\,&\frac{k}{4}\qquad\quad&\,+\,\frac14\,+\,&\,0\qquad&\equiv\frac{k+1}{2}\\
%&&\frac{k+1}{k+3}&+\frac{k^2-1}{4(k+3)}\,+\,&\frac{k(k-2)}{4(k+2)}&\,+\,\frac14\,+\,&\frac{k}{k+2}&\equiv\frac{k+1}{2}
%\end{alignedat}
%\end{equation}
%which matches the spin of the \emph{l.h.s.}~of~\eqref{eq:minimal_model_13_map}

As before, this flow can be explained via a suitable condensate. We assume that the effect of the deformation in the UV can be approximated by the assumption that $\langle\psi_\ell\psi_r\rangle$ picks up a vacuum expectation value. For definiteness, we assume that the condensing field is $\psi_2$, namely $(\boldsymbol2,\boldsymbol2,\boldsymbol1,\ydiagram1)$. The possible condensates are classified by $(\boldsymbol2,\boldsymbol2,\boldsymbol1,\ydiagram1)^2=(\boldsymbol1+\boldsymbol3,\boldsymbol1+\boldsymbol3,\boldsymbol1,\boldsymbol1+\ydiagram2+\ydiagram{1,1})$. The minimal (non-trivial) condensate is $(\boldsymbol1,\boldsymbol1,\boldsymbol1,\ydiagram2)$, which breaks the gauge group as little as possible, namely $SO(k)\to SO(k-1)$. Under this breaking, the vector branches as $\ydiagram1\mapsto\ydiagram1+\boldsymbol1$. Therefore, the quarks become $(\boldsymbol2,\boldsymbol1,\boldsymbol2,\boldsymbol1)+(\boldsymbol2,\boldsymbol2,\boldsymbol1,\ydiagram1)+(\boldsymbol2,\boldsymbol2,\boldsymbol1,\boldsymbol1)$, with the last being massive. Hence, all in all, the effect of the condensation is to break the gauge group from $SU(2)^3\times SO(k)$ to $SU(2)^3\times SO(k-1)$, while leaving the same massless content. In other words, the condensation just decreases $k\to k-1$, which is exactly the effect of the $\phi_{1,3}$ field in the infrared minimal model.

Needless to say, there are many other relevant deformations of minimal models; generically, we can break any $k$ to any other $k'$ as long as $k'<k$. This is easily reproduced by the condensate heuristic, since we can always arrange a symmetry breaking $SO(k)\to SO(k')$, as long as $k'<k$.

It is also interesting to understand the effect of mass terms. In the UV, if either fermion picks up a sufficiently large mass, the theory becomes gapped. Indeed, in this situation we can integrate out the massive quark, and the gauge theory becomes
\begin{equation}
\begin{aligned}
SU(2)^2\times SO(k)+(\boldsymbol2,\boldsymbol2,\ydiagram1)&\equiv SO(4)\times SO(k)+(\ydiagram1,\ydiagram1)\\
SU(2)^2+(\boldsymbol2,\boldsymbol2)&\equiv SO(4)+\ydiagram1
\end{aligned}
\end{equation}
depending on which quark is massive. Both of these are in the list of gapped theories obtained in~\cite{Delmastro:2021otj}.

It is unclear what the fate of the theory is if the masses are small, since in that case we cannot integrate out the quarks reliably. The infrared CFT provides an answer for this. The IR image of the mass terms is as follows:
\begin{equation}
\psi\leftrightarrow \phi_{1,2}\times(\boldsymbol2,\boldsymbol1,\boldsymbol2,\boldsymbol1)+\phi_{2,2}\times(\boldsymbol2,\boldsymbol2,\boldsymbol1,\ydiagram1)\,.
\end{equation}
%Note that the spin of $\phi_{1,2}$ is $\frac{k}{4(k+3)}$ and the spin of $\phi_{2,2}$ is $\frac{3}{4(k+2)(k+3)}$. On the other hand, the vector representation of $SO(n)_k$ has spin $\frac{n-1}{2 (k+n-2)}$. Therefore, the spins of the \emph{r.h.s.}~are
%\begin{equation}\everymath={\displaystyle}
%\begin{array}{ccccccccccc}
%\frac{k}{4(k+3)}&+&\frac{3}{4(k+3)}&+&0&+&\frac{1}{4}&+&0&\equiv&\frac12\\
%\frac{3}{4(k+2)(k+3)}&+&\frac{3}{4(k+3)}&+&\frac{3}{4(k+2)}&+&0&+&\frac{k-1}{2 (k+2)}&\equiv&\frac12
%\end{array}
%\end{equation}
%which agrees with the spin of $\psi$.
Therefore, the original QCD theory deformed by small masses flows at large distances to $\mathcal M_k+m_1 \phi_{1,2}+m_2\phi_{2,2}$. The deformation by $\phi_{1,2}$ is in fact well-understood -- it is integrable -- and it indeed gaps the theory~\cite{Zamolodchikov:1987jf}. On the other hand, the deformation by $\phi_{2,2}$ is not integrable and it is not as well understood as the previous one. The gauge theory makes a concrete prediction for this deformation: it should gap the theory for either sign. We can in fact provide some evidence for this claim: in the Landau-Ginzburg description of the minimal models, where the potential is $V(x)=x^{2(k+1)}$, the deformation $\phi_{2,2}$ corresponds to $\Delta V=m_2 x$. To the extent that this LG description is reliable, it indeed predicts that $m_2$ gaps the spectrum.

\subsection{Marginal deformations}\label{sec:marginal_def}

We close this section with a brief discussion of the effect of exactly marginal deformations, which parametrize the conformal manifold of $\mathcal T_\text{IR}$. In the UV there are usually several marginal operators, but these are typically not marginal in the IR. A general class of operators that are exactly marginal along the flow are bilinears in the currents that generate flavor symmetries. In this section we would like to make a few general remarks about such deformations. This class of deformations is very special and cannot, for obvious reasons, be understood via the tumbling mechanism that describes relevant deformations.

We begin with a very simple example: $SU(N)$ plus one copy of the fundamental representation. Here the flavor symmetry is $F=U(1)_\ell\times U(1)_r$, and we could consider turning on a four-fermi operator in the UV defined as the product of the currents that generate $F$. What is the effect of this deformation in the IR? The infrared theory without deformation is $U(1)_N$ WZW, a compact boson at radius $R=\sqrt N$; hence, the effect of the deformation is simply to change this radius continuously. Interestingly, the deformation generically breaks rationality of the CFT. It is important to stress that this answer is rigorous only when the deformation is small; we do not know whether the four-fermi deformation with a large coefficient still corresponds to changes in the radius of the infrared sigma model, or has a more drastic effect.

More generally, if we consider a QCD theory with flavor symmetry $F=H_\ell\times H_r$ for some group $H$, then the IR will contain a WZW subsector $H_{\ell,k}\times H_{r,k}$, plus a decoupled sector with no flavor symmetry. This theory admits marginal deformations constructed out of the currents that generate $F$. What is the effect of this deformation in the IR? The answer is straightforward: the decoupled sector will not be affected by this deformation (since it commutes with the flavor currents), while the WZW subsector will get deformed by its flavor currents. Such deformations are well-understood~\cite{Forste:2003km,Forste:2003yh}. Roughly speaking, this deformation has the following effect. First, we rewrite the WZW model as $H_k=(H_k/U(1)^r)\times U(1)^r$,\footnote{The branching functions of $H_k/U(1)^r$ are by definition the \emph{string functions} of $H_k$.} where $U(1)^r$ is the maximal torus of $H$ (with $r=\rank(H)$). Then, flavor deformations simply change the radii of the second $U(1)^r$.

A nice, simple example is $U(1)$ plus $N_F$ fermions. The IR CFT is $SU(N_F)_1$, which can be realized as $N_F$ compact bosons propagating on the maximal torus of $SU(N_F)$. Then, marginal four-fermi deformations amount to small deformations of this torus.

\section{Chiral theories}\label{sec:chiral}

It is possible to generalize the analysis of previous sections to chiral theories. Recall that a general QCD theory is defined by a gauge group $G$ and a pair of representations $(R_\ell,R_r)$ characterizing the gauge quantum numbers of the left-moving and right-moving fermions, respectively. The theory is well-defined provided the gauge anomaly cancels, which requires $I(R_\ell)=I(R_r)$, where $I$ denotes the quadratic index. If $\dim R_\ell\neq \dim R_r$, then the theory has a gravitational anomaly $c_\ell-c_r=\frac12(\dim(R_\ell)-\dim(R_r))$, which is preserved along the RG-flow; the infrared CFT will not be invariant under large diffeomorphisims, but rather invariant up to a phase.

The infrared chiral algebras of a chiral theory are given by the general formula~\eqref{eq:intro_conjecture},
\begin{equation}
\mathcal T_\text{IR}=\frac{SO(\dim(R_\ell))_1}{G_{I(R_\ell)}}\times\frac{SO(\dim(R_r))_1}{G_{I(R_r)}}\,.
\end{equation}
Unlike before, the infrared CFT is now chiral, $\mathcal A_\text{IR}\neq\overline{\mathcal A}_\text{IR}$ (these algebras are distinct even if $\dim R_\ell=\dim R_r$: if $R_\ell\neq R_r$, the embedding of $G$ into $SO(\dim R)$ will be different).

Since the criterion for operator decoupling and operator mapping was phrased at the chiral level, the general analysis still applies to chiral theories. A UV operator $\mathcal O_\text{UV}=O_\text{UV}\times\overline O_\text{UV}$ decouples at large distances if and only if the chiral components $O_\text{UV},\overline O_\text{UV}$ are descendants with respect to $G_{I(R_\ell)}$ and $G_{I(R_r)}$, respectively. The mapping of operators is still determined by the branching rules of the embeddings $G_{I(R_\ell)}\subseteq SO(\dim R_\ell)$ and $G_{I(R_r)}\subseteq SO(\dim R_r)$.

The main novelty in the chiral case is that the partition function of $\mathcal T_\text{IR}$ is non-diagonal. Instead, it is given by some modular invariant way to combine the two (distinct!) chiral algebras $\mathcal A_\text{IR},\overline{\mathcal A}_\text{IR}$. The specific invariant depends on the details of the UV theory. A nice general example of a non-diagonal invariant is the following. If the gauge group $G$ admits complex representations, then we can consider a chiral gauge theory with matter $R_\ell=R$, $R_r=\bar R$, where $R$ is some representation of $G$.\footnote{Clearly, this satisfies gauge anomaly cancellation since $I(R)=I(\bar R)$.} The modular invariant in this case is $Z_{\mu\nu}=\delta_{\mu\bar\nu}$, i.e., it corresponds to the charge-conjugation invariant.

Another nice general class of examples are the $2d$ versions of the $4d$ chiral theories studied by Bars-Yankielowicz~\cite{BARS1981159} and Georgi-Glashow~\cite{Georgi:1974sy}, namely, an $SU(N)$ gauge theory with matter $(\ydiagram2,(N+2)\ydiagram1)$ and $(\ydiagram{1,1},(N-2)\ydiagram1)$, respectively. These theories are nonanomalous, cf.~$I(\ydiagram2)=N+2,I(\ydiagram{1,1})=N-2, I(\ydiagram1)=1$. These theories become in the IR the following chiral theories:
\begin{equation}
\frac{U(\frac12N(N+1))_1}{SU(N)_{N+2}}\times\frac{U(N(N+2))_1}{SU(N)_{N+2}}\cong U(1)_{\frac12N(N+1)}\times U(N+2)_N
\end{equation}
and
\begin{equation}
\frac{U(\frac12N(N-1))_1}{SU(N)_{N-2}}\times\frac{U(N(N-2))_1}{SU(N)_{N-2}}\cong U(1)_{\frac12N(N-1)}\times U(N-2)_N
\end{equation}
respectively. The gauge theory predicts a modular invariant for the heterotic CFTs with algebras $\mathcal A=U(1)_{\frac12N(N\pm1)}$ and $\overline{\mathcal A}=U(N\pm2)_N$. It would be nice to exhibit these invariants explicitly.

Perhaps the most interesting class of examples are those related to gapped QCD theories, which turn out to be related to holomorphic invariants. Indeed, let $G+(R,R)$ be a QCD theory that is gapped (i.e., one of the theories in the general classification of~\cite{Delmastro:2021otj}). Then, the chiral QCD theory $G+(R',R)$, with $R'\neq R$, has infrared chiral algebra
\begin{equation}
\mathcal T_\text{IR}=\frac{SO(\dim R')_1}{G_{I(R')}}\times\text{right-moving topological sector,}
\end{equation}
which is purely left-moving. This leads to the following non-trivial prediction: for any $(G,R)$ in the list of gapped theories, and for any $R'$ such that $I(R)=I(R')$, the chiral algebra $SO(\dim R')_1/G_{I(R')}$ has a (fermionic) holomorphic modular invariant!

For example, we know that adjoint QCD is gapped. We can cancel the gauge anomaly by considering $N_F=h$ copies of the fundamental representation, where $h\equiv I(\text{adj})$ is the dual Coxeter number of $G$. In other words, we look at the chiral theory $G+(h\ydiagram1,\text{adj})$. This construction predicts that the coset
\begin{equation}
\frac{SO(\nu Nh)_1}{G_h}
\end{equation}
has a holomorphic modular invariant, where $\nu=1,2,4$ for $G=SO(N),SU(N),Sp(N)$, respectively. By level-rank duality, this predicts a modular invariant for
\begin{equation}
\frac{SO(N(N-2))_1}{SO(N)_{N-2}}\cong SO(N-2)_N,\quad \frac{SO(2N^2)_1}{SU(N)_N}\cong U(N)_N,\quad\frac{SO(4N(N+1))_1}{Sp(N)_N}\cong Sp(N+1)_N
\end{equation}
respectively. For $SO(N)/Sp(N)$ we can also use the other rank-2 representation, which are also gapped, yielding modular invariants for $SO(N+2)_N$ and $Sp(N-1)_N$, respectively.

It is not hard to write down these invariants explicitly. In the adjoint case, these are
\begin{equation}
\begin{aligned}
Z_{\text{NS-}\pm}(q)&=\sum_{\lambda\in\mathcal R}(\pm1)^{h_\lambda}\chi_{\lambda}(q)\\
Z_{\text{R-NS}}(q)&=2^{r/2}\chi_{\hat\rho}(q)\\
Z_\text{R-R}(q)&=0
\end{aligned}
\end{equation}
where $\hat\rho=[1,1,\dots,1]$ is the (affine) Weyl vector of $\mathfrak g$, $r$ is its rank, and~\cite{KAC1988156}
\begin{equation}
\mathcal R:=\{\lambda\,|\,\exists\text{ affine reflection }\hat w\ \text{s.t.}\ \lambda=h\hat \omega_0+(\hat w-1)\hat \rho\}\,,
\end{equation}
with $\hat \omega_0$ the vacuum module of $\hat{\mathfrak g}_h$. (There are analogous formulas for the other rank-2 representations, cf.~\cite[\S 5.1]{Delmastro:2021otj}). These functions are indeed (spin) modular invariant:
\begin{equation}
\begin{split}
S\cdot Z_\text{NS-NS}(q)&=Z_\text{NS-NS}(q)\\
S\cdot Z_\text{NS-R}(q)&=Z_\text{R-NS}(q)\\
S\cdot Z_\text{R-NS}(q)&=Z_\text{NS-R}(q)\\
%S\cdot Z_\text{R-R}(q)&=Z_\text{R-R}(q)
\end{split}\qquad\begin{split}
T\cdot Z_\text{NS-NS}(q)&=e^{-2\pi i c/24}Z_\text{NS-R}(q)\\
T\cdot Z_\text{NS-R}(q)&=e^{-2\pi i c/24}Z_\text{NS-NS}(q)\\
T\cdot Z_\text{R-NS}(q)&=e^{2\pi ic/12}Z_\text{R-NS}(q)\\
%T\cdot Z_\text{R-R}(q)&=Z_\text{R-R}(q)
\end{split}
\end{equation}
where $c$ denotes the central charge.

Other properties of chiral theories closely follow the same analysis of non-chiral ones. For example, if $G+(R,R)$ has emergent $(\mathcal N,\mathcal N)$ supersymmetry, and $G+(R',R')$ has emergent $(\mathcal N',\mathcal N')$, then the chiral theory $G+(R,R')$ has $(\mathcal N,\mathcal N')$ supersymmetry. Naturally, one can also study deformations of chiral theories and other related problems. We will not pursue these investigations any further here.

\section*{Acknowledgments}

We would like to thank Matthew Yu for collaborating with us in the earlier stages of this work. We would also like to thank Davide Gaiotto, Yin-Chen He and specially Alexander Zamolodchikov for very useful discussions. Research at Perimeter Institute is supported in part by the Government of Canada through the Department of Innovation, Science and Economic Development Canada and by the Province of Ontario through the Ministry of Colleges and Universities. Any opinions, findings, and conclusions or recommendations expressed in this material are those of the authors and do not necessarily reflect the views of the funding agencies. 
\vfill\eject

\appendix

\section{$\boldsymbol{3d}$ realization of $\boldsymbol{2d}$ QCD}\label{ap:3d}

In this appendix we realize $2d$ QCD by a $3d$ theory on a slab with suitable boundary conditions. We use this description to provide a complementary perspective on various aspects of $2d$ QCD.\footnote{We thank D. Gaiotto for discussions.}

\begin{figure}
\centering
\begin{tikzpicture}

\draw[thick,pattern=north east lines] 
(3,0) -- (3,3)
     decorate [decoration={snake,amplitude=1.2pt},shorten >= 2pt,decorate] 
     { to (4,3) } -- (4,0)
     decorate [decoration={snake,amplitude=1.2pt},shorten >= 2pt,decorate] 
   { to (3,0) };

\draw[thick] (0,0) -- (3,0) -- (3,3) -- (0,3) -- cycle;
\draw[thick] (4,0) -- (7,0) -- (7,3) -- (4,3) -- cycle;

\node[anchor=east] at (0,1.5) {$\mathcal A$};
\node[anchor=west] at (7,1.5) {$\overline{\mathcal A}$};
\node[anchor=north] at (3.6,2.5+1.6) {gauge $G$};
\draw[->,thick,>=stealth] (3.5,2.5+1) -- (3.5,2.15+1);

\node at (1.5,1.5) {$G_k$ CS};
\node at (3.94+1.5,1.5) {$G_k$ CS};

\draw[double,->,>=stealth,thick] (3.5,-.25) -- (3.5,-.75);
\node[scale=.8] at (4.3,-.5) {RG-flow};

\begin{scope}[shift={(0,-4)}]
\draw[thick] (3.5,0) -- (0,0) -- (0,3) -- (3.5,3);
\draw[thick] (3.5,0) -- (7,0) -- (7,3) -- (3.5,3);
\draw[thick,dashed] (3.5,0) -- (3.5,3);

\node[anchor=east] at (0,1.5) {$\mathcal A$};
\node[anchor=west] at (7,1.5) {$\overline{\mathcal A}$};

\node at (1.5,1.5) {$G_k$ CS};
\node at (3.94+1.5,1.5) {$G_k$ CS};

\node[anchor=west] at (3.5+.5,.5) {$\boldsymbol1$};
\draw[<-,thick,>=stealth] (3.1+.5,.3) -- (3.5+.5,.5);

\end{scope}

\end{tikzpicture}
\caption{Three-dimensional construction of two-dimensional QCD. At the boundaries of the interval we place chiral fermions, and we gauge the $G$ symmetry in the intermediate region. The RG flow corresponds to collapsing this region, which then becomes an interface on $G_{I(R)}$ Chern-Simons. This shows, for example, that the coset algebra $SO(\dim R)_1/G_{I(R)}$ exists at any value of the coupling, since it is localized at the boundaries while the RG flow happens away from the boundary. It also makes manifest that purely chiral operators are protected, as they live entirely within one of the boundaries, and do not know what happens in the bulk.}
\label{fig:2d_from_3d}
\end{figure}
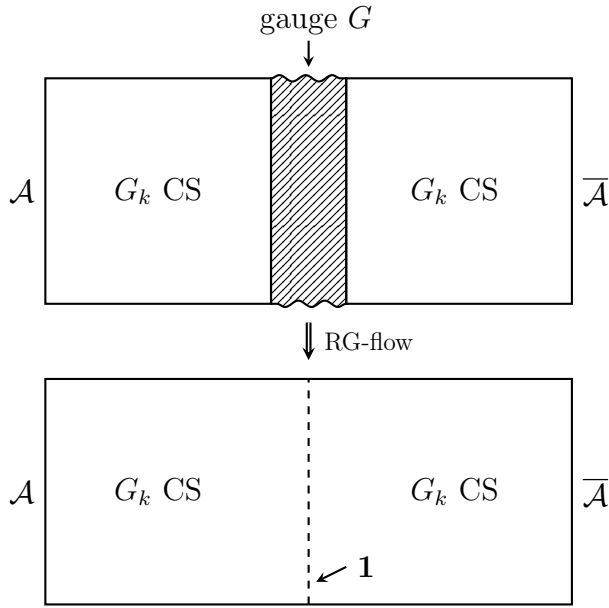

$N_\ell=\dim(R_\ell)$  $2d$ chiral fermions can be represented as $3d$ $G_{I(R_\ell)}$ Chern-Simons theory on a slab, see figure~\ref{fig:2d_from_3d}. On one end we gauge the $G_{I(R_\ell)}\subset SO(N_\ell)_1$ currents by coupling to  $3d$ $G_{I(R_\ell)}$ Chern-Simons theory, while at the other end we impose  a Dirichlet-like Moore-Seiberg boundary condition, which supports a  $G_{I(R_\ell)}$ chiral  algebra. Gauging strips off the $G_{I(R_\ell)}$ currents from the Neumnann-like boundary and leaves behind the commutant chiral algebra ${\mathcal A}=SO(N_\ell)_1/G_{I(R_\ell)}$ and the  chiral $G_{I(R_\ell)}$ global symmetry re-emerges at the Dirichlet-like end. An identical construction can be carried out with $N_r=\dim(R_r)$ fermions of opposite chirality, which now support an antichiral algebra  ${\overline {\mathcal A}}=SO(N_r)_1/G_{I(R_r)}$
at the Neumnann-like boundary and is endowed  with an antichiral  
$G_{I(R_r)}$ global symmetry at the Dirichlet-like end.

$2d$ QCD can then be constructed by connecting the chiral and antichiral slabs with another slab. In the middle slab the global symmetry $G$ is gauged in a non-anomalous fashion, which requires that $I(R_\ell)=I(R_r)$. The gauging occurs near the Dirichlet-like ends, where the global symmetry is supported. This realization demonstrates that the nontrivial RG flow is localized in the intermediate slab, which is non-topological. This perspective makes it clear that  chiral and anti-chiral algebras supported at the Neumnann-like ends exist at any value of the coupling essentially due to locality, as the RG happens in the middle slab. In this description, the claim that the deep IR of QCD is
\begin{equation}
{\mathcal T}_\text{IR}=\frac{SO(\dim(R_\ell))_1}{G_{I(R_\ell)}}\times\frac{SO(\dim(R_r))_1}{G_{I(R_r)}}\,,
\end{equation}
is equivalent to the statement that the RG interface in the intermediate slab is trivial.

\section{Spinning operators}\label{ap:spin_ops}

In this appendix we complete the discussion of section~\ref{sec:qcd_fund} by classifying all operators, the Lorentz scalars and the operators with spin.

We take the theory $U(N)$ with $N_F$ massless Dirac fermions. We assign the following quantum numbers to the quarks:
\begin{equation}\label{eq:ap_q_numbers}
\begin{array}{c|c|c|c|c|c}
&SO(1,1)&\mathbb R_+&U(N)&SU(N_F)_\ell&SU(N_F)_r\\\hline
\psi_\ell&+1/2&1/2&\ydiagram1_{+1}&\ydiagram1&\boldsymbol1\\
\psi_r&-1/2&1/2&\ydiagram1_{+1}&\boldsymbol1&\overline{\ydiagram1}\\
%D_+\psi_L&+3/2&\ydiagram1_{+1}&\ydiagram1&\boldsymbol1\\
%D_-\psi_R&-3/2&\ydiagram1_{+1}&\boldsymbol1&\overline{\ydiagram1}\\
%J_{L/R}&+1&\ydiagram{1+1,1}_{0}&\boldsymbol1&\boldsymbol1\\
%\tilde J_{L/R}&+1&\boldsymbol1_{0}&\boldsymbol1&\boldsymbol1
\end{array}
\end{equation}
Here, $SO(1,1)$ denotes the Lorentz group and $\mathbb R_+$ the dilatation group corresponding to the scaling symmetry in the deep UV; both fermions have scaling dimension $1/2$ with respect to this symmetry. In QCD, we will order the operators in increasing value of their \emph{classical} scaling dimension $\Delta_\text{UV}$. At the infrared fixed point this scaling dimension will generically be renormalized, $\Delta_\text{IR}\neq\Delta_\text{UV}$.

Let us begin by classifying chiral operators constructed entirely in terms of $\psi_\ell$. The right-moving operators follow an identical discussion, and the full spectrum of the theory is obtained by combining the two chiral halves, which we will do momentarily. In order to simplify the notation, here we shall denote $\psi_\ell$ simply as $\psi$, which transforms as $[\ydiagram1,\ydiagram1]$ under $SU(N)\times SU(N_F)$.

At scaling dimension $\Delta_\text{UV}=0$, there is a single operator, namely the identity. At scaling dimension $\Delta_\text{UV}=1/2$ there is also a single operator, namely $\psi$ itself. The first non-trivial operators appear at $\Delta_\text{UV}=1$, to wit, $\{\psi^2\}$ and $\{\psi\psi^\dagger\}$. The quantum numbers of $\{\psi^2\}$ are classified by
\begin{equation}
\wedge^2[\ydiagram1,\ydiagram1]=[\ydiagram2,\ydiagram{1,1}]+[\ydiagram{1,1},\ydiagram2]
\end{equation}
Similarly, the quantum numbers of $\{\psi\psi^\dagger\}$ are classified by
\begin{equation}
[\ydiagram1,\ydiagram1]\times[\overline{\ydiagram1},\overline{\ydiagram1}]=[\boldsymbol1,\boldsymbol1]+[\ydiagram{1+1,1},\boldsymbol1]+[\boldsymbol1,\ydiagram{1+1,1}]+[\ydiagram{1+1,1},\ydiagram{1+1,1}]
\end{equation}
where $\ydiagram{1+1,1}$ denotes the adjoint representation. From these operators we would like to remove the gauge currents $J^a=\psi^\dagger t^a\psi$, inasmuch as these decouple at large distances. It is rather easy to identify these currents. The $SU(N)$ current transforms under the adjoint of the gauge group, and it is a singlet of flavor, hence its quantum numbers are $[\ydiagram{1+1,1},\boldsymbol1]$. On the other hand, the $U(1)$ current is a singlet of both $SU(N)$ and $SU(N_F)$, hence its quantum numbers are $[\boldsymbol1,\boldsymbol1]$. Therefore, at $\Delta_\text{UV}=1$ there are four operators that do not contain gauge currents, and two that do contain currents:
\begin{equation}
\begin{alignedat}{4}
&\{\psi^2\}&&\colon\quad [\ydiagram2,\ydiagram{1,1}]+[\ydiagram{1,1},\ydiagram2] &&\quad \text{do not contain currents}\\
&\{\psi\psi^\dagger\}&&\colon\quad [\boldsymbol1,\ydiagram{1+1,1}]+[\ydiagram{1+1,1},\ydiagram{1+1,1}]&&\quad \text{do not contain currents}\\
& &&\hphantom{\colon}\quad[\boldsymbol1,\boldsymbol1]+[\ydiagram{1+1,1},\boldsymbol1]&&\quad \text{do contain currents}
\end{alignedat}
\end{equation}
The claim is that the operators that do not contain currents survive the infrared limit, while those that do contain currents decouple.

If we continue this way, and classify higher order operators, we obtain the following table:
\begin{alignat}{3}
&1&&\quad[\boldsymbol1,\boldsymbol1]&&\quad[\boldsymbol1,\boldsymbol1]\\\hline
&\psi&&\quad[\ydiagram1,\ydiagram1]&&\quad[\ydiagram1,\ydiagram1]\label{eq:c_psi}\\\hline
&\{\psi^2\}&&\quad\wedge^2\![\ydiagram1,\ydiagram1]&&\quad[\ydiagram{1,1},\ydiagram2]+[\ydiagram2,\ydiagram{1,1}]\label{eq:c_psi2}\\\hline
&\{\psi\psi^\dagger\}&&\quad[\ydiagram1,\ydiagram1]\times[\overline{\ydiagram1},\overline{\ydiagram1}]&&\quad{\color{red}[\boldsymbol1,\boldsymbol1]}+{\color{red}[\ydiagram{1+1,1},\boldsymbol1]}+[\boldsymbol1,\ydiagram{1+1,1}]+[\ydiagram{1+1,1},\ydiagram{1+1,1}]\label{eq:c_psi_psi*}\\\hline
&\{\psi^2\psi^\dagger\}&&\quad\wedge^2\![\ydiagram1,\ydiagram1]\times[\overline{\ydiagram1},\overline{\ydiagram1}]&&\quad{\color{red}2[\ydiagram1,\ydiagram1]}+{\color{red}[\ydiagram{1+2,1},\ydiagram1]}+{\color{red}[\ydiagram{1+1,1+1,1},\ydiagram1]}+[\ydiagram1,\ydiagram{1+1,1+1,1}]\label{eq:c_psi2_psi*}\\
& && &&\quad[\ydiagram{1+2,1},\ydiagram{1+1,1+1,1}]+[\ydiagram1,\ydiagram{1+2,1}]+[\ydiagram{1+1,1+1,1},\ydiagram{1+2,1}]\notag\\\hline
&\{\partial\psi\}&&\quad[\ydiagram1,\ydiagram1]&&\quad[\ydiagram1,\ydiagram1]\label{eq:c_Dpsi}\\\hline
&\{\psi^3\}&&\quad\wedge^3\![\ydiagram1,\ydiagram1]&&\quad[\ydiagram{1,1,1},\ydiagram3]+[\ydiagram{2,1},\ydiagram{2,1}]+[\ydiagram3,\ydiagram{1,1,1}]\label{eq:c_psi3}\\\hline
&\{\psi^2\psi^{\dagger2}\}&&\quad\wedge^2\![\ydiagram1,\ydiagram1]\times\wedge^2[\overline{\ydiagram1},\overline{\ydiagram1}]&&\quad{\color{red}2[\boldsymbol1,\boldsymbol1]}+{\color{red}2[\ydiagram{1+1,1},\boldsymbol1]}+{\color{red}2[\boldsymbol1,\ydiagram{1+1,1}]}+{\color{red}4[\ydiagram{1+1,1},\ydiagram{1+1,1}]}\label{eq:c_psi2_psi2*}\\
& && &&\quad {\color{red}[\ydiagram{1+1,1+1,1,1},\boldsymbol1]}+{\color{red}[\ydiagram{2+2,2},\boldsymbol1]}+{\color{red}[\ydiagram{1+1,1+1,1,1},\ydiagram{1+1,1}]}+{\color{red}[\ydiagram{2+2,2},\ydiagram{1+1,1}]}\notag\\
& && &&\quad {\color{red}[\ydiagram{1+2,1,1},\ydiagram{1+1,1}]}+{\color{red}[\ydiagram{2+1,2+1,2},\ydiagram{1+1,1}]}+[\ydiagram{1+1,1},\ydiagram{1+2,1,1}]+[\ydiagram{1+1,1},\ydiagram{2+1,2+1,2}]\notag\\
& && &&\quad[\boldsymbol1,\ydiagram{1+1,1+1,1,1}]+[\boldsymbol1,\ydiagram{2+2,2}]+[\ydiagram{1+1,1},\ydiagram{1+1,1+1,1,1}]+[\ydiagram{1+1,1},\ydiagram{2+2,2}]\notag\\
& && &&\quad [\ydiagram{2+1,2+1,2},\ydiagram{1+2,1,1}]+[\ydiagram{1+2,1,1},\ydiagram{2+1,2+1,2}]+[\ydiagram{1+1,1+1,1,1},\ydiagram{2+2,2}]+[\ydiagram{2+2,2},\ydiagram{1+1,1+1,1,1}]\notag\\\hline
&\{\psi \partial\psi^\dagger\} &&\quad [\ydiagram1,\ydiagram1]\times[\overline{\ydiagram1},\overline{\ydiagram1}]&&\quad [\boldsymbol1,\boldsymbol1]+[\boldsymbol1,\ydiagram{1+1,1}]+[\ydiagram{1+1,1},\boldsymbol1]+[\ydiagram{1+1,1},\ydiagram{1+1,1}]\label{eq:c_psi_Dpsi*}
\end{alignat}
In red we single out the operators that contain gauge currents, i.e., the operators that become trivial at large distances. %\note{In~\eqref{eq:c_psi2_psi2*} and~\eqref{eq:c_psi_Dpsi*} I am not sure that I identified the currents correctly.}

%\note{
%We also have $\psi^3\psi^\dagger$ classified by
%\begin{equation}\label{eq:psi3_*psi}
%\begin{aligned}
%\wedge^3[\ydiagram1,\ydiagram1]\times[\overline{\ydiagram1},\overline{\ydiagram1}]&={\color{blue}2[\ydiagram{1,1},\ydiagram2]}+{\color{blue}2[\ydiagram2,\ydiagram{1,1}]}+{\color{blue}[\ydiagram{1,1},\ydiagram{1,1}]}+{\color{blue}[\ydiagram2,\ydiagram2]}\\
%&+{\color{blue}[\ydiagram{1+1,1+1,1+1,1},\ydiagram2]}+{\color{blue}[\ydiagram{1+2,1+1,1},\ydiagram{1,1}]}+{\color{blue}[\ydiagram{1+2,1+1,1},\ydiagram2]}+{\color{blue}[\ydiagram{1+3,1},\ydiagram{1,1}]}\\
%&+[\ydiagram{1+1,1+1,1+1,1},\ydiagram{1+3,1}]+[\ydiagram2,\ydiagram{1+2,1+1,1}]+[\ydiagram{1,1},\ydiagram{1+2,1+1,1}]+[\ydiagram2,\ydiagram{1+1,1+1,1+1,1}]\\
%&+[\ydiagram{1+2,1+1,1},\ydiagram{1+2,1+1,1}]+[\ydiagram{1+3,1},\ydiagram{1+1,1+1,1+1,1}]+[\ydiagram{1,1},\ydiagram{1+3,1}]
%\end{aligned}
%\end{equation}
%
%}

Next, we combine the two chiral halves to obtain the full spectrum of operators, up to $\Delta_\text{UV}\le3$. We also include their spin $s$, i.e., their quantum numbers under $SO(1,1)$. We restrict attention to $s\ge0$; operators with negative $s$ can be obtained from those with positive $s$ by applying a parity transformation. A generic operator will have quantum numbers under $U(N)\times SU(N_F)_\ell\times SU(N_F)_r$; we only count the gauge singlets, i.e., the operators that transform trivially under $U(N)$. The quantum numbers under $SU(N_F)_\ell\times SU(N_F)_r$ are denoted as $(R_\ell,R_r)$, with $R$ an $SU(N_F)$ representation.

\paragraph{Scaling dimension $\boldsymbol{\Delta_\text{UV}=0}$.} The only operator is the identity, with quantum numbers $(\boldsymbol1,\boldsymbol1)$.

\paragraph{Scaling dimension $\boldsymbol{\Delta_\text{UV}=1}$.} The operators are as follows:
\begin{itemize}
\item $s=0$. We have $\{\psi_\ell\psi^\dagger_r\}$ (plus conjugate), classified by
\begin{equation}
\eqref{eq:c_psi}\times\eqref{eq:c_psi}=(\ydiagram1,\ydiagram1)
\end{equation}
\item $s=1$. We have $\{\psi_\ell\psi_\ell^\dagger\}$, classified by
\begin{equation}
\eqref{eq:c_psi_psi*}=(\ydiagram{1+1,1},\boldsymbol1)
\end{equation}

\end{itemize}

\paragraph{Scaling dimension $\boldsymbol{\Delta_\text{UV}=2}$.} The operators are as follows:
\begin{itemize}
\item $s=0$. We have $\{\psi_\ell\psi_\ell\psi^\dagger_r\psi^\dagger_r\}$ (plus conjugate), classified by
\begin{equation}
\eqref{eq:c_psi2}\times\eqref{eq:c_psi2}=(\ydiagram2,\ydiagram2)+(\ydiagram{1,1},\ydiagram{1,1})
\end{equation}
We also have operators $\{\psi_\ell\psi^\dagger_\ell\psi_r\psi^\dagger_r\}$, classified by
\begin{equation}
\eqref{eq:c_psi_psi*}\times \eqref{eq:c_psi_psi*}=2(\ydiagram{1+1,1},\ydiagram{1+1,1})
\end{equation}

\item $s=1$. We have $\{\psi_\ell\psi_\ell\psi_\ell^\dagger\psi^\dagger_r\}$ (plus conjugate), classified by
\begin{equation}
\eqref{eq:c_psi2_psi*}\times\eqref{eq:c_psi}=(\ydiagram{1+1,1+1,1},\ydiagram1),(\ydiagram{1+2,1},\ydiagram1)
\end{equation}
We also have $\{D_+\psi_\ell\psi^\dagger_r\}$ (plus conjugate), classified by
\begin{equation}
\eqref{eq:c_Dpsi}\times \eqref{eq:c_psi}=(\ydiagram1,\ydiagram1)
\end{equation}

\item $s=2$. We have $\{\psi_\ell\psi_\ell\psi^\dagger_\ell\psi^\dagger_\ell\}$, classified by
\begin{equation}
\eqref{eq:c_psi2_psi2*}=(\ydiagram{1+1,1+1,1,1},\boldsymbol1)+(\ydiagram{2+2,2},\boldsymbol1)
\end{equation}
We also have $\{D_+\psi_\ell\psi_\ell^\dagger\}$, classified by
\begin{equation}\label{eq:EM_ops}
\eqref{eq:c_psi_Dpsi*}=(\boldsymbol1,\boldsymbol1)+(\ydiagram{1+1,1},\boldsymbol1)
\end{equation}
We should also consider $\{\psi_\ell D_+\psi_\ell^\dagger\}$, which has the same quantum numbers. Actually, the two copies of $(\boldsymbol1,\boldsymbol1)$ are not independent, since their difference is (the derivative of) the $U(1)$ gauge current: $D_+J^{(1)}\equiv \text{flavor singlet in }(D_+\psi_\ell\psi_\ell^\dagger+\psi_\ell D_+\psi_\ell^\dagger)$. Therefore, the derivative operators at this scaling dimension, modulo gauge currents, are classified by
\begin{equation}
(\boldsymbol1,\boldsymbol1)+2\,(\ydiagram{1+1,1},\boldsymbol1)
\end{equation}
\end{itemize}

\paragraph{Scaling dimension $\boldsymbol{\Delta_\text{UV}=3}$.} The operators are as follows:
\begin{itemize}
\item $s=0$. We have $\{\psi_\ell\psi_\ell\psi_\ell\psi^\dagger_r\psi^\dagger_r\psi^\dagger_r\}$ (plus conjugate), classified by
\begin{equation}
\eqref{eq:c_psi3}\times\eqref{eq:c_psi3}=(\ydiagram{1,1,1},\ydiagram{1,1,1})+(\ydiagram{2,1},\ydiagram{2,1})+(\ydiagram3,\ydiagram3)
\end{equation}
We also have $\{\psi_\ell\psi_\ell\psi^\dagger_\ell\psi_r\psi^\dagger_r\psi^\dagger_r\}$ (plus conjugate), classified by
\begin{equation}
\eqref{eq:c_psi2_psi*}\times\eqref{eq:c_psi2_psi*}=2(\ydiagram{1+1,1+1,1},\ydiagram{1+1,1+1,1})+2(\ydiagram{1+2,1},\ydiagram{1+2,1})+(\ydiagram{1+1,1+1,1},\ydiagram{1+2,1})+(\ydiagram{1+2,1},\ydiagram{1+1,1+1,1})
\end{equation}
We also have $\{D_+\psi^\dagger_\ell D_-\psi_r\}$ (plus conjugate), classified by
\begin{equation}
\eqref{eq:c_psi}\times\eqref{eq:c_psi}=(\ydiagram1,\ydiagram1)
\end{equation}
We also have $\{\psi_\ell\psi_\ell\psi^\dagger_\ell D_-\psi^\dagger_r\}$ (plus conjugate, plus $\ell\leftrightarrow r$), classified by
\begin{equation}
\eqref{eq:c_psi2_psi*}\times\eqref{eq:c_Dpsi}=(\ydiagram{1+1,1+1,1},\ydiagram1)+(\ydiagram{1+2,1},\ydiagram1)
\end{equation}

\item $s=1$. We have $\{\psi_\ell\psi_\ell\psi_\ell\psi_\ell^\dagger\psi_r^\dagger\psi_r^\dagger\}$ (plus conjugate), classified by
\begin{equation}
\bigl(\wedge^3[\ydiagram1,\ydiagram1]\times[\overline{\ydiagram1},\overline{\ydiagram1}]\bigr)\times\eqref{eq:c_psi2}=(\ydiagram{1+2,1+1,1},\ydiagram{1,1})+(\ydiagram{1+1,1+1,1+1,1},\ydiagram{1,1})+(\ydiagram{1+3,1},\ydiagram2)+(\ydiagram{1+2,1+1,1},\ydiagram2)
\end{equation}
We also have $\{\psi_\ell\psi_\ell\psi_\ell^\dagger\psi_\ell^\dagger\psi_r\psi_r^\dagger\}$, classified by
\begin{equation}
\eqref{eq:c_psi2_psi2*}\times\eqref{eq:c_psi_psi*}=2\,(\ydiagram{1+1,1+1,1,1},\ydiagram{1+1,1})+2\,(\ydiagram{2+2,2},\ydiagram{1+1,1})+(\ydiagram{1+2,1,1},\ydiagram{1+1,1})+(\ydiagram{2+1,2+1,2},\ydiagram{1+1,1})
\end{equation}
Similarly, we have $\{\psi_\ell D_+\psi_\ell\psi^\dagger_r\psi_r^\dagger\}$, classified by (the gauge singlets in) $[\ydiagram1,\ydiagram1]^2\times\wedge^2[\overline{\ydiagram1},\overline{\ydiagram1}]$, 
\begin{equation}
(\ydiagram2,\ydiagram{1,1})+(\ydiagram2,\ydiagram2)+(\ydiagram{1,1},\ydiagram{1,1})+(\ydiagram{1,1},\ydiagram2)
\end{equation}
and $\{\psi_\ell D_+\psi_\ell^\dagger\psi_r\psi_r^\dagger\}$, classified by
\begin{equation}
\eqref{eq:c_psi_Dpsi*}\times\eqref{eq:c_psi2}=2\,(\boldsymbol1,\ydiagram{1+1,1})+2\,(\ydiagram{1+1,1},\ydiagram{1+1,1})
\end{equation}
We should also count $\{\psi^\dagger_\ell D_+\psi_\ell\psi_r\psi_r^\dagger\}$, which has the same quantum numbers. Actually, we should not count $(\boldsymbol1,\ydiagram{1+1,1})$ separately since this corresponds to (the derivative of) the $U(1)$ current, as in~\eqref{eq:EM_ops}. With this in mind, these last two operators are classified by
\begin{equation}
2\,(\boldsymbol1,\ydiagram{1+1,1})+4\,(\ydiagram{1+1,1},\ydiagram{1+1,1})
\end{equation}

\item $s=2$. We have $\{\psi_\ell\psi_\ell\psi_\ell\psi_\ell^\dagger\psi_\ell^\dagger\psi^\dagger_r\}$ (plus conjugate). Let us look at the left-movers. We will eventually contract with $\psi_r^\dagger$, which is $\overline{\ydiagram1}$ under the gauge group; hence, we need the $[\ydiagram1,\cdots]$ in $\{\psi_\ell^3\psi_\ell^{\dagger2}\}$. These are
\begin{equation}
\wedge^3[\ydiagram1,\ydiagram1]\times\wedge^2[\overline{\ydiagram1},\overline{\ydiagram1}]=[\ydiagram1,\ydiagram3\times\overline{\ydiagram2}]+[\ydiagram1,\ydiagram{2,1}\times\overline{\ydiagram2}]+[\ydiagram1,\ydiagram{2,1}\times\overline{\ydiagram{1,1}}]+[\ydiagram1,\ydiagram{1,1,1}\times\overline{\ydiagram{1,1}}]+\cdots
\end{equation}
Finally, we drop the gauge currents, so for example $\ydiagram3\times\overline{\ydiagram2}=\ydiagram{2+3,2}+\cdots$. The end result, after contracting with $\psi_r^\dagger$, is
\begin{equation}
(\ydiagram{2+3,2},\ydiagram1)+(\ydiagram{2+2,2+1,2},\ydiagram1)+(\ydiagram{1+2,1+1,1,1},\ydiagram1)+(\ydiagram{1+1,1+1,1+1,1,1},\ydiagram1)
\end{equation}
We also have $\{\psi_\ell\psi_\ell D_+\psi^\dagger_\ell\psi^\dagger_r\}$ (plus conjugate), classified by
\begin{equation}
\eqref{eq:c_psi2_psi*}\times\eqref{eq:c_psi}=(\ydiagram{1+1,1+1,1},\ydiagram1)+(\ydiagram{1+2,1},\ydiagram1)
\end{equation}
Similarly, we look at $\{\psi_\ell\psi_\ell^\dagger D_+\psi_\ell\psi^\dagger_r\}$ (plus conjugate). As before, we need the operators with $[\ydiagram1,\cdots]$ in $\{\psi_\ell\psi_\ell^\dagger D_+\psi_\ell\}$, which are given by 
\begin{equation}
[\ydiagram1,\ydiagram1]\times[\overline{\ydiagram1},\overline{\ydiagram1}]\times[\ydiagram1,\ydiagram1]=4[\ydiagram1,\ydiagram1]+2[\ydiagram1,\ydiagram{1+1,1+1,1}]+2[\ydiagram1,\ydiagram{1+2,1}]+\cdots
\end{equation}
Two copies of $[\ydiagram1,\ydiagram1]$ are gauge currents, and therefore we drop them. The end result is
\begin{equation}
2\,(\ydiagram1,\ydiagram1)+2\,(\ydiagram{1+1,1+1,1},\ydiagram1)+2\,(\ydiagram{1+2,1},\ydiagram1)
\end{equation}
Finally, we consider $\{D^2_+\psi_\ell\psi^\dagger_r\}$ (plus conjugate), classified by
\begin{equation}
\eqref{eq:c_psi}\times\eqref{eq:c_psi}=(\ydiagram1,\ydiagram1)
\end{equation}

\item $s=3$. We will not look at these operators here.

\end{itemize}

These operators all survive the macroscopic limit, and they become non-trivial operators in the infrared CFT. We next show that this agrees with the spectrum of operators of the coset $U(NN_F)_1/U(N)_{N_F}$. By level-rank duality, this coset is identical to the WZW model $SU(N_F)_N$, whose spectrum of operators is straightforward. We will make use of the chiral characters of this algebra:
\begin{equation}\label{ap:su_chars}
\begin{aligned}
\chi_{\boldsymbol1}(q)&=\boldsymbol1+\ydiagram{1+1,1}\,q+(\boldsymbol1+ 2\,\ydiagram{1+1,1} + \ydiagram{1+1,1+1,1,1}+\delta_{N\ge2}\,\ydiagram{2+2,2}\,)q^2+\cdots\\
\chi_{\ydiagram1}(q)&=\ydiagram1+(\ydiagram1+\ydiagram{1+1,1+1,1}+\delta_{N\ge2}\,\ydiagram{1+2,1}\,)q\\
&+\bigl(
%2 n + 2 1/2 (-2 + n) n (1 + n) + 1/2 (-1 + n) n (2 + n) + 1/12 (-4 + n) (-1 + n) n^2 (1 + n)
%+\delta_{k\ge2}(n + 1/2 (-2 + n) n (1 + n) + 2 1/2 (-1 + n) n (2 + n) +  1/6 (-3 + n) (-1 + n) n (1 + n) (2 + n) +  1/6 (-2 + n) (-1 + n) n (1 + n) (3 + n))
%+\delta_{\ge3} 1/12 (-1 + n) n^2 (1 + n) (4 + n)
2\,\ydiagram1 + 2\,\ydiagram{1+1,1+1,1} + \ydiagram{1+2,1} + \ydiagram{1+1,1+1,1+1,1,1}
+\delta_{N\ge2}\bigl(\ydiagram1 + \ydiagram{1+1,1+1,1} + 2\, \ydiagram{1+2,1} +\ydiagram{1+2,1+1,1,1} +\ydiagram{2+2,2+1,2}\,\bigr)\\
&+\delta_{N\ge3}\, \ydiagram{2+3,2}\,
%(2+\delta_{k\ge2})\ydiagram1 + (2+\delta_{k\ge2})\ydiagram{1+1,1+1,1} +(1+2\delta_{k\ge2}) \ydiagram{1+2,1}\\
%&+ \ydiagram{1+1,1+1,1+1,1,1}
%+\delta_{k\ge2}\bigl(\,\ydiagram{1+2,1+1,1,1} +\ydiagram{2+2,2+1,2}\,\bigr)
%+\delta_{\ge3} \ydiagram{2+3,2}\,
\bigr)q^2+\cdots\\
\chi_{\ydiagram{1,1}}(q)&=\ydiagram{1,1}+\bigl(\,\ydiagram{1,1} +\ydiagram2 +\ydiagram{1+1,1+1,1+1,1}+\delta_{N\ge2}\,\ydiagram{1+2,1+1,1}\,\bigr)q+\cdots\\
\chi_{\ydiagram2}(q)&=\delta_{N\ge2}\bigg(\ydiagram2+(\,\ydiagram{1,1} +\ydiagram2+\ydiagram{1+2,1+1,1}+\delta_{N\ge3}\,\ydiagram{1+3,1}\,)q+\cdots\biggr)\\
\chi_{\ydiagram{1+1,1}}(q)&=\delta_{N\ge2}\bigg(\ydiagram{1+1,1}+(\boldsymbol1+2\,\ydiagram{1+1,1}+\ydiagram{1+1,1+1,1,1}+\ydiagram{1+2,1,1}+\ydiagram{2+1,2+1,2}+\delta_{N\ge3}\,\ydiagram{2+2,2}\,)q+\cdots\biggr)
%\chi_\ell(q)&=\wedge^\ell\ydiagram1+\cdots
\end{aligned}
\end{equation}
(We omit the global power $q^{h-c/24}$ to simplify the notation.)

In order to compare the infrared operators with the ultraviolet ones, we label the former by their classical scaling dimension $\Delta_\text{UV}\equiv\lim_{N_F\to\infty}\Delta_\text{IR}$. The reason for this is that the renormalized scaling dimension agrees with the classical one in the large $N_F$ limit, i.e., $\Delta_\text{IR}=\Delta_\text{UV}+\mathcal O(1/N_F)$. For primaries, $\Delta_\text{UV}\equiv r$, where $r$ is the rank of the representation (that is, the number of boxes in its Young diagram).

The classification of infrared operators goes as follows.

\paragraph{Scaling dimension $\boldsymbol{\Delta_\text{UV}=0}$.} We only have the identity operator, namely
\begin{equation}
[q^0\bar q^0]|\chi_{\boldsymbol1}|^2=(\boldsymbol1,\boldsymbol1)
\end{equation}

\paragraph{Scaling dimension $\boldsymbol{\Delta_\text{UV}=1}$.}
\begin{itemize}
\item $s=0$. The operators are in
\begin{equation}
[q^0\bar q^0]|\chi_{\ydiagram1}|^2+\text{c.c.}=(\ydiagram1,\ydiagram1)+\text{c.c.}
\end{equation}
\item $s=1$. The operators are in
\begin{equation}
[q^1\bar q^0]|\chi_{\boldsymbol1}|^2=(\ydiagram{1+1,1},\boldsymbol1)
\end{equation}

\end{itemize}

\paragraph{Scaling dimension $\boldsymbol{\Delta_\text{UV}=2}$.}
\begin{itemize}
\item $s=0$. The operators are 
\begin{equation}\label{eq:sing_sym}
\begin{aligned}
[q^0\bar q^0]\bigl(|\chi_{\ydiagram{1,1}}|^2+|\chi_{\ydiagram2}|^2+\text{c.c.}\bigr)&=(\ydiagram{1,1},\ydiagram{1,1})+\delta_{N\ge2}(\ydiagram{2},\ydiagram{2})+\text{c.c}\\
[q^0\bar q^0]|\chi_{\ydiagram{1+1,1}}|^2&=\delta_{N\ge2}(\ydiagram{1+1,1},\ydiagram{1+1,1})\\
[q^1\bar q^1]|\chi_{\boldsymbol1}|^2&=(\ydiagram{1+1,1},\ydiagram{1+1,1})
\end{aligned}
\end{equation}

\item $s=1$. The operators are
\begin{equation}
[q^1\bar q^0]|\chi_{\ydiagram1}|^2+\text{c.c.}=(\ydiagram1+\ydiagram{1+1,1+1,1}+\delta_{N\ge2}\,\ydiagram{1+2,1},\ydiagram1)+\text{c.c}
\end{equation}

\item $s=2$. The operators are
\begin{equation}
[q^2\bar q^0]|\chi_{\boldsymbol1}|^2=(\boldsymbol1+ 2\,\ydiagram{1+1,1} + \ydiagram{1+1,1+1,1,1}+\delta_{N\ge2}\,\ydiagram{2+2,2},\boldsymbol1)
\end{equation}

\end{itemize}

\paragraph{Scaling dimension $\boldsymbol{\Delta_\text{UV}=3}$.}
\begin{itemize}
\item $s=0$. The operators are
\begin{equation}
\hspace*{-1cm}\begin{aligned}
[q^0&\bar q^0]\bigl(|\chi_{\ydiagram{1,1,1}}|^2+|\chi_{\ydiagram{1+1,1+1,1}}|^2+|\chi_{\ydiagram{2,1}}|^2+|\chi_{\ydiagram3}|^2+|\chi_{\ydiagram{1+2,1}}|^2+\text{c.c.}\bigr)=\\
&=(\ydiagram{1,1,1},\ydiagram{1,1,1})+\delta_{N\ge2}\biggl(\!(\ydiagram{1+1,1+1,1},\ydiagram{1+1,1+1,1})+(\ydiagram{2,1},\ydiagram{2,1})\!\biggr)+\delta_{N\ge3}\biggl(\!(\ydiagram3,\ydiagram3)+(\ydiagram{1+2,1},\ydiagram{1+2,1})\!\biggr)+\text{c.c.}\\
[q^1&\bar q^1]|\chi_{\ydiagram1}|^2+\text{c.c.}=(\ydiagram1+\ydiagram{1+1,1+1,1}+\delta_{N\ge2}\,\ydiagram{1+2,1},\ydiagram1+\ydiagram{1+1,1+1,1}+\delta_{N\ge2}\,\ydiagram{1+2,1})+\text{c.c.}
\end{aligned}
\end{equation}
\item $s=1$. The operators are
\begin{equation}
\hspace*{-1.5cm}\begin{aligned}
[q^1\bar q^0]\bigl(|\chi_{\ydiagram{1,1}}|^2+|\chi_{\ydiagram2}|^2+\text{c.c.}\bigr)&=(\ydiagram{1,1} +\ydiagram2 +\ydiagram{1+1,1+1,1+1,1}+\delta_{N\ge2}\,\ydiagram{1+2,1+1,1},\ydiagram{1,1})\\
&+\delta_{N\ge2}(\,\ydiagram{1,1} +\ydiagram2+\ydiagram{1+2,1+1,1}+\delta_{N\ge3}\,\ydiagram{1+3,1},\ydiagram2)+\text{c.c}\\
[q^1\bar q^0]|\chi_{\ydiagram{1+1,1}}|^2&=\delta_{N\ge2}(\boldsymbol1+2\,\ydiagram{1+1,1}+\ydiagram{1+1,1+1,1,1}+\ydiagram{1+2,1,1}+\ydiagram{2+1,2+1,2}+\delta_{N\ge3}\,\ydiagram{2+2,2},\ydiagram{1+1,1})\\
[q^2\bar q^1]|\chi_{\boldsymbol1}|^2&=(\boldsymbol1+ 2\,\ydiagram{1+1,1} + \ydiagram{1+1,1+1,1,1}+\delta_{N\ge2}\,\ydiagram{2+2,2},\ydiagram{1+1,1})
\end{aligned}
\end{equation}
\item $s=2$. The operators are
\begin{equation}
\hspace*{-1.5cm}\begin{aligned}
[q^2\bar q^0]|\chi_{\ydiagram1}|^2+\text{c.c.}&=(2\,\ydiagram1 + 2\,\ydiagram{1+1,1+1,1} + \ydiagram{1+2,1} + \ydiagram{1+1,1+1,1+1,1,1}
+\delta_{N\ge2}\bigl(\ydiagram1 + \ydiagram{1+1,1+1,1} + 2\, \ydiagram{1+2,1} +\ydiagram{1+2,1+1,1,1} +\ydiagram{2+2,2+1,2}\,\bigr)\\
&+\delta_{N\ge3}\, \ydiagram{2+3,2},\ydiagram1)+\text{c.c.}
\end{aligned}
\end{equation}
\item $s=3$. The operators are $[q^3\bar q^0]|\chi_{\boldsymbol1}|^2$, which we do not compute here.

\end{itemize}

Unsurprisingly, this classification agrees with the ultraviolet classification from before. Comparing quantum numbers, we get the following map of operators
\begin{equation}
\begin{aligned}
1&\rightsquigarrow(\boldsymbol1,\boldsymbol1)\\
\psi_\ell\psi^\dagger_r&\rightsquigarrow(\ydiagram1,\ydiagram1)\\
\psi_\ell\psi_\ell^\dagger&\rightsquigarrow \mathcal J_{-1}(\boldsymbol1,\boldsymbol1)\\
\psi^2_\ell\psi^{\dagger2}_r&\rightsquigarrow(\ydiagram2,\ydiagram2)+(\ydiagram{1,1},\ydiagram{1,1})\\
\psi_\ell\psi^\dagger_\ell\psi_r\psi^\dagger_r&\rightsquigarrow (\ydiagram{1+1,1},\ydiagram{1+1,1})+\mathcal J_{-1}\bar{\mathcal J}_{-1}(\boldsymbol1,\boldsymbol1)\\
\psi^2_\ell\psi_\ell^\dagger\psi^\dagger_r+D_+\psi_\ell\psi^\dagger_r&\rightsquigarrow \mathcal J_{-1}(\ydiagram1,\ydiagram1)\\
\psi^2_\ell\psi^{\dagger2}_\ell+D_+\psi_\ell\psi_\ell^\dagger&\rightsquigarrow \mathcal J_{-2}(\boldsymbol1,\boldsymbol1)+\mathcal J_{-1}^2(\boldsymbol1,\boldsymbol1)\\
\psi^3_\ell\psi^{\dagger3}_r&\rightsquigarrow(\ydiagram{1,1,1},\ydiagram{1,1,1})+(\ydiagram{2,1},\ydiagram{2,1})+(\ydiagram3,\ydiagram3)\\
\psi^2_\ell\psi^\dagger_\ell\psi_r\psi^{\dagger2}_r+D_+\psi^\dagger_\ell D_-\psi_r+\psi^2_\ell\psi^\dagger_\ell D_-\psi^\dagger_r&\rightsquigarrow(\ydiagram{1+1,1+1,1},\ydiagram{1+1,1+1,1})+(\ydiagram{1+2,1},\ydiagram{1+2,1})+\mathcal J_{-1}\bar{\mathcal J}_{-1}(\ydiagram1,\ydiagram1)\\
\psi^3_\ell\psi_\ell^\dagger\psi_r^{\dagger2}+\psi_\ell D_+\psi_\ell\psi^{\dagger2}_r&\rightsquigarrow \mathcal J_{-1}(\ydiagram{1,1},\ydiagram{1,1})+\mathcal J_{-1}(\ydiagram2,\ydiagram2)\\
\psi^2_\ell\psi_\ell^{\dagger2}\psi_r\psi_r^\dagger+\psi_\ell D_+\psi_\ell^\dagger\psi_r\psi_r^\dagger&\rightsquigarrow \mathcal J_{-1}(\ydiagram{1+1,1},\ydiagram{1+1,1})+\mathcal J_{-2}\bar{\mathcal J}_{-1}(\boldsymbol1,\boldsymbol1)+\mathcal J_{-1}^2\bar{\mathcal J}_{-1}(\boldsymbol1,\boldsymbol1)\\
\psi_\ell^3\psi_\ell^{\dagger2}\psi^\dagger_r+\psi_\ell\psi_\ell^\dagger D_+\psi_\ell\psi^\dagger_r&\rightsquigarrow \mathcal J_{-2}\mathcal J_{-1}(\ydiagram1,\ydiagram1)\\
\psi_\ell^2D_+\psi^\dagger_\ell\psi^\dagger_r+D^2_+\psi_\ell\psi^\dagger_r&\rightsquigarrow \mathcal J_{-3}(\ydiagram1,\ydiagram1)\\
&\ \ \vdots
\end{aligned}
\end{equation}
where $\mathcal J_{-n}$ are the flavor currents (i.e., they are the ladder operators with respect to $SU(N_F)_N$).

We remark that, for low values of the level $N$, some of the operators above are actually null and thus not part of the spectrum; they are to be removed from the list. This is indicated in~\eqref{ap:su_chars} via a Kronecker delta. This vanishing of operators for low values of $N$ has a nice interpretation in the original UV variables: it simply indicates that the corresponding fermion contraction is zero by Fermi statistics. Consider, for example, the operators $\psi_\ell^2\psi_r^{\dagger2}$ at $(\Delta_\text{UV},s)=(2,0)$; these map in the IR to the primaries $(\ydiagram2,\ydiagram2)+(\ydiagram{1,1},\ydiagram{1,1})$. The former $(\ydiagram2,\ydiagram2)$ is a singular operator at $N=1$, cf.~\eqref{eq:sing_sym}. In the UV theory, this singular operator corresponds to the contraction where $\psi_\ell^2$ is symmetric under flavor and anti-symmetric under color (and similarly for $\psi^{\dagger2}_r$). Clearly, for $N=1$, this contraction vanishes, since we cannot anti-symmetrize two color indices that can only take one value. One can check that the rest of singular WZW operators all follow the same pattern, where the decoupling in the UV is always manifest and follows by anti-symmetrization of color indices. See also~\cite{He:2021xvg} for a similar discussion.

\subsection{Generating function}\label{ap:gen_fun}

There exists another method to classify the operators of a gauge theory modulo gauge currents. We describe this for $U(N)$ with $N_F$ fundamental quarks, although similar considerations hold for other groups. The idea is to write down a generating function for these operators.

The space of operators of QCD is generated by $\psi_\ell,\psi_r$ and their complex conjugates, and their derivatives. The quantum numbers of these generators are given by~\eqref{eq:ap_q_numbers}. We introduce fugacities for these quantum numbers as follows: $t$ for the scaling dimension, $\eta$ for the spin, $g$ for the gauge group $U(N)$, and $\mu_\ell,\mu_r$ for the $SU(N_F)_\ell\times SU(N_F)_r$ flavor symmetry. We also denote $q=t\eta$ and $\bar q=t/\eta$. With this, the ``single letter'' partition function is
\begin{equation}
f(q,\bar q,g,\mu_\ell,\mu_r)=\frac{q^{1/2}}{1-q}\tr(g)\tr(\mu_\ell)+\frac{\bar q^{1/2}}{1-\bar q}\tr(g)\tr(\mu_r)^\dagger+\text{c.c}
\end{equation}
Furthermore, we have a constraint $J_\ell=J_r=0$, whose partition function is
\begin{equation}
f_\text{c}(q,\bar q,g)=\biggl(\frac{q}{1-q}+\frac{\bar q}{1-\bar q}\biggr)\tr(g)\tr(g)^\dagger
\end{equation}
Finally, the generating function of QCD operators modulo gauge currents reads
\begin{equation}
Z(q,\bar q,\mu_\ell,\mu_r)=\int_G \exp\biggl[\,\sum_{n\ge1}\frac{(-1)^{n+1}f(q^n,\bar q^n,g^n,\mu_\ell^n,\mu_r^n)-f_\text{c}(q,\bar q,g)}{n}\biggr]
\end{equation}
where the integration is over $G=U(N)$ over the Haar measure (this simply projects to the space of gauge singlets). The factor of $(-1)^{n+1}$ is due to the fermionic nature of the generators, while the negative sign in front of $f_\text{c}$ is there because it imposes a constraint on the generators. This generating function is known as a \emph{plethystic exponential}. A known fact about this exponential is that, if $f(x)=\sum_k a_k x^k$, then $\exp(\sum_kf(x^k)/k)\equiv \prod_k(1-x^k)^{-a_k}$; this also allows us to write the generating function above in terms of the Pochhammer symbol $(a;q):=\prod_k(1-aq^k)$, namely
\begin{equation}
%Z(q,\bar q,\mu_\ell,\mu_r)=\int_G \prod_{\substack{w\in\ydiagram1\\\hat w\in\ydiagram1}}(-g^w\mu_\ell^{\hat w}q^{1/2};q)(-g^{-w}\mu_\ell^{-\hat w}q^{1/2};q)\times\prod_{w\in\text{adj}}(g^wq;q)\times\begin{pmatrix}\ell\leftrightarrow r\\q\leftrightarrow\bar q\end{pmatrix}
Z(q,\bar q,\mu_\ell,\mu_r)=\int_G\ \prod_{\substack{w\in\ydiagram1\\\hat w\in\ydiagram1}}|(-g^w\mu_\ell^{\hat w}q^{1/2};q)|^2\times\prod_{w\in\text{adj}}(g^wq;q)\times\begin{pmatrix}\ell\leftrightarrow r\\q\leftrightarrow\bar q\end{pmatrix}
\end{equation}
where $w,\hat w$ denote weights of $U(N)$ and $SU(N_F)$, respectively. By expanding this generating function in powers of $q,\bar q$ we obtain the list of all infrared operators of QCD, labeled by their quantum numbers.

Since the infrared CFT of this QCD theory is the WZW model $SU(N_F)_N$, we can adapt the formulas above to obtain the characters of $SU(N_F)_N$. This is done simply by projecting to a given $G$-representation before performing the integral. In particular, we find
\begin{equation}
\chi_\lambda(q,\mu)=q^{h-c/24-r/2}\int_G\chi_{\bar\lambda^t}(g)\prod_{\substack{w\in\ydiagram1\\ \hat w\in \ydiagram1}}|(-g^w\mu^{\hat w}q^{1/2};q)|^2\prod_{w\in \text{adj}}(g^w q;q)
\end{equation}
where $\lambda$ is an $SU(N_F)_N$ integrable representation, and $\bar\lambda^t$ is the conjugate and transposed representation of $U(N)$ (recall that level-rank duality interchanges a representation with its transpose); and $r$ denotes the rank of the representation (the number of boxes in its Young diagram). In this equation, $\chi_\lambda(q,\mu)$ is an affine character of $SU(N_F)_N$ while $\chi_{\bar\lambda^t}(g)$ is a regular (finite) character of $U(N)$. To the best of our knowledge, this formula for $\chi_\lambda(q,\mu)$ is new.

\clearpage
\printbibliography
\end{document}